\pdfoutput=1
\documentclass[runningheads,10.5pt]{llncs} 

\usepackage[hmargin=4.7cm,top=3cm,bottom=3cm]{geometry}
\usepackage{times}
\usepackage{inconsolata}

\usepackage[utf8]{inputenc}
\usepackage[T1]{fontenc}
\usepackage[american]{babel}

\usepackage{amssymb}
\usepackage{stmaryrd}
\usepackage{mathtools}
\usepackage{xcolor}
\usepackage{paralist}
\usepackage{enumitem}
\usepackage{hyperref}
\usepackage{xspace}
\usepackage{xifthen}
\usepackage{stackengine}
\usepackage{booktabs}
\usepackage{wrapfig}
\usepackage{subcaption}
\usepackage{float}
\usepackage{tabularx}
\usepackage{comment}

\usepackage{marvosym}
\usepackage{orcidlink}

\usepackage{wasysym}
\usepackage{empheq}

\usepackage{stmaryrd}
\usepackage{cite}

\usepackage{cleveref}
\crefname{section}{Sect.}{Sect.}
\Crefname{section}{Section}{Section}
\crefname{figure}{Fig.}{Fig.}
\Crefname{figure}{Figure}{Figure}

\usepackage{listings}
\lstdefinelanguage{SPL}{
  morekeywords={acc, method, struct,if,else,returns,procedure,requires,ensures,:=,var,
    new,old,free,implicit,modifies,call,locals,assume,assert,choose,havoc,ghost,
    predicate,function,invariant,while, return,atomic, split, type, field, result,
    define, datatype, domain, axiom, val},
  deletekeywords={union,int},
  numbers=left,
  xleftmargin=2em,
  escapeinside={@}{@},
  numberstyle=\tiny,
  basicstyle=\footnotesize\ttfamily,
  columns=flexible,
  morecomment=[s][\color{green!60!black}]{/*}{*/},
  morecomment=[l][\color{green!60!black}]{//},
  moredelim=**[is][\color{purple}]{|<}{>|},
  mathescape=true,
}

\lstdefinestyle{realcode}{ 
  language=Java,
  basicstyle=\fontsize{7}{11}\selectfont\ttfamily,
  basewidth = {0.5em,0.5em},
  keywordstyle=\color{teal}\ttfamily,
  commentstyle=\color{gray}\ttfamily,
  morekeywords={bool,shared,atomic,assert,struct,NULL,default,auto,lock,unlock,is},
  tabsize=3,
  numberbychapter=false,
  keepspaces=true,
  numbers=left,
  numberstyle=\fontsize{5.75}{12}\selectfont,
  numbersep=7pt,
  firstnumber=last,
  xleftmargin=5pt,
  moredelim=[is][\underbar]{`}{'},
}
\lstdefinestyle{proof}{
  language=Java,
  basicstyle=\linespread{1.15}\footnotesize\ttfamily,
  commentstyle=\color{gray}\ttfamily,
  tabsize=2,
  numberbychapter=false,
  keepspaces=true,
  numbers=left,
  numberstyle=\scriptsize,
  numbersep=7pt,
  firstnumber=last,
  mathescape=true,
  keywords={},
  xleftmargin=3pt,
}
\lstdefinestyle{inline}{
  basicstyle=\relscale{.9}\ttfamily,
  keywords={},
}
\usepackage{relsize}
\newcommand{\code}[2][]{\lstinline[style=inline,#1]!#2!}

\newtheorem{fact}[theorem]{Fact}

\usepackage{tikz}
\usetikzlibrary{matrix,arrows,positioning,calc,fit,backgrounds}

\tikzset{%
  array/.style={matrix of nodes,nodes={draw, minimum size=5mm, anchor=center},column sep=-\pgflinewidth, row sep=-\pgflinewidth, nodes in empty cells,anchor=center},
  ptr/.style={*->, shorten <=-(1.8pt+1.4\pgflinewidth)},
  edge/.style={->, thick},
  dedge/.style={<->, dashed},
  fedge/.style={->, dashed},
  unode/.style={circle, draw=black, thick, minimum size=8mm},
  mnode/.style={circle, draw=black, thick, fill=gray!20, minimum size=8mm, font=\scriptsize},
  stackVar/.style={circle, fill=none, inner sep=0pt, minimum size=8mm, font=\normalsize},
  gnode/.style={circle, draw=black, thick, minimum size=8mm},
  pnode/.style={circle, draw=black, thick, minimum size=8mm},
  rnode/.style={draw=black, thick, minimum size=8mm},
  lbl/.style={circle, fill=none, inner sep=0pt, minimum size=8mm},
  dnode/.style={circle, draw=black, thick, dotted, minimum size=8mm},
  inflow/.style={circle, fill=none, inner sep=0pt, minimum size=5mm, font=\normalsize},
  phantomNode/.style={circle, fill=none, inner sep=0pt, minimum
    size=0pt}
}

\usetikzlibrary{decorations}
\tikzset{snake it/.style={decorate, decoration={snake, amplitude=.25mm, segment length=1mm, post length=1.2mm}}}

\usepackage{pgfplots}
\usetikzlibrary{patterns}
\pgfplotsset{compat=1.7}
\pgfplotsset{
    /pgfplots/area legend/.style={
        /pgfplots/legend image code/.code={
            \fill[##1] (0cm,0.6em) rectangle (0.9em,-0.3em);
}, },
}

\definecolor{colorUnklar}{RGB}{150,30,190}
\definecolor{colorTodo}{RGB}{200,30,30}
\definecolor{colorHighlight}{RGB}{30,30,200}
\definecolor{colorMyGreen}{RGB}{80,170,0}
\definecolor{colorMyRed}{RGB}{180,20,20}
\definecolor{colorMyBlue}{RGB}{40,40,220}
\definecolor{colorMyPink}{RGB}{220,40,220}
\newcommand{\makePurple}[1]{\textcolor{purple}{#1}}
\newcommand{\makeBlue}[1]{\textcolor{colorMyBlue}{#1}}

\newcommand{\makePink}[1]{\textcolor{colorMyPink}{#1}}

\newcommand{\makeTeal}[1]{\textcolor{teal}{#1}}

\usepackage{mathpartir}
\makeatletter
\newcommand{\rulelabel}[2]{%
   \protected@write \@auxout {}{\string \newlabel {#1}{{#2}{\thepage}{#2}{#1}{}} }%
   \hypertarget{#1}{}
}
\makeatother
\newcommand{\mkrulelabel}[1]{\rulelabel{rule:#1}{\textsc{(#1)}}}
\newcommand{\infrule}[3][]{\ifthenelse{\equal{#1}{}}{\inferrule{#2}{#3}}{\inferrule[\textsc{(#1)}\mkrulelabel{#1}]{#2}{#3}}}

\usepackage{stackengine}
\stackMath

\newcommand{\smartparagraph}[1]{\subsubsection*{#1}}

\def\prallspacing{\mskip 2mu plus 2mu minus 3mu}
\newcommand{\prall}[1]{{\prallspacing#1\prallspacing}}

\renewcommand{\emptyset}{\varnothing}

\newcommand{\set}[1]{\{\,#1\,\}}

\newcommand{\setcond}[2]{\set{#1\;\mid\;#2}}

\newcommand{\powerset}[1]{\mathbb{P}(#1)}
\newcommand{\nat}{\mathbb{N}}
\newcommand{\ZZ}{\mathbb{Z}}
\newcommand{\BB}{\mathbb{B}}

\newcommand{\defeq}{\triangleq}

\newcommand{\setstates}{\Sigma}

\newcommand{\discup}{\uplus}

\newcommand{\statemult}{\mathop{*}}

\newcommand{\statemultdef}{\mathop{\#}}

\newcommand{\mstar}{\mathop{*}}
\DeclareMathOperator*{\bigmstar}{\scalerel*{\ast}{\sum}}
\newcommand{\emp}{\mathsf{emp}}

\newcommand{\aform}{A}

\newcommand{\afform}{F}

\newcommand{\join}{\sqcup}
\newcommand{\bigjoin}{\bigsqcup}

\newcommand{\givename}[1]{\framebox{\textnormal{\textsc{#1}}}\,}

\newcommand{\annot}[2][]{\makeTeal{\ifthenelse{\equal{#1}{}}{}{\givename{#1}}\left\{\,\begin{aligned}#2\end{aligned}\,\right\}}}
\newcommand{\MC}[2][]{\makeBlue{\ifthenelse{\equal{#1}{}}{#2}{\underbracket{#2}_{\text{\textnormal{#1}}}}}}
\newcommand{\MCS}[2][]{\makeBlue{\ifthenelse{\equal{#1}{}}{#2}{\smash{\underbracket{#2}_{\text{\textnormal{#1}}}}}}}
\newcommand{\MF}[2][]{\makePurple{\ifthenelse{\equal{#1}{}}{#2}{\underbracket{#2}_{\text{\textnormal{#1}}}}}}
\newcommand{\MH}[2][]{\makePink{\ifthenelse{\equal{#1}{}}{#2}{\underbracket{#2}_{\text{\textnormal{#1}}}}}}

\newcommand{\amonoid}{\mathbb{M}}
\newcommand{\monadd}{+}
\newcommand{\monbigadd}{\sum}
\newcommand{\monunit}{0}
\newcommand{\amonval}{\mathit{m}}
\newcommand{\amonvalp}{\mathit{n}}
\newcommand{\amonvalpp}{\mathit{o}}

\newcommand{\setnodes}{\mathit{X}}
\newcommand{\setnodesp}{\mathit{Y}}
\newcommand{\setnodespp}{\mathit{Z}}
\newcommand{\anode}{\mathit{x}}
\newcommand{\anodep}{\mathit{y}}
\newcommand{\anodepp}{\mathit{z}}

\newcommand{\edges}{\mathit{E}}

\newcommand{\edgesatof}[3]{\edges_{(#1, #2)}(#3)}
\newcommand{\rhs}{\mathit{rhs}}

\newcommand{\rhsatof}[2]{\rhs_{#1}(#2)}
\newcommand{\aflowconstraint}{\mathit{h}}
\newcommand{\setflowconstraints}{\mathit{FG}}
\newcommand{\fval}{\mathit{flow}}
\newcommand{\fvalof}[1]{\fval(#1)}
\newcommand{\inflow}{\mathit{in}}
\newcommand{\inflowof}[2][]{\inflow_{#1}(#2)}
\newcommand{\outflow}{\mathit{out}}
\newcommand{\outflowof}[2][]{\outflow_{#1}(#2)}

\newcommand{\transformerof}[1]{\mathit{tf}(#1)}

\newcommand{\lfpof}[1]{\mathit{lfp}.#1}
\newcommand{\pairingof}[2]{\langle #1, #2\rangle}

\newcommand{\myid}{\mathit{id}}
\newcommand{\myidof}[1]{\myid_{#1}}

\newcommand{\cval}{\mathit{cval}}

\newcommand{\aflowgraph}{\mathit{h}}

\newcommand{\setvars}{\mathit{Y}}

\newcommand{\pointsto}[1]{\mapsto_{#1}}

\usepackage{scalerel}

\newcommand{\flow}{\mathit{fl}}

\newcommand{\somenodes}{\mathit{N}}

\newcommand{\atoolname}[1]{\code{#1}\xspace}

\newcommand{\plankton}{\atoolname{plankton}}
\newcommand{\krill}{\atoolname{krill}}

\usepackage{pifont}

\newcommand{\twocases}[3]{\begin{cases}#2&\text{if }#1\\#3&\text{otherwise}\end{cases}}

\newcommand{\restrictto}[2]{#1|_{#2}}

\newcommand{\contfunof}[1]{\mathit{ContFun}(#1)}

\newcommand{\distfunof}[1]{\mathit{DistFun}(#1)}
\newcommand{\distdecfunof}[1]{\mathit{DistDecFun}(#1)}
\newcommand{\achain}{K}

\newcommand{\firstof}[1]{\mathit{first}(#1)}
\newcommand{\lastof}[1]{\mathit{last}(#1)}
\newcommand{\setpathof}[3]{\mathit{Paths}(#1, #2, #3)}
\newcommand{\setpathp}{\mathit{Q}}
\newcommand{\setpath}{\mathit{P}}
\newcommand{\setsimplepathof}[3]{\mathit{SimplePaths}(#1, #2, #3)}
\newcommand{\apath}{p}
\newcommand{\apathp}{q}
\newcommand{\ahvar}{x}

\newcommand{\val}{\sigma}
\newcommand{\valof}[1]{\val(#1)}
\newcommand{\atfun}{f}
\newcommand{\atfunof}[1]{\atfun(#1)}

\newcommand{\aninflow}{\mathit{in}}

\newcommand{\outof}[2]{\mathit{odif}_{#1, #2}}
\newcommand{\outpof}[2]{\mathit{odif}^{\#}(#1, #2)}

\newcommand{\tfailof}[3]{\mathit{tfail}_{#1, #2}(#3)}

\newcommand{\extend}[2]{\mathit{ext}_{#1, #2}}
\newcommand{\extendof}[3]{\extend{#1}{#2}(#3)}
\newcommand{\afootprint}{\mathit{F}}

\newcommand{\setflowfootprintof}[2]{\mathit{FFP}(#1, #2)}

\DeclareMathOperator{\ctxequiv}{\ =_{\mathit{ctx}}\ }
\newcommand{\pc}{\cdot}

\begin{document}
	\title{Make flows small again: revisiting the flow framework}
	\author{%
		Roland Meyer\inst{1}\,\orcidlink{0000-0001-8495-671X}
		\and
		Thomas Wies\inst{2}\,\orcidlink{0000-0003-4051-5968}
		\and
		Sebastian Wolff\inst{2}\textsuperscript{(\Letter)}\,\orcidlink{0000-0002-3974-7713}
	}
	\authorrunning{Roland Meyer, Thomas Wies, and Sebastian Wolff}
	\institute{
		TU Braunschweig, Braunschweig, Germany, \email{roland.meyer@tu-bs.de}
		\and
		New York University, New York, USA, \email{\{wies,sebastian.wolff\}@cs.nyu.edu}
	}
	\maketitle

\begin{abstract}
  We present a new flow framework for separation logic reasoning about programs that manipulate general graphs. The framework overcomes problems in earlier developments: it is based on standard fixed point theory, guarantees least flows, rules out vanishing flows, and has an easy to understand notion of footprint as needed for soundness of the frame rule. In addition, we present algorithms for automating the frame rule, which we evaluate on graph updates extracted from linearizability proofs for concurrent data structures. The evaluation demonstrates that our algorithms help to automate key aspects of these proofs that have previously relied on user guidance or heuristics.
  \par\addvspace\baselineskip
  \noindent\textbf{Conference Version:}\enspace\ignorespaces
  Meyer, R., Wies, T., Wolff, S. (2023).
  Make flows small again: revisiting the flow framework.
  In: Sankaranarayanan, S., Sharygina, N. (eds)
  Tools and Algorithms for the Construction and Analysis of Systems. TACAS 2023.
  Lecture Notes in Computer Science, vol 13993. Springer.
  \url{https://doi.org/10.1007/978-3-031-30823-9_32}
\end{abstract}


\section{Introduction}
\label{sec:intro}

The flow framework~\cite{DBLP:journals/pacmpl/KrishnaSW18,DBLP:conf/esop/KrishnaSW20} is an abstraction mechanism based on separation logic~\cite{DBLP:conf/csl/OHearnRY01,DBLP:conf/lics/Reynolds02, DBLP:conf/lics/CalcagnoOY07} that enables reasoning about global inductive invariants of general graphs in a local manner.
The framework has proved useful to verify intricate algorithms that are difficult to handle by other techniques, such as the Priority Inheritance Protocol, object-oriented design patterns, and complex concurrent data structures~\cite{DBLP:conf/esop/KrishnaSW20, DBLP:conf/pldi/KrishnaPSW20, DBLP:journals/pacmpl/PatelKSW21,oopsla}.
However, these efforts have also exposed some rough corners in the underlying meta theory that either limit expressivity or automation. 
In this paper, we propose a new meta theory for the flow framework that aims to strike a balance between these conflicting requirements. In addition, we present algorithms that aid proof automation.

\smartparagraph{Background.}
The central notion of the flow framework is that of a \emph{flow}.
Given a commutative monoid $(\amonoid,\monadd,\monunit)$ (e.g. natural numbers with addition), and a graph with nodes $\setnodes$ and an \emph{edge function} $\edges \colon \setnodes \times \setnodes \to \amonoid \to \amonoid$, a flow is a function $\flow \colon \setnodes \to \amonoid$ that satisfies the \emph{flow equation}:
\[
  \forall \anode \in \setnodes.\quad \flow(\anode) = \inflow_\anode \monadd {\textstyle \monbigadd_{\anodep \in \setnodes}}~ \edgesatof{\anodep}{\anode}{\flow(\anodep)}\enspace.
\]
That is, $\flow$ is a fixed point of the function that assigns every node $\anode$ an initial value $\inflow_\anode \in \amonoid$, its \emph{inflow}, and then propagates these values through the graph according to the edge function.
This is akin to a forward data flow analysis where the monoid operation $\monadd$ is used as the join.
By choosing an appropriate flow monoid, inflow, and edge function, one can express inductive properties of graphs (reachability, sortedness, etc.) in terms of conditions that refer only to each node's flow value $\flow(x)$.

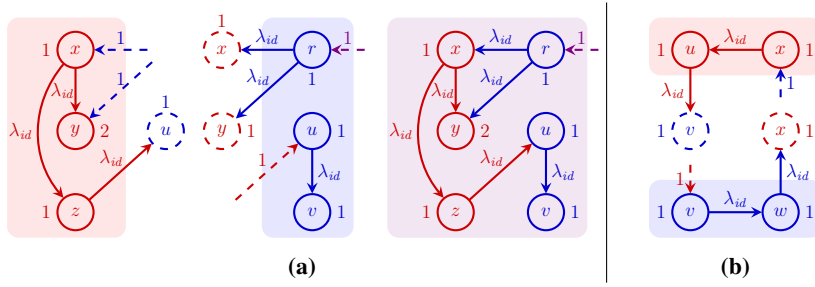
\begin{figure}
  \centering
  \begin{subfigure}[t]{.65\textwidth}
  \begin{tikzpicture}[>=stealth, scale=0.6, every node/.style={scale=0.6,font=\large}]

    \def\xsep{2}
    \def\ysep{-1.8}
    \def\xshift{2.5}
    \def\sd{-.25}
    \def\s{.4}
    \def\t{.8}

    \node[unode,red!80!black] (x) {$x$};
    \node[lbl,red!80!black] (flowx) at ($(x) + (-.65, 0)$) {$ 1$};
    \node[stackVar] (r) at ($(x) + (\xsep, 0)$) {};
    \node[unode,red!80!black] (y) at ($(x) + (0, \ysep)$) {$y$};
    \node[lbl,red!80!black] (flowy) at ($(y) + (.65, 0)$) {$ 2$};
    \node[unode, blue!80!black, dashed] (u) at ($(r) + (0, \ysep)$) {$u$};
    \node[lbl, blue!80!black] (flowu) at ($(u) + (0, .65)$) {$ 1$};
    \node[unode,red!80!black] (z) at ($(y) + (0, \ysep)$) {$z$};
    \node[lbl,red!80!black] (flowz) at ($(z) + (-.65, 0)$) {$ 1$};

    \draw[edge, blue!80!black, dashed] (r) to node[above]{$ 1$} (x);
    \draw[edge, blue!80!black, dashed] (r) to node[above]{$ 1$} (y);
    \draw[edge,red!80!black] (x) to node[left,xshift=1mm] {$\lambda_{\mathit{id}}$} (y);
    \draw[edge,red!80!black] (x) to[bend right=40] node[left] {$\lambda_{\mathit{id}}$} (z);
    \draw[edge,red!80!black] (z) to node[above,xshift=-2mm] {$\lambda_{\mathit{id}}$} (u);

    \begin{scope}[on background layer] 
      \draw[draw=red!10, rounded corners, thick, fill=red!10]
      ($(x.north west) + (-\t-.4, \s)$)
      -- ($(x.north east) + (\t, \s)$)
      -- ($(z.south east) + (\t, \sd)$)
      -- ($(z.south west) + (-\t-.4, \sd)$) -- cycle;
    \end{scope}
  \end{tikzpicture}\;\;
  \begin{tikzpicture}[>=stealth, scale=0.6, every node/.style={scale=0.6,font=\large}]

    \def\xsep{2}
    \def\ysep{-1.8}
    \def\xshift{2.5}
    \def\sd{-.25}
    \def\s{.4}
    \def\t{.8}

    \node[unode, red!80!black, dashed] (x) {$x$};
    \node[lbl,red!80!black] (flowx) at ($(x) + (0,.65)$) {$ 1$};
    \node[unode, blue!80!black] (r) at ($(x) + (\xsep, 0)$) {$r$};
    \node[unode,inner sep=0pt,minimum size=0pt] (rin) at ($(r) + (\xsep/1.7,0)$) {};
    \node[lbl, blue!80!black] (flowr) at ($(r) + (0, -.65)$) {$ 1$};
    \node[unode,red!80!black, dashed] (y) at ($(x) + (0, \ysep)$) {$y$};
    \node[lbl,red!80!black] (flowy) at ($(y) + (.65, 0)$) {$ 1$};
    \node[unode, blue!80!black] (u) at ($(r) + (0, \ysep)$) {$u$};
    \node[lbl, blue!80!black] (flowu) at ($(u) + (.65,0)$) {$ 1$};
    \node[stackVar] (z) at ($(y) + (0, \ysep)$) {};
    \node[unode, blue!80!black] (v) at ($(u) + (0, \ysep)$) {$v$};
    \node[lbl, blue!80!black] (flowv) at ($(v) + (.65, 0)$) {$ 1$};

    \draw[edge, blue!80!black] (r) to node[above]{$\lambda_{\mathit{id}}$} (x);
    \draw[edge, blue!80!black] (r) to node[above,xshift=-2mm]{$\lambda_{\mathit{id}}$} (y);
    \draw[edge, blue!80!black] (u) to node[right] {$\lambda_{\mathit{id}}$} (v);
    \draw[edge, red!80!black, dashed] (z) to node[above,xshift=-1mm] {$ 1$} (u);
    \draw[edge, violet, dashed] (rin) to node[above] {$ 1$} (r);

    \begin{scope}[on background layer] 
      \draw[draw=blue!10, rounded corners, thick, fill=blue!10]
      ($(r.north west) + (-\t, \s)$)
      -- ($(r.north east) + (\t-.2, \s)$)
      -- ($(v.south east) + (\t-.2, \sd)$)
      -- ($(v.south west) + (-\t, \sd)$) -- cycle;
    \end{scope}
  \end{tikzpicture}\;\;
  \begin{tikzpicture}[>=stealth, scale=0.6, every node/.style={scale=0.6,font=\large}]

    \def\xsep{2}
    \def\ysep{-1.8}
    \def\xshift{2.5}
    \def\sd{-.25}
    \def\s{.4}
    \def\t{.8}

    \node[unode,red!80!black] (x) {$x$};
    \node[lbl,red!80!black] (flowx) at ($(x) + (-.65, 0)$) {$ 1$};
    \node[unode, blue!80!black] (r) at ($(x) + (\xsep, 0)$) {$r$};
    \node[unode,inner sep=0pt,minimum size=0pt] (rin) at ($(r) + (\xsep/1.7,0)$) {};
    \node[lbl, blue!80!black] (flowr) at ($(r) + (0, -.65)$) {$ 1$};
    \node[unode,red!80!black] (y) at ($(x) + (0, \ysep)$) {$y$};
    \node[lbl,red!80!black] (flowy) at ($(y) + (.65, 0)$) {$ 2$};
    \node[unode, blue!80!black] (u) at ($(r) + (0, \ysep)$) {$u$};
    \node[lbl, blue!80!black] (flowu) at ($(u) + (.65,0)$) {$ 1$};
    \node[unode,red!80!black] (z) at ($(y) + (0, \ysep)$) {$z$};
    \node[lbl,red!80!black] (flowz) at ($(z) + (-.65, 0)$) {$ 1$};
    \node[unode, blue!80!black] (v) at ($(u) + (0, \ysep)$) {$v$};
    \node[lbl, blue!80!black] (flowv) at ($(v) + (.65, 0)$) {$ 1$};

    \draw[edge, blue!80!black] (r) to node[above]{$\lambda_{\mathit{id}}$} (x);
    \draw[edge, blue!80!black] (r) to node[above,xshift=-2mm]{$\lambda_{\mathit{id}}$} (y);
    \draw[edge, blue!80!black] (u) to node[right] {$\lambda_{\mathit{id}}$} (v);
    \draw[edge,red!80!black] (x) to node[left,xshift=1mm] {$\lambda_{\mathit{id}}$} (y);
    \draw[edge,red!80!black] (x) to[bend right=40] node[left] {$\lambda_{\mathit{id}}$} (z);
    \draw[edge,red!80!black] (z) to node[above,xshift=-2mm] {$\lambda_{\mathit{id}}$} (u);
    \draw[edge, violet, dashed] (rin) to node[above] {$ 1$} (r);

    \begin{scope}[on background layer] 
      \draw[draw=violet!10, rounded corners, thick, fill=violet!10]
      ($(x.north west) + (-\t-.4, \s)$)
      -- ($(r.north east) + (\t-.2, \s)$)
      -- ($(v.south east) + (\t-.2, \sd)$)
      -- ($(z.south west) + (-\t-.4, \sd)$) -- cycle;
    \end{scope}
  \end{tikzpicture}
  \caption{\label{fig:flow-graph-comp}}
  \end{subfigure}
  \vrule
  \begin{subfigure}[t]{.28\linewidth}
    \centering
    \begin{tikzpicture}[>=stealth, scale=0.6, every node/.style={scale=0.6,font=\large}]

    \def\xsep{2}
    \def\ysep{-1.8}
    \def\xshift{2.5}
    \def\sd{-.25}
    \def\s{.4}
    \def\t{.8}

    \node[unode,red!80!black] (u) {$u$};
    \node[lbl,red!80!black] (flowu) at ($(u) + (-.65, 0)$) {$ 1$};
    \node[unode,red!80!black] (x) at ($(u) + (\xsep, 0)$) {$x$};
    \node[lbl,red!80!black] (flowx) at ($(x) + (.65, 0)$) {$ 1$};
    \node[unode, blue!80!black, dashed] (v) at ($(u) + (0, \ysep)$) {$v$};
    \node[lbl,blue!80!black] (flowv) at ($(v) + (-.65, 0)$) {$ 1$};
    \node[stackVar, inner sep=0pt,minimum size=0pt] (w) at ($(x) + (0, \ysep/1.7)$) {};
    \node[unode, red!80!black, dashed] (x1) at ($(x) + (0, \ysep)$) {$x$};
    \node[lbl,red!80!black] (flowx) at ($(x1) + (.65, 0)$) {$ 1$};
    \node[unode,blue!80!black] (v1) at ($(v) + (0, \ysep)$) {$v$};
    \node[stackVar, inner sep=0pt,minimum size=0pt] (u1) at ($(v1) - (0, \ysep/1.7)$) {};
    \node[lbl,blue!80!black] (flowv1) at ($(v1) + (-.65, 0)$) {$ 1$};
    \node[unode,blue!80!black] (w1) at ($(x1) + (0, \ysep)$) {$w$};
    \node[lbl,blue!80!black] (floww) at ($(w1) + (.65, 0)$) {$ 1$};

    \draw[edge,blue!80!black] (v1) to node[above]{$\lambda_{\mathit{id}}$} (w1);
    \draw[edge,red!80!black] (x) to node[above]{$\lambda_{\mathit{id}}$} (u);
    \draw[edge,red!80!black,dashed] (u1) to node[left]{$ 1$} (v1);
    \draw[edge,blue!80!black] (w1) to node[right]{$\lambda_{\mathit{id}}$} (x1);
    \draw[edge,blue!80!black,dashed] (w) to node[right]{$ 1$} (x);
    \draw[edge,red!80!black] (u) to node[left]{$\lambda_{\mathit{id}}$} (v);

    \begin{scope}[on background layer] 
      \draw[draw=blue!10, rounded corners, thick, fill=blue!10]
      ($(v1.north west) + (-\t+.2, \s)$)
      -- ($(w1.north east) + (\t-.2, \s)$)
      -- ($(w1.south east) + (\t-.2, \sd)$)
      -- ($(v1.south west) + (-\t+.2, \sd)$) -- cycle;
    \end{scope}
    \begin{scope}[on background layer] 
      \draw[draw=red!10, rounded corners, thick, fill=red!10]
      ($(u.north west) + (-\t+.2, \s)$)
      -- ($(x.north east) + (\t-.2, \s)$)
      -- ($(x.south east) + (\t-.2, \sd)$)
      -- ($(u.south west) + (-\t+.2, \sd)$) -- cycle;
    \end{scope}
    \end{tikzpicture}
    \caption{\label{fig:flow-graph-vanishing}}
  \end{subfigure}
  \caption{\textbf{(\subref*{fig:flow-graph-comp})} Two flow graphs $\aflowgraph_1$ with nodes $\aflowgraph_1.\setnodes \prall= \set{x,y,z}$ (left) and $\aflowgraph_2$ with nodes $\aflowgraph_2.\setnodes = \set{r,u,v}$ (center) for the flow monoid of natural numbers with addition. The edge label $\lambda_{\mathit{id}}$ stands for the identity function. Omitted edges are labeled by the constant $0$ function. Dashed edges represent the inflows. Nodes are labeled by their flow, respectively, outflow.
    The right side shows the composition $\aflowgraph = \aflowgraph_1 \mstar \aflowgraph_2$.
    \textbf{(\subref*{fig:flow-graph-vanishing})} Two flow graphs $\aflowgraph_1$ with $\aflowgraph_1.X = \set{u,x}$ (top) and $\aflowgraph_2$ with $\aflowgraph_2.X = \set{v,w}$ (bottom) whose composition is undefined due to vanishing flows.
    \label{fig:path-counting-flow}}
\end{figure}

A graph endowed with an inflow and associated flow is a \emph{flow graph}. 
An example flow graph $\aflowgraph$ is shown on the right-hand side of \cref{fig:flow-graph-comp}.
Here, the flow value $\flow(w)$ for a node $w$ counts the number of paths from  $r$ to $w$.
A flow graph can be partial and have edges to nodes outside of $\setnodes$ like the node $u$ for $\aflowgraph_1$ in \cref{fig:flow-graph-comp}.
If we include these nodes in the computation of the flow, then their flow values constitute the \emph{outflow} of the flow graph. For instance, the outflow of $\aflowgraph_1$ for $u$ is $1$.

Flow graphs are equipped with a notion of disjoint composition, $\aflowgraph = \aflowgraph_1 \mstar \aflowgraph_2$. An example is given in \cref{fig:flow-graph-comp}.
The composition is only defined if the union of the flows of $\aflowgraph_1$ and $\aflowgraph_2$ is again a flow of $\aflowgraph$. This may not always be the case. For instance, the inflows and outflows of $\aflowgraph_1$ and $\aflowgraph_2$ may be mutually incompatible such as $\aflowgraph_1$ sending outflow $2$ to $u$ whereas the inflow to $u$ in $\aflowgraph_2$ is only $1$.

Flow graph composition yields a \emph{separation algebra}.
That is, if we use flow graphs as an abstraction of program states (e.g., the heap), then we can use separation logic to reason locally about properties of programs that are expressed in terms of the induced flow graphs.
For example, suppose the program updates the flow graph $\aflowgraph$ in \cref{fig:flow-graph-comp} to a new flow graph $\aflowgraph'$ by inserting a new edge labeled $\lambda_{\mathit{id}}$ between the nodes $r$ and $u$. This increases the flow of $u$ and $v$ from $1$ to $2$.
We can break this update down as follows.
First, we decompose $\aflowgraph$ into $\aflowgraph_1$ and $\aflowgraph_2$.
Next, we obtain $\aflowgraph_2'$ from $\aflowgraph_2$ by inserting the edge and updating the flow of $u$ and $v$ to $2$.
Finally, we compose $\aflowgraph_2'$ again with $\aflowgraph_1$ to obtain $\aflowgraph'$.
Note that the composition $\aflowgraph_1 \mstar \aflowgraph_2'$ is still defined. This means that any property expressed over the flow in the $\aflowgraph_1$-portion of $\aflowgraph$ still holds in $\aflowgraph'$.
This is the well-known \emph{frame rule} of separation logic, instantiated for flow graphs.

The crux in applying the frame rule is to show that the composition $\aflowgraph_1 \mstar \aflowgraph_2'$ is indeed defined. One can do this locally by showing that the update $\aflowgraph_2 \leadsto \aflowgraph_2'$ is \emph{frame-preserving}, i.e., for \emph{any} $\aflowgraph_1$ such that $\aflowgraph_1 \mstar \aflowgraph_2$ is defined, $\aflowgraph_1 \mstar \aflowgraph_2'$ is also defined.

Typically, the flow subgraphs involved in a frame-preserving update $\aflowgraph_2 \leadsto \aflowgraph_2'$ include more nodes than those immediately affected by the update.
For instance, consider the subgraphs of $\aflowgraph$ and $\aflowgraph'$ in our example that consist only of the nodes $\{r,u\}$ directly affected by inserting the edge.
These subgraphs do not constitute a frame-preserving update because inserting the edge between $r$ and $u$ also changes the outflow to $v$ from $1$ to $2$.
Hence, the updated subgraph for $\{r,u\}$ would no longer compose with the rest of $\aflowgraph$ where $v$'s flow is still $1$ instead of $2$.
We refer to a set of nodes such as $\{r,u,v\}$ that identifies a frame-preserving update as the update's \emph{footprint}.

\smartparagraph{Meta theories of flow graphs.}
In addition to ensuring that flow graph composition yields a separation algebra, there are two desiderata that one has to take into consideration when designing a meta theory of flow graphs:
\begin{compactitem}
  \item \emph{Obtaining unique flows.} When encoding inductive properties using flows, one is often interested in a particular flow, most commonly the least fixed point of the flow equation for a given inflow. One therefore needs a way to focus the reasoning on the particular flow of interest.
  \item \emph{Identifying frame-preserving updates.} In order to enable the application of the frame rule, one needs a way to effectively compute candidate footprints and check whether they identify frame-preserving updates.
\end{compactitem}
The first subgoal is crucial for expressivity and the second one for proof automation.
Achieving one subgoals makes it more difficult to achieve the other.
Specifically, consider the meta theory proposed in~\cite{DBLP:conf/esop/KrishnaSW20}.
It requires that the flow monoid $(\amonoid, \monadd, 0)$ is also cancellative ($\amonval \prall{\monadd} \amonvalp_1 \prall{=} \amonvalpp$ and $\amonval \prall{\monadd} \amonvalp_2 \prall{=} \amonvalpp$ implies $\amonvalp_1 \prall{=} \amonvalp_2$).
Requiring cancellativity has the advantage that it is easy to check if an update $\aflowgraph \leadsto \aflowgraph'$ is frame-preserving: it suffices to show that $\aflowgraph$ and $\aflowgraph'$ have the same inflow and outflow.
Cancellativity also ensures that for each flow $\flow$, there exists a unique inflow that produces $\flow$.
Hence, it is sufficient to track only $\flow$ since the inflow is a derived quantity. However, the converse does not hold.

In fact, obtaining unique flows for cancellative $\amonoid$ becomes more difficult.
A natural requirement that one would like to impose on $\amonoid$ is that the pre-order induced by $\monadd$ forms a complete partial order (cpo) or even a complete lattice.
This way, one can focus on the least flow, which is guaranteed to exist if one applies standard fixed point theorems, imposing only mild assumptions on the edge functions.
However, cancellativity is inherently incompatible with standard domain-theoretic prerequisites.
For instance, the only ordered cancellative commutative monoid that is a directed cpo is the trivial one: $\amonoid_0=\{0\}$.
Similarly, $\amonoid_0$ is the only such monoid that has a greatest element.

For cases where unique flows are desired, \cite{DBLP:conf/esop/KrishnaSW20} imposes additional requirements on the edge functions (nil-potent) or the graph structure (effectively acyclic).
The former is quite restrictive in terms of expressivity. The latter again complicates the computation of frame-preserving updates: one now has to ensure that no cycles are introduced when the updated graph $\aflowgraph_2'$ is composed with its frame $\aflowgraph_1$.
In fact, for the effectively acyclic case, \cite{DBLP:conf/esop/KrishnaSW20} only provides a sufficient condition that a given footprint yields a frame-preserving update but it gives no algorithm for computing such a footprint.

\smartparagraph{Contributions.}
In this paper, we propose a new meta theory of flows based on flow monoids that form $\omega$-cpos (but need not be cancellative). The cpo requirement yields the desired least fixed point semantics. The differences in the requirements on the flow monoid necessitate a new notion of flow graph composition. In particular, for a least fixed point semantics of flows, $\aflowgraph = \aflowgraph_1 \mstar \aflowgraph_2$ is only defined if the flows of $\aflowgraph_1$ and $\aflowgraph_2$ do not vanish. An example of such a situation is shown in \cref{fig:flow-graph-vanishing}, where the flows in $\aflowgraph_1$ and $\aflowgraph_2$ would vanish to $0$ in $\aflowgraph_1 \mstar \aflowgraph_2$ because the created cycle has no external inflow. Moreover, an update $\aflowgraph \leadsto \aflowgraph'$ is frame-preserving if $\aflowgraph$ and $\aflowgraph'$ route inflows to outflows in the same way. We formalize this condition using a notion of contextual equivalence of the graphs' \emph{transfer functions}, which are the least fixed points of the flow equation, parameterized by the inflows and restricted to the nodes outside the graphs. We then identify conditions on the edge functions that are commonly satisfied in practice and that allow us to effectively check contextual equivalence of transfer functions. This result is remarkable because the flow monoid can have infinite ascending chains and the flow graphs can be cyclic. Building on this equivalence check, we propose an iterative algorithm for computing footprints of updates. This algorithm enables the automation of the frame rule for reasoning about programs manipulating flow graphs. We evaluate the presented algorithms on a benchmark suite of flow graph updates that are extracted from linearizability proofs for concurrent search structures constructed by the tool \plankton~\cite{oopsla,plankton}. The evaluation demonstrates that our algorithms help to automate key aspects of these proofs that have previously relied on user guidance or heuristics.


\section{Flow Graph Separation Algebra}\label{Section:Flows}

We start with the presentation of our new separation algebra of flow graphs.

Given a commutative monoid $(\amonoid, \monadd, \monunit)$, we define the binary relation $\leq$ on $\amonoid$ by $\amonvalp \leq \amonval$ if there is $\amonvalpp\in\amonoid$ with $\amonval = \amonvalp+\amonvalpp$.
Flow values are drawn from a \emph{flow monoid}, a commutative monoid for which the relation $\leq$ is an $\omega$-cpo. That is, $\leq$ is a partial order and every ascending chain $\achain=\amonval_0\leq \amonval_1\leq \ldots$ in $\amonoid$ has a least upper bound, denoted $\bigjoin \achain$.
We require addition to be compatible with upper bounds, $\amonvalp+\bigjoin\achain=\bigjoin(\amonvalp+\achain)$.
In the following, we fix a flow monoid $(\amonoid, \monadd, \monunit)$.

Let $\contfunof{\amonoid\to\amonoid}$ be the continuous functions in $\amonoid\to\amonoid$. 
Recall that a function $\atfun:\amonoid\rightarrow\amonoid$ is \emph{continuous}~\cite{Scott70} if it commutes with limits of ascending chains, $\atfunof{\bigjoin \achain}=\bigjoin \atfunof{\achain}$ for every chain $\achain$ in $\amonoid$. 
We lift $\monadd$ and $\leq$ to functions $\amonoid \to \amonoid$ in the expected way.
An empty iterated sum $\sum_{i\in \emptyset} \amonval_i$ is defined to be~$\monunit$.

\begin{lemma}\label{Lemma:FunctionDomains}
$(\contfunof{\amonoid\to\amonoid}, \circ, \myid)$ is a monoid. Moreover, if $(\amonoid, \leq)$ is an $\omega$-cpo, so is $(\contfunof{\amonoid\to\amonoid}, \leq)$. 
\end{lemma}

A \emph{flow graph} is a tuple $\aflowconstraint=(\setnodes, \edges, \inflow)$ consisting of a finite set of nodes $\setnodes\subseteq \nat$, a set of edges $\edges:\setnodes\times\nat\rightarrow\contfunof{\amonoid\to\amonoid}$ labeled by continuous functions, and an \emph{inflow} $\inflow: (\nat\setminus\setnodes)\times\setnodes\rightarrow\amonoid$. We use $\setflowconstraints$ for the set of all flow graphs and denote the empty flow graph by $\aflowconstraint_{\emptyset}\defeq (\emptyset, \emptyset, \emptyset)$.

We define two derived functions for flow graphs.
First, the \emph{flow} is the least function $\fval:\setnodes\rightarrow\amonoid$ satisfying the flow equation: $\fvalof{\anode} = \inflow_\anode+\rhsatof{\anode}{\fval}$, for all $\anode\in\setnodes$.
Here, $\inflow_\anode\defeq\sum_{\anodep\in(\nat\setminus\setnodes)}\inflowof{\anodep, \anode}$ is a monoid value and $\rhs_{\anode} \defeq \sum_{\anodep\in \setnodes}\edges_{(\anodep, \anode)}$ is a function of type $\contfunof{(\setnodes\rightarrow \amonoid)\rightarrow \amonoid}$.
Finally, we also define the \emph{outflow} $\outflow:\setnodes\times(\nat\setminus\setnodes)\rightarrow\amonoid$ by $\outflowof{\anode, \anodep}\defeq\edges_{(\anode, \anodep)}(\fvalof{\anode})$.

\begin{example}
  \label{ex-keyset-flow}
  For linearizability proofs of concurrent search structures one can use a flow that labels every data structure node $x$ with its \emph{inset}, the set of keys~$k'$ such that a thread searching for~$k'$ may traverse the node~$x$~\cite{DBLP:journals/pacmpl/KrishnaSW18,DBLP:conf/pldi/KrishnaPSW20}. Translated to our setting, the relevant flow monoid is the powerset of keys, $\powerset{\ZZ \cup \set{-\infty,\infty}}$, with set union as addition. \Cref{fig-keyset-flow} shows two keyset flow graphs that abstract potential states of a concurrent set implementation based on sorted linked lists. When a key $k$ is removed from the set, the node $\anode$ that stores $k$ is first marked to indicate that $\anode$ has been logically deleted. In a second step, $\anode$ is then physically unlinked from the list. The idea of the abstraction is that an edge leaving a node~$\anode$ that stores a key~$k$ is labeled by the function~$\lambda_k$ if~$\anode$ is unmarked and otherwise by~$\lambda_{-\infty}$. This is because a search for $k' \in \ZZ$ will traverse the edge leaving~$\anode$ iff $k < k'$ or~$\anode$ is marked. In the figure,~$l$ and $r$ are assumed to be unmarked, storing keys~$6$ and~$8$, respectively. Node~$t$ is assumed to be marked. Flow graph $\aflowconstraint_2$ is obtained from $\aflowconstraint_1$ by physically unlinking the marked node $t$. Using the keyset flow one can then express the crucial data structure invariants that are needed for a linearizability proof based on local reasoning (e.g., the invariant that the logical contents of a node is always a subset of its inset).

  We note that the inflow of the global flow graph that abstracts the program state can be used in the specification. In the example, one lets $\inflow_r=\ZZ$ for the root $r$ of the data structure and $\inflow_x = \emptyset$ for all other nodes to indicate that all searches start at $r$.
  \hfill\qed
  
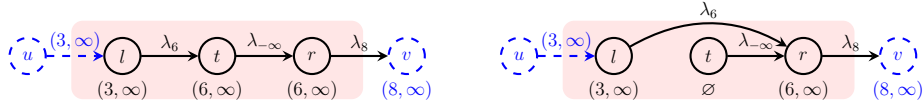
\begin{figure}[t]
    \centering
  \begin{tikzpicture}[>=stealth, scale=0.7, every node/.style={scale=0.6,font=\large}]

    \def\xsep{1.8}
    \def\ysep{-1.8}
    \def\xshift{2.5}
    \def\sd{-.55}
    \def\s{.4}
    \def\t{.7}

    \node[unode,blue,dashed] (u) {$u$};
    \node[unode] (l) at ($(u) + (\xsep, 0)$) {$l$};
    \node[lbl] (flowl) at ($(l) + (0, -.65)$) {$(3,\infty)$};
    \node[unode] (t) at ($(l) + (\xsep, 0)$) {$t$};
    \node[lbl] (flowt) at ($(t) + (0, -.65)$) {$(6,\infty)$};
    \node[unode] (r) at ($(t) + (\xsep, 0)$) {$r$};
    \node[lbl] (flowr) at ($(r) + (0, -.65)$) {$(6,\infty)$};
    \node[unode,blue,dashed] (v) at ($(r) + (\xsep, 0)$) {$v$};
    \node[lbl,blue] (flowv) at ($(v) + (0, -.65)$) {$(8,\infty)$};

    \draw[edge, blue, dashed] (u) to node[above]{$(3,\infty)$} (l);
    \draw[edge] (l) to node[above]{$\lambda_6$} (t);
    \draw[edge] (t) to node[above]{$\lambda_{-\infty}$} (r);
    \draw[edge] (r) to node[above]{$\lambda_8$} (v);

    \begin{scope}[on background layer] 
      \draw[draw=red!10, rounded corners, thick, fill=red!10]
      ($(l.north west) + (-\t, \s)$)
      -- ($(r.north east) + (\t, \s)$)
      -- ($(r.south east) + (\t, \sd)$)
      -- ($(l.south west) + (-\t, \sd)$) -- cycle;
    \end{scope}
  \end{tikzpicture}\hfill
    \begin{tikzpicture}[>=stealth, scale=0.7, every node/.style={scale=0.6,font=\large}]

    \def\xsep{1.8}
    \def\ysep{-1.8}
    \def\xshift{2.5}
    \def\sd{-.55}
    \def\s{.4}
    \def\t{.7}

    \node[unode,blue,dashed] (u) {$u$};
    \node[unode] (l) at ($(u) + (\xsep, 0)$) {$l$};
    \node[lbl] (flowl) at ($(l) + (0, -.65)$) {$(3,\infty)$};
    \node[unode] (t) at ($(l) + (\xsep, 0)$) {$t$};
    \node[lbl] (flowt) at ($(t) + (0, -.65)$) {$\emptyset$};
    \node[unode] (r) at ($(t) + (\xsep, 0)$) {$r$};
    \node[lbl] (flowr) at ($(r) + (0, -.65)$) {$(6,\infty)$};
    \node[unode,blue,dashed] (v) at ($(r) + (\xsep, 0)$) {$v$};
    \node[lbl,blue] (flowv) at ($(v) + (0, -.65)$) {$(8,\infty)$};

    \draw[edge, blue, dashed] (u) to node[above]{$(3,\infty)$} (l);
    \draw[edge] (l) to[bend left=30] node[above]{$\lambda_6$} (r);
    \draw[edge] (t) to node[above]{$\lambda_{-\infty}$} (r);
    \draw[edge] (r) to node[above]{$\lambda_8$} (v);

    \begin{scope}[on background layer] 
      \draw[draw=red!10, rounded corners, thick, fill=red!10]
      ($(l.north west) + (-\t, \s)$)
      -- ($(r.north east) + (\t, \s)$)
      -- ($(r.south east) + (\t, \sd)$)
      -- ($(l.south west) + (-\t, \sd)$) -- cycle;
    \end{scope}
  \end{tikzpicture}
  \caption{Two flow graphs $\aflowconstraint_1$ (left) and $\aflowconstraint_2$ (right) with $\aflowconstraint_1.\setnodes = \aflowconstraint_2.\setnodes = \set{l,t,r}$ for the keyset flow monoid $\powerset{\ZZ \cup \set{-\infty,\infty}}$. The edge label $\lambda_k$ for a key $k$ denotes the function $\lambda \amonval.\, (m \setminus [-\infty, k])$.\label{fig-keyset-flow}}
  \end{figure}
\end{example}

\smartparagraph{Composition without vanishing flows.}
To define the composition of flow graphs, $\aflowconstraint_1 \mstar \aflowconstraint_2$, we proceed in two steps.
We first define an auxiliary composition that may suffer from \emph{vanishing} flows, local flows that disappear in the composition.
That is, this composition is defined for the flow graphs shown in \cref{fig:flow-graph-vanishing}.
In the composed graph the flow of each node is $0$ where it was $1$ before the composition---the flow vanishes. This means that the auxiliary composition does not allow to lift lower bounds on the flow values from the individual components to the composed graph. Hence, the actual composition restricts the auxiliary composition to rule out such vanishing flows.
Definedness of the auxiliary composition requires disjointness of the nodes in $\aflowconstraint_1$ and $\aflowconstraint_2$.
Moreover, the outflow of one flow graph has to match the inflow expectations of the other:
\[
\aflowconstraint_1\statemultdef\!\statemultdef\aflowconstraint_2\quad\text{if}\quad  \setnodes_1\cap \setnodes_2=\emptyset\;\;\wedge\;\;\forall \anode \in\setnodes_1,\, \anodep\in\setnodes_2.\;
\begin{aligned}[t]
& \outflowof[1]{\anode, \anodep}=\inflowof[2]{\anode, \anodep} \wedge {} \\
& \outflowof[2]{\anodep, \anode}=\inflowof[1]{\anodep, \anode}\ .
\end{aligned}
\]
The auxiliary composition $\aflowconstraint_1\discup\aflowconstraint_2$ removes the inflow provided by the other component: 
\[
\aflowconstraint_1\discup\aflowconstraint_2 \quad \defeq \quad (\setnodes_1\discup \setnodes_2, \edges_1\discup \edges_2, \restrictto{(\inflow_1\discup\inflow_2)}{(\nat\setminus(\setnodes_1\discup \setnodes_2))\times(\setnodes_1\discup \setnodes_2)}) \enspace.
\]

To rule out vanishing flows, we incorporate a suitable equality on the flows into the definedness for the composition:
\[
\aflowconstraint_1\statemultdef\aflowconstraint_2\quad\text{if}\quad  \aflowconstraint_1\statemultdef\statemultdef\aflowconstraint_2\;\;\wedge\;\;\aflowconstraint_1.\fval\discup\aflowconstraint_2.\fval = (\aflowconstraint_1\discup\aflowconstraint_2).\fval \enspace.
\]
Only if the latter equality holds, do we have the composition $\aflowconstraint_1\mstar\aflowconstraint_2\defeq\aflowconstraint_1\discup\aflowconstraint_2$.
It is worth noting that $\aflowconstraint_1.\fval\discup\aflowconstraint_2.\fval\geq (\aflowconstraint_1\discup\aflowconstraint_2).\fval$ always holds.
What definedness really asks for is the reverse inequality.

Recall from~\cite{DBLP:conf/lics/CalcagnoOY07} that a \emph{separation algebra} is a partial commutative monoid $(\setstates, \statemult, \emp)$ with a set of units $\emp \subseteq \setstates$.

\begin{lemma}\label{Lemma:FlowAlgebra}
$(\setflowconstraints, \mstar, \set{\aflowconstraint_{\emptyset}})$ is a separation algebra.
\end{lemma}


\section{Frame-Preserving Updates}
\label{Section:Framing}

Since flow graphs form a separation algebra, we can use separation logic assertions to describe sets of flow graphs as in~\cite{DBLP:conf/esop/KrishnaSW20} and then use them to prove separation logic Hoare triples. A key proof rule used in such proofs is the frame rule. Given separation logic assertions $P_1$ and $P_2$, and a command $c$, the frame rule states:
if the Hoare triple $\{P_1\}\,c\,\{P_2\}$ is valid, then so is $\{P_1\mstar F\}\,c\,\{P_2\mstar F\}$ for any \emph{frame} $F$. The remainder of the paper focuses on developing algorithms for automating this proof rule.

The flow graphs described by an assertion may have unbounded size (e.g., due to the use of \textit{iterated separating conjunctions}).
We only consider bounded flow graphs in the following; the unbounded case is known to be a challenge for which orthogonal techniques are being developed (cf. \cref{sec:related}).
However, even if the flow graphs have bounded size, there may still be infinitely many of them because the inflows and edge functions are encoded symbolically in a logical theory of the flow monoid. For pedagogy, we present our algorithms in terms of concrete flow graphs rather than symbolic ones. However, our development readily extends to symbolic representations assuming the underlying flow monoid theory is decidable. In fact, our implementation discussed in \cref{sec:eval} works with symbolic flow graphs.

The soundness of the frame rule relies on the assumption that the state update induced by the command $c$ satisfies a certain locality condition. In our setting, this condition amounts to checking that the update of $P_1$ under $c$ is \emph{frame-preserving} with respect to flow graph composition.
For the flow graphs $\aflowconstraint_1$ described by $P_1$ and all flow graphs $\aflowconstraint_2$ in the post image of $\aflowconstraint_1$ under $c$, this means that $\aflowconstraint_1 \statemultdef \aflowconstraint$ implies $\aflowconstraint_2 \statemultdef \aflowconstraint$ for all $\aflowconstraint$.
Intuitively, $\aflowconstraint_2 \statemultdef \aflowconstraint$ still holds if $\aflowconstraint_1$ and $\aflowconstraint_2$ transfer inflows to outflows in the same way.

Formally, for a flow graph $\aflowconstraint$ we define its \emph{transfer function} \(\transformerof{\aflowconstraint}\) mapping inflows to outflows, \(
  \transformerof{\aflowconstraint}:((\nat\setminus\setnodes)\times\setnodes\rightarrow \amonoid)\rightarrow\setnodes\times(\nat\setminus\setnodes)\rightarrow \amonoid,
\) by \[
  \transformerof{\aflowconstraint}(\inflow') \defeq \aflowconstraint[\inflow \mapsto \inflow'].\outflow
  \ .
\]
For a given inflow $\inflow$, we also write $\transformerof{\aflowconstraint_1} =_{\inflow} \transformerof{\aflowconstraint_2}$ to mean that for all inflows $\inflow' \leq \inflow$, $\transformerof{\aflowconstraint_1}(\inflow') = \transformerof{\aflowconstraint_2}(\inflow')$.

\begin{definition}\label{def:ctx-eq}
  Flow graphs $\aflowconstraint_1, \aflowconstraint_2$ are \emph{contextually equivalent}, denoted $\aflowconstraint_1 \ctxequiv \aflowconstraint_2$, if we have 
  $\aflowconstraint_1.\setnodes = \aflowconstraint_2.\setnodes$, $\aflowconstraint_1.\inflow=\aflowconstraint_2.\inflow$, and $\transformerof{\aflowconstraint_1}=_{\aflowconstraint_1.\inflow}\transformerof{\aflowconstraint_2}$. 
\end{definition}

\begin{theorem}[Frame Preservation]\label{Theorem:FramePreserving}
  For all flow graphs $\aflowconstraint_1 \ctxequiv \aflowconstraint_2$ and $\aflowconstraint$,
  $\aflowconstraint_1\statemultdef \aflowconstraint$ if and only if $\aflowconstraint_2\statemultdef\aflowconstraint$ and, in case of definedness, $\aflowconstraint_1\statemult\aflowconstraint\ctxequiv\aflowconstraint_2\statemult\aflowconstraint$.
\end{theorem}

To automate the frame rule for a command $c$ and a precondition $P$, we need to identify a decomposition $P = P_1 \mstar F$ so as to infer $\{P_1\}\,c\,\{P_2\}$ and then apply the frame rule to derive $\{P\}\,c\,\{Q\}$ for the postcondition $Q=P_2\mstar F$. This is closely related to the \emph{frame inference problem}\cite{DBLP:conf/aplas/BerdineCO05}.
When a command modifies a flow graph $\aflowgraph_1$ to $\aflowgraph_2$, our goal is to identify a (hopefully small) set of nodes $Y$ in $\aflowgraph_1$ that are affected by this update, the \emph{flow footprint}.
That is, $Y$ captures the difference between the flow graphs before and after the update and the complement of $Y$ defines the frame.
To make this formal, we need the restriction of flow graphs to subsets of nodes, which then gives us a notion of flow graph decomposition. 
Towards this, consider $\aflowconstraint$ and $\setnodesp\subseteq\nat$.  
We define \[
  \restrictto{\aflowconstraint}{\setnodesp}
  ~~\defeq~~
  (\aflowconstraint.\setnodes\cap\setnodesp, \restrictto{\aflowconstraint.\edges}{(\aflowconstraint.\setnodes\cap\setnodesp)\times\nat}, \inflow)
\]
such that the inflow $\inflow$ satisfies
$\inflow(\anodepp, \anodep)\defeq\aflowconstraint.\inflow(\anodepp, \anodep)$ for all $\anodepp\in\nat\setminus\aflowconstraint.\setnodes$, $\anodep\in\aflowconstraint.\setnodes\cap\setnodesp$ and $\inflow(\anode, \anodep)\defeq\aflowconstraint.\edges_{(\anode, \anodep)}(\aflowconstraint.\fvalof{\anode})$ for all $\anode\in\aflowconstraint.\setnodes\setminus\setnodesp$,  $\anodep\in\aflowconstraint.\setnodes\cap\setnodesp$.

\begin{definition}\label{def:footprint}
  Consider $\aflowconstraint_1$ and $\aflowconstraint_2$ with $\setnodes\defeq\aflowconstraint_1.\setnodes=\aflowconstraint_2.\setnodes$ and $\aflowconstraint_1.\inflow=\aflowconstraint_2.\inflow$.
  A \emph{flow footprint} for the difference between $\aflowconstraint_1$ and $\aflowconstraint_2$ is a subset of nodes $\setnodesp\subseteq\setnodes$ so that $\restrictto{\aflowconstraint_1}{\setnodesp}\ctxequiv\restrictto{\aflowconstraint_2}{\setnodesp}$ and $\restrictto{\aflowconstraint_1}{\setnodes\setminus\setnodesp}=\restrictto{\aflowconstraint_2}{\setnodes\setminus\setnodesp}$.
  The set of all such footprints is $\setflowfootprintof{\aflowconstraint_1}{\aflowconstraint_2}$.
\end{definition}

Flow graphs over different sets of nodes or inflows never have a flow footprint.
The former requirement merely simplifies the presentation.
To that end, we assume that all nodes that will be allocated during program execution are already present in the initial flow graph.
This assumption can be lifted.
The latter requirement is motivated by the fact that the global inflow is part of the specification as noted earlier in \Cref{ex-keyset-flow}.

Before we proceed with the problem of how to compute flow footprints, we highlight some of their properties.

\begin{lemma}[Footprint Monotonicity]
  \label{Lemma:FootprintMonotonicity}
  If $\setnodespp\in \setflowfootprintof{\aflowconstraint_1}{\aflowconstraint_2}$ and $\setnodespp\subseteq\setnodesp\subseteq\aflowconstraint_1.\setnodes$, then $\setnodesp\in \setflowfootprintof{\aflowconstraint_1}{\aflowconstraint_2}$. 
\end{lemma}

A consequence of monotonicity is the existence of a canonical flow footprint: if there is a flow footprint at all, then the set of all nodes will work as a footprint.
Of course this canonical footprint is undesirably large.
It corresponds to the case where one reasons about flow graph updates globally, forgoing the application of the frame rule.
Unfortunately, an inclusion-minimal flow footprint does not exist. 

\begin{proposition}[Canonical Footprints]\label{Proposition:CanonicalFootprints}
  We have: $\setflowfootprintof{\aflowconstraint_1}{\aflowconstraint_2}\neq\emptyset$ if and only if $\aflowconstraint_1.\setnodes\in \setflowfootprintof{\aflowconstraint_1}{\aflowconstraint_2}$. There is no inclusion-minimal flow footprint; in particular, the set $\setflowfootprintof{\aflowconstraint_1}{\aflowconstraint_2}$ is not closed under intersection. 
\end{proposition}

The proof of monotonicity requires a better understanding of the restriction operator, as provided by the following lemma.  

\begin{lemma}[Restriction]\label{Lemma:Restriction}
  Consider $\aflowconstraint$ and $\setnodesp, \setnodespp\subseteq\nat$. 
  Then
    (i)~$\restrictto{\aflowconstraint}{\setnodesp}.\fval = \restrictto{\aflowconstraint.\fval}{\setnodesp}$,
    (ii)~$\restrictto{\aflowconstraint}{\setnodesp}\statemultdef\restrictto{\aflowconstraint}{\setnodes\setminus\setnodesp}$ and $\restrictto{\aflowconstraint}{\setnodesp}\statemult\restrictto{\aflowconstraint}{\setnodes\setminus\setnodesp}=\aflowconstraint$, and 
    (iii)~$\restrictto{(\restrictto{\aflowconstraint}{\setnodesp})}{\setnodespp}=\restrictto{\aflowconstraint}{\setnodesp\cap\setnodespp}$.
\end{lemma}

Since flow footprints are defined via restriction, the lemma also shows that flow footprints are well-behaved.
For example, the restriction to the footprint $\setnodesp$ does not change the flow of a node $\anodep\in\setnodesp$ nor that of a node $\anode\in\aflowconstraint.\setnodes\setminus\setnodesp$.
More formally, this means $\restrictto{\aflowconstraint}{\setnodesp}.\fvalof{\anodep}=\aflowconstraint.\fvalof{\anodep}$ and $\restrictto{\aflowconstraint}{\setnodes\setminus\setnodesp}.\fvalof{\anode}=\aflowconstraint.\fvalof{\anode}$,  by Lemma~\ref{Lemma:Restriction}(i). 

For our development, it will be convenient to have a more operational formulation of the transfer function.
Towards this, we understand the flow graph as a function that takes an inflow as a parameter and yields a transformer of flow approximants:
\begin{align*}
  &\aflowconstraint ~:~ ((\nat\setminus\setnodes)\times\setnodes\rightarrow \amonoid)\rightarrow (\setnodes\rightarrow \amonoid)\rightarrow \setnodes\rightarrow \amonoid
  \\
  \text{defined by}\qquad
  &\aflowconstraint[\inflow](\val)(\anode) ~=~ \inflow_{\anode}+\rhs_{\anode}(\val)\;.
\end{align*} 
Recall $\inflow_{\anode}\defeq\sum_{\anodep\in\nat\setminus\setnodes}\inflow(\anodep, \anode)$ and $\rhs_{\anode}(\sigma)=\sum_{\anodep\in\setnodes}\edges_{(\anodep, \anode)}(\valof{\anodep})$.
The least fixed point of $\aflowconstraint[\inflow]$ is $\bigjoin_{i\in\nat}\aflowconstraint[\inflow]^i(\bot)$ with $\aflowconstraint^0=\myidof{\setnodes\rightarrow\amonoid}$ and $\aflowconstraint^{i+1}=\aflowconstraint^i\circ \aflowconstraint$, by Kleene's theorem.
Define $\outflow:(\setnodes\prall{\rightarrow}\amonoid)\prall{\rightarrow}\setnodes\times(\nat\setminus\setnodes)\prall{\rightarrow}\amonoid$ by $\outflow(\val)(\anodep, \anodepp)\defeq\edges_{(\anodep, \anodepp)}(\valof{\anodep})$.
This yields the following characterization of transfer functions and flows.

\begin{lemma}[Transfer]\label{Lemma:Transfer}
  For all flow graphs $\aflowconstraint$ we have
    (i)~$\transformerof{\aflowconstraint} = \outflow\circ (\mathit{lfp}.\aflowconstraint[-])$ and
    (ii)~$\mathit{lfp}.\aflowconstraint[\aflowconstraint.\inflow])=\aflowconstraint.\fval$. 
\end{lemma}


\section{Computing Footprints}
\label{Section:Footprints}

We present an algorithm for computing a footprint for the difference between two given flow graphs.
We proceed in two steps.
We first give a high-level description of the algorithm that ignores computability problems.
In a second step, we show how to solve the computability problems.
Throughout the development, we will assume to have flow graphs $\aflowconstraint_1$ and $\aflowconstraint_2$ over the same nodes ${\setnodes\defeq\aflowconstraint_1.\setnodes=\aflowconstraint_2.\setnodes}$ and with the same inflow ${\aflowconstraint_1.\inflow=\aflowconstraint_2.\inflow}$.
If this assumption fails, a flow footprint does not exist by definition.

\subsection{Algorithm}
\label{Section:Footprints:Algo}

We compute the flow footprint as a fixed point.
We start with the footprint candidate $\setnodespp$ consisting of the nodes whose outgoing edges differ in $\aflowconstraint_1$ and $\aflowconstraint_2$.
Then, we iteratively add the nodes whose outflow leaving the current footprint candidate $\setnodespp$ differs in $\restrictto{\aflowconstraint_1}{\setnodespp}$ and $\restrictto{\aflowconstraint_2}{\setnodespp}$.
That the outflow differs means that the transfer functions $\transformerof{\restrictto{\aflowconstraint_1}{\setnodespp}}$ and $\transformerof{\restrictto{\aflowconstraint_2}{\setnodespp}}$ differ and thus the candidate $\setnodespp$ is not a footprint.
In turn, if all outflows match, the transfer functions coincide and $\setnodespp$ is a footprint as desired.

Technically, we compute the fixed point over the powerset lattice of nodes endowed with a distinguished top element:
${(\powerset{\setnodes}^\top\!,\, \sqsubseteq)}$ with ${\powerset{\setnodes}^{\top}\!\defeq\powerset{\setnodes}\uplus\set{\top}}$.
Element $\top$ indicates a failure of the footprint computation.
This may arise if the footprint is not covered by $\setnodes$, i.e., extends beyond the flow graphs $\aflowconstraint_1,\aflowconstraint_2$.

Our fixed point computation starts from $\setnodespp=\outof{\aflowconstraint_1}{\aflowconstraint_2}\subseteq\setnodes$ as defined by
\[
	\outof{\aflowconstraint_1}{\aflowconstraint_2}
	~\defeq~
	\setcond{\anode\in\setnodes}{\exists \anodepp\in\nat.\aflowconstraint_1.\edges(\anode, \anodepp)\neq\aflowconstraint_2.\edges(\anode, \anodepp)}
	\ .
\]
The fixed point then proceeds to extend $\setnodespp$ as long as the transfer functions associated with $\restrictto{\aflowconstraint_1}{\setnodespp}$ and $\restrictto{\aflowconstraint_2}{\setnodespp}$ do not match.
To define the extension, we let the \emph{transfer failure} of $\setnodespp\subseteq\setnodes$ be the successor nodes of $\setnodespp$ that may receive different outflow from $\aflowconstraint_1$ and $\aflowconstraint_2$:
\[
	\tfailof{\aflowconstraint_1}{\aflowconstraint_2}{\setnodespp}
	~\defeq~
	\left\{
		\anode \in \nat\setminus\setnodespp
	~\middle|~
		\begin{aligned}
			&\exists\, \aninflow\leq\restrictto{\aflowconstraint_1}{\setnodespp}.\inflow~~
			\exists\, \anodepp\in\setnodespp.~~~
			\\
			&\qquad[\transformerof{\restrictto{\aflowconstraint_1}{\setnodespp}}(\aninflow)](\anodepp, \anode)\neq [\transformerof{\restrictto{\aflowconstraint_2}{\setnodespp}}(\aninflow)](\anodepp,\anode)
		\end{aligned}
	\right\}
	\ .
\]
This set is the \textit{reason} why the current footprint candidate $\setnodespp$ is not a footprint, that is, $\setnodespp\notin\setflowfootprintof{\aflowconstraint_1}{\aflowconstraint_2}$.
Extending $\setnodespp$ with the transfer failure yields a new candidate.
We check that the new candidate is covered by $\setnodes$ (i.e., does not include nodes outside of $\aflowconstraint_1,\aflowconstraint_2$).
If the check fails, the new candidate is $\set{\top}$ to indicate that no footprint could be computed.
The following definition makes the extension procedure precise.

\begin{definition}
	The function $\extend{\aflowconstraint_1}{\aflowconstraint_2}:\powerset{\setnodes}^{\top}\rightarrow\powerset{\setnodes}^\top$ is defined by
	\[
		\extendof{\aflowconstraint_1}{\aflowconstraint_2}{\setnodespp}
		~\defeq~
		\twocases{
			\tfailof{\aflowconstraint_1}{\aflowconstraint_2}{\setnodespp}\not\subseteq\setnodes
		}{
			\top
		}{
			\setnodespp ~\sqcup~ \outof{\aflowconstraint_1}{\aflowconstraint_2} ~\sqcup~ \tfailof{\aflowconstraint_1}{\aflowconstraint_2}{\setnodespp}
		}
		\ .
	\]
\end{definition}

Iteratively extending the candidate $\setnodespp$ with the transfer failure eventually produces a footprint for the difference of $\aflowconstraint_1$ and $\aflowconstraint_2$, or fails with $\top$. The approach is sound.

\begin{theorem}[Soundness]
	\label{Theorem:Soundness}
	Let $\afootprint\defeq\mathit{lfp}.\extend{\aflowconstraint_1}{\aflowconstraint_2}$.
	If $\afootprint\prall{\neq}\top$, then $\afootprint\prall{\in}\setflowfootprintof{\aflowconstraint_1}{\aflowconstraint_2}$.
\end{theorem}

\begin{example}
	For an illustration of the proposed approach consider \cref{fig:fixed-point}.
	There, we apply the fixed point computation to find a footprint for the difference of the flow graphs $\aflowconstraint$ and $\aflowconstraint'$.
	As alluded to in \cref{sec:intro}, $\aflowconstraint'$ is the result of inserting into $\aflowconstraint$ a new edge between nodes $r$ and $u$ labeled with $\lambda_{\mathit{id}}$.


\begin{figure}[tb]
	\def\xsep{2}
	\def\ysep{-1.8}
	\def\xshift{2.5}
	\def\sd{-.25}
	\def\sdd{-.4}
	\def\s{.4}
	\def\t{.95}
	\newcommand{\tpspacing}{\hfill}
	\begin{tikzpicture}[>=stealth, scale=0.6, every node/.style={scale=0.6,font=\large},baseline=(r)]
		\useasboundingbox (-1.7,1.35) rectangle (3.7,-4.4);
		\node[unode] (x) {$x$};
		\node[lbl] (flowx) at ($(x) + (-.65, 0)$) {$ 1$};
		\node[unode] (r) at ($(x) + (\xsep, 0)$) {$r$};
		\node[stackVar] (rin) at ($(r) + (\xsep, 0)$) {};
		\node[lbl] (flowr) at ($(r) + (0, -.65)$) {$ 1$};
		\node[unode] (y) at ($(x) + (0, \ysep)$) {$y$};
		\node[lbl] (flowy) at ($(y) + (.65, 0)$) {$ 2$};
		\node[unode] (u) at ($(r) + (0, \ysep)$) {$u$};
		\node[unode] (z) at ($(y) + (0, \ysep)$) {$z$};
		\node[lbl] (flowz) at ($(z) + (-.65, 0)$) {$ 1$};
		\node[unode] (v) at ($(u) + (0, \ysep)$) {$v$};
		\draw[edge] (r) to node[above]{$\lambda_{\mathit{id}}$} (x);
		\draw[edge] (r) to node[above,xshift=-5pt]{$\lambda_{\mathit{id}}$} (y);
		\draw[edge] (u) to node[right] {$\lambda_{\mathit{id}}$} (v);
		\draw[edge] (x) to node[left,xshift=2pt] {$\lambda_{\mathit{id}}$} (y);
		\draw[edge] (x) to[bend right=45] node[left] {$\lambda_{\mathit{id}}$} (z);
		\draw[edge] (z) to node[above,xshift=-5pt] {$\lambda_{\mathit{id}}$} (u);
		\draw[edge, red, dashed] (rin) to node[above] {$ 1$} (r);

		\node[lbl] (flowu) at ($(u) + (.85,0)$) {$ 1{\color{blue}{+}1}$};
		\node[lbl] (flowv) at ($(v) + (.85, 0)$) {$ 1{\color{blue}{+}1}$};
		\draw[edge,draw=blue,snake it] (r) to[bend left=30] node[right,blue]{$\lambda_{\mathit{id}}$} (u);

		\node[] at ($(x.north) + (1, .65)$) {\scalebox{1.2}{$\aflowconstraint$\:/\:\textcolor{blue}{$\aflowconstraint'$}}};
		\begin{scope}[on background layer] 
		\draw[draw=violet!10, rounded corners, thick, fill=violet!10]
			($(x.north west) + (-\t-.4, \s)$)
			-- ($(r.north east) + (\t, \s)$)
			-- ($(v.south east) + (\t, \sd)$)
			-- ($(z.south west) + (-\t-.4, \sd)$) -- cycle;
		\end{scope}
	\end{tikzpicture}
	\tpspacing
	\begin{tikzpicture}[>=stealth, scale=0.6, every node/.style={scale=0.6,font=\large},baseline=(r)]
		\useasboundingbox (-.8,1.35) rectangle (3.7,-4.4);
		\node[unode, red, dashed] (x) {$x$};
		\node[lbl,red] (flowx) at ($(x) + (-.65, 0)$) {$ 1$};
		\node[unode] (r) at ($(x) + (\xsep, 0)$) {$r$};
		\node[stackVar] (rin) at ($(r) + (\xsep, 0)$) {};
		\node[lbl] (flowr) at ($(r) + (0, -.65)$) {$ 1$};
		\node[unode, red, dashed] (y) at ($(x) + (0, \ysep)$) {$y$};
		\node[lbl,red] (flowy) at ($(y) + (.65, 0)$) {$ 1$};
		\node[unode, red, dashed] (u) at ($(r) + (0, \ysep)$) {$u$};
		\draw[edge] (r) to node[above]{$\lambda_{\mathit{id}}$} (x);
		\draw[edge] (r) to node[above,yshift=5pt]{$\lambda_{\mathit{id}}$} (y);
		\draw[edge, red, dashed] (rin) to node[above] {$ 1$} (r);

		\node[lbl,red] (flowu) at ($(u) + (.85,0)$) {$ 0{\color{blue}{+}1}$};
		\draw[edge,draw=blue,snake it] (r) to[bend left=30] node[right,blue]{$\lambda_{\mathit{id}}$} (u);

		\node[] at ($(r.north) + (0, .65)$) {\scalebox{1.2}{$\restrictto{\aflowconstraint}{\setnodespp_0}$\:/\:\textcolor{blue}{$\restrictto{\aflowconstraint'}{\setnodespp_0}$}}};
		\begin{scope}[on background layer] 
			\draw[draw=blue!10, rounded corners, thick, fill=blue!10]
			($(r.north west) + (-\t, \s)$)
			-- ($(r.north east) + (\t, \s)$)
			-- ($(r.south east) + (\t, \sdd)$)
			-- ($(r.south west) + (-\t, \sdd)$) -- cycle;
		\end{scope}
	\end{tikzpicture}
	\tpspacing
	\begin{tikzpicture}[>=stealth, scale=0.6, every node/.style={scale=0.6,font=\large},baseline=(r)]
		\useasboundingbox (-.8,1.35) rectangle (3.7,-4.4);
		\node[unode, red, dashed] (x) {$x$};
		\node[lbl,red] (flowx) at ($(x) + (-.65, 0)$) {$ 1$};
		\node[unode] (r) at ($(x) + (\xsep, 0)$) {$r$};
		\node[stackVar] (rin) at ($(r) + (\xsep, 0)$) {};
		\node[lbl] (flowr) at ($(r) + (0, -.65)$) {$ 1$};
		\node[unode, red, dashed] (y) at ($(x) + (0, \ysep)$) {$y$};
		\node[lbl,red] (flowy) at ($(y) + (.65, 0)$) {$ 1$};
		\node[unode] (u) at ($(r) + (0, \ysep)$) {$u$};
		\node[stackVar] (z) at ($(y) + (0, \ysep)$) {};
		\node[unode, red, dashed] (v) at ($(u) + (0, \ysep)$) {$v$};
		\draw[edge] (r) to node[above]{$\lambda_{\mathit{id}}$} (x);
		\draw[edge] (r) to node[above,yshift=5pt]{$\lambda_{\mathit{id}}$} (y);
		\draw[edge] (u) to node[right] {$\lambda_{\mathit{id}}$} (v);
		\draw[edge, red, dashed] (z) to node[above] {$ 1$} (u);
		\draw[edge, red, dashed] (rin) to node[above] {$ 1$} (r);

		\node[lbl] (flowu) at ($(u) + (.85,0)$) {$ 1{\color{blue}{+}1}$};
		\node[lbl,red] (flowv) at ($(v) + (.85, 0)$) {$ 1{\color{blue}{+}1}$};
		\draw[edge,draw=blue,snake it] (r) to[bend left=30] node[right,blue]{$\lambda_{\mathit{id}}$} (u);

		\node[] at ($(r.north) + (0, .65)$) {\scalebox{1.2}{$\restrictto{\aflowconstraint}{\setnodespp_1}$\:/\:\textcolor{blue}{$\restrictto{\aflowconstraint'}{\setnodespp_1}$}}};
		\begin{scope}[on background layer] 
			\draw[draw=blue!10, rounded corners, thick, fill=blue!10]
			($(r.north west) + (-\t, \s)$)
			-- ($(r.north east) + (\t, \s)$)
			-- ($(u.south east) + (\t, \sd)$)
			-- ($(u.south west) + (-\t, \sd)$) -- cycle;
		\end{scope}
	\end{tikzpicture}
	\tpspacing
	\begin{tikzpicture}[>=stealth, scale=0.6, every node/.style={scale=0.6,font=\large},baseline=(r)]
		\useasboundingbox (-.8,1.35) rectangle (3.7,-4.4);
		\node[unode, red, dashed] (x) {$x$};
		\node[lbl,red] (flowx) at ($(x) + (-.65, 0)$) {$ 1$};
		\node[unode] (r) at ($(x) + (\xsep, 0)$) {$r$};
		\node[stackVar] (rin) at ($(r) + (\xsep, 0)$) {};
		\node[lbl] (flowr) at ($(r) + (0, -.65)$) {$ 1$};
		\node[unode, red, dashed] (y) at ($(x) + (0, \ysep)$) {$y$};
		\node[lbl,red] (flowy) at ($(y) + (.65, 0)$) {$ 1$};
		\node[unode] (u) at ($(r) + (0, \ysep)$) {$u$};
		\node[stackVar] (z) at ($(y) + (0, \ysep)$) {};
		\node[unode] (v) at ($(u) + (0, \ysep)$) {$v$};
		\draw[edge] (r) to node[above]{$\lambda_{\mathit{id}}$} (x);
		\draw[edge] (r) to node[above]{$\lambda_{\mathit{id}}$} (y);
		\draw[edge] (u) to node[right] {$\lambda_{\mathit{id}}$} (v);
		\draw[edge, red, dashed] (z) to node[above] {$ 1$} (u);
		\draw[edge, red, dashed] (rin) to node[above] {$ 1$} (r);

		\node[lbl] (flowu) at ($(u) + (.85,0)$) {$ 1{\color{blue}{+}1}$};
		\node[lbl] (flowv) at ($(v) + (.85, 0)$) {$ 1{\color{blue}{+}1}$};
		\draw[edge,draw=blue,snake it] (r) to[bend left=30] node[right,blue]{$\lambda_{\mathit{id}}$} (u);

		\node[] at ($(r.north) + (0, .65)$) {\scalebox{1.2}{$\restrictto{\aflowconstraint}{\setnodespp_2}$\:/\:\textcolor{blue}{$\restrictto{\aflowconstraint'}{\setnodespp_2}$}}};
		\begin{scope}[on background layer] 
			\draw[draw=blue!10, rounded corners, thick, fill=blue!10]
			($(r.north west) + (-\t, \s)$)
			-- ($(r.north east) + (\t, \s)$)
			-- ($(v.south east) + (\t, \sdd)$)
			-- ($(v.south west) + (-\t, \sdd)$) -- cycle;
		\end{scope}
	\end{tikzpicture}
	\caption{%
		Computing a footprint for the difference of $\aflowconstraint$ and $\aflowconstraint'$ iterates through the sets $\setnodespp_0\defeq\set{r}$, $\setnodespp_1\defeq\set{r,u}$, and $\setnodespp_2\defeq\set{r,u,v}$.
		The latter is the least fixed point of $\extend{\aflowconstraint}{\aflowconstraint'}$ and a footprint as desired, $\setnodespp_2\in\setflowfootprintof{\aflowconstraint}{\aflowconstraint'}$.
		\label{fig:fixed-point}
	}
\end{figure}
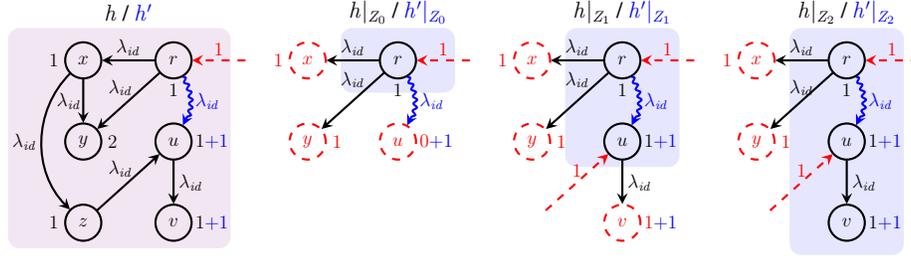

	The fixed point computation starts from $\setnodespp_0\defeq\set{r}=\outof{H}{H'}$ as it is the only node whose outgoing edges have changed.
	Next, we compute $\tfailof{\aflowconstraint}{\aflowconstraint'}{\setnodespp_0}$.
	This yields $\set{u}$ because $u$ receives $0$ from $\setnodespp_0$ in $\aflowconstraint$ but $1$ in $\aflowconstraint'$ due to the new edge.
	The outflow from $\setnodespp_0$ to the remaining nodes coincides in $\aflowconstraint$ and $\aflowconstraint'$.
	Hence, the extension of $\setnodespp_0$ with the transfer failure yields $\setnodespp_1\defeq\extend{\aflowconstraint}{\aflowconstraint'}(\setnodespp_0)=\set{u,r}$.
	Similarly, we compute $\tfailof{\aflowconstraint}{\aflowconstraint'}{\setnodespp_1}$ and obtain $\setnodespp_2\defeq\extend{\aflowconstraint}{\aflowconstraint'}(\setnodespp_1)=\set{r,u,v}$.
	Since $v$ has no outgoing edges, $\setnodespp_2$ is the least fixed point of $\extend{\aflowconstraint}{\aflowconstraint'}$.
	Because $\setnodespp_2$ is a subset of the nodes of $\aflowconstraint$ and $\aflowconstraint'$, it is a footprint, $\setnodespp_2\in\setflowfootprintof{\aflowconstraint}{\aflowconstraint'}$.
	\hfill\qed
\end{example}

To obtain \Cref{Theorem:Soundness}, we have to prove that the fixed point $\afootprint\defeq\mathit{lfp}.\extend{\aflowconstraint_1}{\aflowconstraint_2}$ is indeed a footprint if $\afootprint\neq\top$.
That is, we have to establish the following two properties according to \Cref{def:footprint}:
	(i) $\restrictto{\aflowconstraint_1}{\afootprint}\ctxequiv\restrictto{\aflowconstraint_2}{\afootprint}$ and
	(ii) $\restrictto{\aflowconstraint_1}{\setnodes\setminus\afootprint}=\restrictto{\aflowconstraint_2}{\setnodes\setminus\afootprint}$.

To see the latter one, note that the graph structures (the nodes and edges) of $\restrictto{\aflowconstraint_1}{\setnodes\setminus\afootprint}$ and $\restrictto{\aflowconstraint_2}{\setnodes\setminus\afootprint}$ coincide because $\outof{\aflowconstraint_1}{\aflowconstraint_2}\subseteq\afootprint$.
The inflows coincide as well because they are, intuitively, comprised of the flow graph's overall inflow $\aflowconstraint_1.\inflow=\aflowconstraint_2.\inflow$ and the outflow of the footprint, which is equal in both flow graphs due to $\restrictto{\aflowconstraint_1}{\afootprint}\ctxequiv\restrictto{\aflowconstraint_2}{\afootprint}$.

The interesting part of the soundness proof is to establish property (i), the contextual equivalence $\restrictto{\aflowconstraint_1}{\afootprint}\ctxequiv\restrictto{\aflowconstraint_2}{\afootprint}$.
Since $\afootprint$ is a fixed point of $\extend{\aflowconstraint_1}{\aflowconstraint_2}$, we know that $\tfailof{\aflowconstraint_1}{\aflowconstraint_2}{\setnodespp}=\emptyset$ and thus the transfer functions of $\restrictto{\aflowconstraint_1}{\afootprint}$ and $\restrictto{\aflowconstraint_2}{\afootprint}$ coincide.
Hence, it suffices to establish $\restrictto{\aflowconstraint_1}{\afootprint}.\inflow=\restrictto{\aflowconstraint_2}{\afootprint}.\inflow$ to obtain the desired contextual equivalence, \Cref{def:ctx-eq}.
This key step in the proof is obtained with the help of the following \namecref{Lemma:Inflow}.

\begin{lemma}\label{Lemma:Inflow}
	Let $\outof{\aflowconstraint_1}{\aflowconstraint_2}\prall{\subseteq}\afootprint\prall{\subseteq}\setnodes$
	with $\tfailof{\aflowconstraint_1}{\aflowconstraint_2}{\afootprint}\prall{=}\emptyset$.
	Then $\restrictto{\aflowconstraint_1}{\afootprint}.\inflow\prall{=}\restrictto{\aflowconstraint_2}{\afootprint}.\inflow$.
\end{lemma}

To establish the \namecref{Lemma:Inflow} one has to show that the inflow into $\afootprint$ from the non-footprint part $\setnodesp\defeq\setnodes\setminus\afootprint$ coincides in $\aflowconstraint_1$ and $\aflowconstraint_2$.
This is challenging due to a cyclic dependency in the flow: the inflow from $\setnodesp$ depends on the outflow of $\afootprint$, which in turn depends on the inflow from~$\setnodesp\!$.
To tackle this challenge, we rephrase the flow equation for $\aflowconstraint_i$ as a pairing of the two separate flow equations for $\restrictto{\aflowconstraint_i}{\afootprint}$ and $\restrictto{\aflowconstraint_i}{\setnodesp}$, for $i\in\set{1,2}$.
Intuitively, the pairings compute the flow locally in $\restrictto{\aflowconstraint_i}{\afootprint}$ and $\restrictto{\aflowconstraint_i}{\setnodesp}$ for a fixed inflow (initially $\aflowconstraint_i.\inflow$).
Then, the inflow to $\restrictto{\aflowconstraint_i}{\afootprint}$ is updated to the inflow from outside $\aflowconstraint_i$ and the inflow from $\restrictto{\aflowconstraint_i}{\setnodesp}$, and similarly for the inflow to $\restrictto{\aflowconstraint_i}{\setnodesp}$.
This is repeated until a fixed point is reach.
Technically, we rely on Beki\'c's Lemma~\cite{DBLP:conf/ibm/Bekic84e} to compute the pairings.
Then, we observe $\transformerof{\restrictto{\aflowconstraint_1}{\afootprint}}=\transformerof{\restrictto{\aflowconstraint_2}{\afootprint}}$ because $\tfailof{\aflowconstraint_1}{\aflowconstraint_2}{\afootprint}=\emptyset$ as well as $\transformerof{\restrictto{\aflowconstraint_1}{\setnodesp}}=\transformerof{\restrictto{\aflowconstraint_2}{\setnodesp}}$ because $\outof{\aflowconstraint_1}{\aflowconstraint_2}\subseteq\afootprint$.
Roughly, this means that the flow pairings for $\aflowconstraint_1$ and $\aflowconstraint_2$ must also coincide as the individual parts propagate the same values.
Phrased differently, the updated inflow for $\restrictto{\aflowconstraint_1}{\afootprint}$ and $\restrictto{\aflowconstraint_2}{\afootprint}$ as well as $\restrictto{\aflowconstraint_1}{\setnodesp}$ and $\restrictto{\aflowconstraint_2}{\setnodesp}$ coincide in each iteration.
Overall, this means $\restrictto{\aflowconstraint_1}{\afootprint}.\inflow=\restrictto{\aflowconstraint_2}{\afootprint}.\inflow$.


\begin{figure}[tb]
	\def\xsep{1.8}
	\def\ysep{-1.8}
	\def\xshift{2.5}
	\def\sd{-.25}
	\def\sdd{-.4}
	\def\s{.5}
	\def\t{.95}
	\def\l{.65}
	\def\ins{1.8}
	\def\inss{2.4}
	\def\infvar{in}
	\begin{tikzpicture}[>=stealth, scale=0.6, every node/.style={scale=0.6,font=\large},baseline=(r)]
		\useasboundingbox (-1.4,1.35) rectangle (3.4,-2.7);
		\node[] at ($(current bounding box.north) + (0, -.2)$) {\scalebox{1.2}{$\aflowconstraint_1$}};
		\node[unode] (x) {$x$};
		\node[unode] (y) at ($(x) + (\xsep, 0)$) {$y$};
		\node[unode] (z) at ($(x) + (0, \ysep)$) {$z$};
		\node[unode] (u) at ($(z) + (\xsep, 0)$) {$u$};
		\node[stackVar] (xin) at ($(x) + (-\ins, 0)$) {};
		\node[stackVar] (yout) at ($(y) + (\ins, 0)$) {};

		\draw[edge] (x) to node[left]{$\lambda_{\mathit{id}}$} (z);
		\draw[edge] (y) to node[left,xshift=1.5mm,yshift=3mm]{$\lambda_{\mathit{id}}$} (z);
		\draw[edge] (z) to node[above]{$\lambda_{\mathit{id}}$} (u);
		\draw[edge] (u) to node[right]{$\lambda_{\mathit{id}}$} (y);
		\draw[edge, red, dashed] (xin) to node[above] {$k$} (x);
		\draw[edge, red, dashed] (y) to node[above] {$k$} (yout);

		\node[lbl] at ($(x) + (0, \l)$) {$k$};
		\node[lbl] at ($(y) + (0, \l)$) {$k$};
		\node[lbl] at ($(z) + (-\l, 0)$) {$k$};
		\node[lbl] at ($(u) + (\l, 0)$) {$k$};

		\begin{scope}[on background layer] 
		\draw[draw=violet!10, rounded corners, thick, fill=violet!10]
			($(x.north west) + (-\l, \s)$)
			-- ($(z.south west) + (-\l, -\s)$)
			-- ($(u.south east) + (\l, -\s)$)
			-- ($(y.north east) + (\l, \s)$)
			-- cycle;
		\end{scope}
	\end{tikzpicture}
	\hfill
	\begin{tikzpicture}[>=stealth, scale=0.6, every node/.style={scale=0.6,font=\large},baseline=(r)]
		\useasboundingbox (-1.4,1.35) rectangle (3.4,-2.7);
		\node[] at ($(current bounding box.north) + (0, -.2)$) {\scalebox{1.2}{$\restrictto{\aflowconstraint_1}{\setnodespp}$}};
		\node[unode] (x) {$x$};
		\node[unode] (y) at ($(x) + (\xsep, 0)$) {$y$};
		\node[unode] (z) at ($(x) + (0, \ysep)$) {$z$};
		\node[unode, red, dashed] (u) at ($(z) + (\xsep, 0)$) {$u$};
		\node[stackVar] (xin) at ($(x) + (-\ins, 0)$) {};
		\node[stackVar] (yout) at ($(y) + (\ins, 0)$) {};

		\draw[edge] (x) to node[left] {$\lambda_{\mathit{id}}$} (z);
		\draw[edge] (y) to node[left,xshift=1.5mm,yshift=3mm]{$\lambda_{\mathit{id}}$} (z);
		\draw[edge, red, dashed] (z) to node[above]{$k$} (u);
		\draw[edge, red, dashed] (u) to node[right]{$\infvar$} (y);
		\draw[edge, red, dashed] (xin) to node[above] {$k$} (x);
		\draw[edge, red, dashed] (y) to node[above] {$\infvar$} (yout);

		\node[lbl] at ($(x) + (0, \l)$) {$k$};
		\node[lbl] at ($(y) + (0, \l)$) {$\infvar$};
		\node[lbl] at ($(z) + (-\l, 0)$) {$k$};

		\begin{scope}[on background layer] 
		\draw[draw=red!10, rounded corners, thick, fill=red!10]
			($(x.north west) + (-\l, \s)$)
			-- ($(z.south west) + (-\l, -\s)$)
			-- ($(z.south east) + (\l, -\s)$)
			-- ($(z.north east) + (\l, \s)$)
			|- ($(y.south east) + (\l, -\s)$)
			-- ($(y.north east) + (\l, \s)$)
			-- cycle;
		\end{scope}
	\end{tikzpicture}
	\hfill
	\begin{tikzpicture}[>=stealth, scale=0.6, every node/.style={scale=0.6,font=\large},baseline=(r)]
		\useasboundingbox (-1.4,1.35) rectangle (3.4,-2.7);
		\node[] at ($(current bounding box.north) + (0, -.2)$) {\scalebox{1.2}{$\aflowconstraint_2$}};
		\node[unode] (x) {$x$};
		\node[unode] (y) at ($(x) + (\xsep, 0)$) {$y$};
		\node[unode] (z) at ($(x) + (0, \ysep)$) {$z$};
		\node[unode] (u) at ($(z) + (\xsep, 0)$) {$u$};
		\node[stackVar] (xin) at ($(x) + (-\ins, 0)$) {};
		\node[stackVar] (yout) at ($(y) + (\ins, 0)$) {};

		\draw[edge] (x) to node[above]{$\lambda_{\mathit{id}}$} (y);
		\draw[edge] (y) to node[left,xshift=1.5mm,yshift=3mm]{$\lambda_{\mathit{id}}$} (z);
		\draw[edge] (z) to node[above]{$\lambda_{\mathit{id}}$} (u);
		\draw[edge] (u) to node[right]{$\lambda_{\mathit{id}}$} (y);
		\draw[edge, red, dashed] (xin) to node[above] {$k$} (x);
		\draw[edge, red, dashed] (y) to node[above] {$k$} (yout);

		\node[lbl] at ($(x) + (0, \l)$) {$k$};
		\node[lbl] at ($(y) + (0, \l)$) {$k$};
		\node[lbl] at ($(z) + (-\l, 0)$) {$k$};
		\node[lbl] at ($(u) + (\l, 0)$) {$k$};

		\begin{scope}[on background layer] 
		\draw[draw=violet!10, rounded corners, thick, fill=violet!10]
			($(x.north west) + (-\l, \s)$)
			-- ($(z.south west) + (-\l, -\s)$)
			-- ($(u.south east) + (\l, -\s)$)
			-- ($(y.north east) + (\l, \s)$)
			-- cycle;
		\end{scope}
	\end{tikzpicture}
	\hfill
	\begin{tikzpicture}[>=stealth, scale=0.6, every node/.style={scale=0.6,font=\large},baseline=(r)]
		\useasboundingbox (-1.4,1.35) rectangle (3.8,-2.7);
		\node[] at ($(current bounding box.north) + (0, -.2)$) {\scalebox{1.2}{$\restrictto{\aflowconstraint_2}{\setnodespp}$}};
		\node[unode] (x) {$x$};
		\node[unode] (y) at ($(x) + (\xsep, 0)$) {$y$};
		\node[unode] (z) at ($(x) + (0, \ysep)$) {$z$};
		\node[unode, red, dashed] (u) at ($(z) + (\xsep, 0)$) {$u$};
		\node[stackVar] (xin) at ($(x) + (-\ins, 0)$) {};
		\node[stackVar] (yout) at ($(y) + (\inss, 0)$) {};

		\draw[edge] (x) to node[above]{$\lambda_{\mathit{id}}$} (y);
		\draw[edge] (y) to node[left,xshift=1.5mm,yshift=3mm]{$\lambda_{\mathit{id}}$} (z);
		\draw[edge, red, dashed] (z) to node[above]{$k$} (u);
		\draw[edge, red, dashed] (u) to node[right]{$\infvar$} (y);
		\draw[edge, red, dashed] (xin) to node[above] {$k$} (x);
		\draw[edge, red, dashed] (y) to node[above] {$\max k,\infvar$} (yout);

		\node[lbl] at ($(x) + (0, \l)$) {$k$};
		\node[lbl] at ($(y) + (0, \l)$) {$\max k,\infvar$};
		\node[lbl] at ($(z) + (-\l, 0)$) {$k$};

		\begin{scope}[on background layer] 
		\draw[draw=blue!10, rounded corners, thick, fill=blue!10]
			($(x.north west) + (-\l, \s)$)
			-- ($(z.south west) + (-\l, -\s)$)
			-- ($(z.south east) + (\l, -\s)$)
			-- ($(z.north east) + (\l, \s)$)
			|- ($(y.south east) + (\l, -\s)$)
			-- ($(y.north east) + (\l, \s)$)
			-- cycle;
		\end{scope}
	\end{tikzpicture}
	\caption{%
		Counterexample to completeness using the monoid $(\nat \cup \{\infty\},\max,0)$.
		While the set $\{x,y,z,u\}$ is a footprint for the difference between flow graphs $\aflowconstraint_1$ and $\aflowconstraint_2$, our fixed point will produce the candidates $\{x\}$ and $\setnodespp\defeq\{x,y,z\}$ and then fail with $\{\top\}$.
		\label{fig:incomplete}
	}
\end{figure}
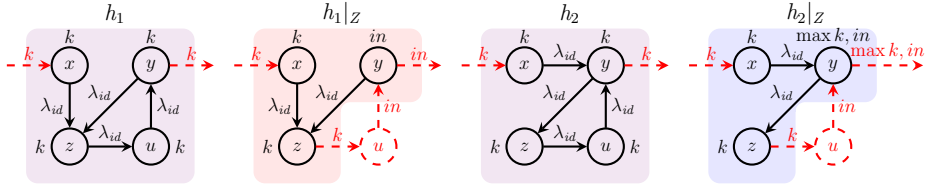

Our computation of a flow footprint is forward, it starts from the nodes where the flow graphs differ and follows the edges.
It may therefore fail if predecessor nodes of an iterate $\setnodespp$ need to be considered to determine a flow footprint.
For an example refer to \cref{fig:incomplete}.
Using the monoid $(\nat\cup\{\infty\},\max,0)$, it is easy to see that the set $\set{x,y,z,u}$ is a footprint for the difference between $\aflowconstraint_1$ and $\aflowconstraint_2$.
Our fixed point, however, will start with $\set{x}$ and extend this to $\setnodespp\defeq\set{x,y,z}$.
Let $v$ be the node outside the flow graphs that $y$ is pointing to.
Then, the next transfer failure is $\tfailof{\aflowconstraint_1}{\aflowconstraint_2}{\setnodespp}=\set{v}$ because for $in<k$ the outflow of $y$ to $v$ differs in $\restrictto{\aflowconstraint_1}{\setnodespp}$ and $\restrictto{\aflowconstraint_2}{\setnodespp}$.
Our approach fails to compute a footprint.

\begin{fact}[Incompleteness]
	\label{Fact:Completeness}
	There are flow graphs $\aflowconstraint_1$ and $\aflowconstraint_2$ for which our algorithm is not able to determine a flow footprint although one exists.
\end{fact}

\subsection{Comparing Transfer Functions}

\looseness=-1
When implementing the above fixed point computation, the challenge is to prove the equivalence between given transfer functions in order to obtain the transfer failure: $[\transformerof{\restrictto{\aflowconstraint_1}{\setnodespp}}(-)](-, \anode) = [\transformerof{\restrictto{\aflowconstraint_2}{\setnodespp}}(-)](-, \anode)$?
Already the comparison of two functions is known to be difficult to do algorithmically.
What adds to the problem is that transfer functions are defined as least fixed points, meaning we do not have a closed-form representation of the functions to compare.

Our approach is to impose additional requirements on the set of edge functions.
The requirements are met in all our experiments, and so do not mean a limitation for the applicability of our approach.
We show that if the edge functions are not only continuous but also distributive, then the transfer functions can be understood in terms of paths through the underlying flow graphs.
If the edge functions are additionally decreasing and the underlying monoid's addition is idempotent, then acyclic paths are sufficient.
Both results do not hold for merely continuous edge functions.

\smartparagraph{Distributivity.}
Our first additional assumption is that the edge functions $\atfun:\amonoid\rightarrow\amonoid$ are not only continuous, but also \emph{distributive} in that $\atfunof{\amonval+\amonvalp}=\atfunof{\amonval}+\atfunof{\amonvalp}$ for all $\amonval, \amonvalp\in\amonoid$ and $\atfunof{0}=0$.
We use $\distfunof{\amonoid}$ to refer to the set of all continuous and distributive functions over $\amonoid$.
The properties formulated in \Cref{Lemma:FunctionDomains} carry over.


For continuous and distributive transfer functions, we can understand $\aflowconstraint[\inflow]^i$ in terms of the paths through $\aflowconstraint[\inflow]$ of length $i$.
For example, $i=3$ yields
\begin{align*}
	[\aflowconstraint[\inflow]^3](\bot)(\anodepp)
	\;&=\;
	\inflow_{\anodepp}+\sum_{\anodep\in\setnodes}\edges_{(\anodep, \anodepp)}(\; \inflow_{\anodep}+\sum_{\anode\in\setnodes}\edges_{(\anode, \anodep)}(\inflow_{\anode}+\sum_{u\in\setnodes}\edges_{(u, \anode)}(\bot(u))\; )
	\\
	\;&=\;
	\inflow_{\anodepp}+\sum_{\anodep\in\setnodes}\edges_{(\anodep, \anodepp)}(\inflow_{\anodep})+\sum_{\anodep\in\setnodes}\sum_{\anode\in\setnodes}\edges_{(\anodep, \anodepp)}( \edges_{(\anode, \anodep)}(\inflow_{\anode}) )
	\; .
\end{align*}
The first equality is by definition, the second is where distributivity comes in.
In particular, $\bot(u)=0$ and so $\edges_{(\anodep, \anodepp)}(\; \edges_{(\anode, \anodep)}(\; \edges_{(u, \anode)}(\; \bot(u)\; )\; )=0$.
The last term shows that we forward the inflow given at a node $\anode$ to an intermediary node $\anodep$ and from there to the node~$\anodepp$ of interest.
For higher powers of $\aflowconstraint[\inflow]$, we take longer paths.
For $\aflowconstraint[\inflow]^*$, we thus obtain the sum over all nodes $\anode$ and all paths from $\anode$ to $\anodepp$ through the flow graph.
We need some definitions to make this precise.

A \emph{path} $\apath$ through flow graph $\aflowconstraint$ is a finite, non-empty sequence of nodes all of which belong to the flow graph except the last which lies outside: \[
	\apath
	~~=~~
	\anode_0 \pc \ldots \pc \anode_n \pc \anodepp
	~~\in~~
	\setnodes^+ \pc (\nat\setminus\setnodes)
\]
where $\pc$ denotes path concatenation.
We use $\firstof{\apath}=\anode_0$ resp. $\lastof{\apath}=\anode_n$ to extract the first resp. last node from within the flow graph $\aflowconstraint$.
By $\setpathof{\aflowconstraint}{\anode}{\anodep, \anodepp}$ we denote the set of all paths through flow graph $\aflowconstraint$ that start in node $\firstof{\apath}=\anode$ and leave $\aflowconstraint$ from node $\lastof{\apath}=\anodep$ to move to $\anodepp\in\nat\setminus\setnodes$.
Given a set of nodes $\setnodes'\subseteq \setnodes$, we use $\setpathof{\aflowconstraint}{\setnodes'}{\anodep, \anodepp}$ for the union over all $\anode\in\setnodes'$ of the sets $\setpathof{\aflowconstraint}{\anode}{\anodep, \anodepp}$.
The path induces the function $\edges_{\apath}:\amonoid\rightarrow\amonoid$ that composes the edge functions along the path: \[
	\edges_{\anode}~=~\myid
	\qquad\qquad\qquad
	\edges_{\anode.\apath}~=~\edges_{\apath}\circ \edges_{(\anode, \firstof{\apath})}
	\ .
\]
Together with Lemma~\ref{Lemma:Transfer}, the above analysis yields the first closed-form representation of a flow graph's transfer function, which so far has involved a fixed point computation.

\begin{theorem}[Closed-Form Representation]
	\label{Theorem:ClosedForm}
	If $\aflowconstraint$ is labeled over $\distfunof{\amonoid}$, then:
	\[
		[\transformerof{\aflowconstraint}(\aninflow)](\anodep, \anodepp)
		\quad = \quad
		{\textstyle\sum_{\anode\,\in\,\setnodes}~ \sum_{\apath\,\in\,\setpathof{\aflowconstraint}{\anode}{\anodep, \anodepp}}}~~ \edges_{\apath}(\aninflow_{\anode})
		\ .
	\]
\end{theorem}

\Cref{Theorem:ClosedForm} pushes the fixed point computation of transfer functions into the sets $\setpathof{\aflowconstraint}{\anode}{\anodep, \anodepp}$ which are themselves defined inductively and potentially infinite.
In the following, we alleviate this problem without requiring acyclicity of the flow graph.

\smartparagraph{Idempotence.}
Our second assumption is that addition in the monoid is idempotent, meaning $\amonval+\amonval=\amonval$ for all $\amonval\in\amonoid$.
Idempotence ensures the addition degenerates to a join for comparable elements: $\amonval\prall{+}\amonvalp\prall{=}\amonval\prall{\join}\amonvalp\prall{=}\amonvalp$ for all $\amonval\leq\amonvalp\in\amonoid$.
Unless stated otherwise, we hereafter assume an idempotent addition.

With Theorem~\ref{Theorem:ClosedForm}, it remains to compare sums over paths.
With idempotence, we show that we can further reduce the problem and reason over single paths rather than sums.
We show that every path in $\aflowconstraint_1$ can be replaced by a set of paths in $\aflowconstraint_2$, and vice versa.
Even more, we only have to consider the paths from nodes where the edges changed.
The precise formulation of the path replacement condition is the following.

\begin{definition}
	\label{Definition:PathReplacement}
	The \emph{path replacement condition} for flow graphs $\aflowconstraint_1$ by $\aflowconstraint_2$ over the same set of nodes $\setnodes$ and labeled by $\distdecfunof{\amonoid}$ requires that for every $\anode\in \outof{\aflowconstraint_1}{\aflowconstraint_2}$, for every $\anodep\in\setnodes$, and for every $\anodepp\in\nat\setminus\setnodes$ we have \[
		\forall\, \apath\in\setpathof{\aflowconstraint_1}{\anode}{\anodep, \anodepp}~~
		\exists\, \setpath\subseteq \setpathof{\aflowconstraint_2}{\anode}{\anodep, \anodepp}.~~~
		\edges_{\apath}\; \leq\; \edges_{\setpath}\; \defeq\; {\textstyle\sum_{\apathp\in\setpath}}~ \edges_{\apathp}
		\ .
	\]
\end{definition}

\begin{example}
	For the flow graphs $\aflowconstraint_1$ and $\aflowconstraint_2$ from \cref{fig:incomplete}, we have path replacement of $\aflowconstraint_1$ by $\aflowconstraint_2$, and vice versa.
	To see this, consider the path $\apath \defeq x \pc z \pc u \pc y \pc v$ in $\aflowconstraint_1$ and $\apathp \defeq x \pc y \pc v$ in $\aflowconstraint_2$, where $v$ is the node outside of $\aflowconstraint_1,\aflowconstraint_2$ that $y$ points to.
	Since all edges are labeled with $\lambda_\mathit{id}$, we have $\edges_{\apath}=\lambda_\mathit{id}=\edges_{\apathp}$.
	It is worth noting that, in this example, we can ignore the cycles in $\aflowconstraint_1$ and $\aflowconstraint_2$.
	In a moment, we will introduce restrictions on edge functions in order to do avoid cycles in general.

	Similarly, we have path replacement for the flow graphs from \cref{fig-keyset-flow}.
	To be precise, $\edges_\apath = \lambda_8 = \edges_\apathp$ for the paths $\apath \defeq l \pc t \pc r \pc v$ in $\aflowconstraint_1$ and $\apathp \defeq l \pc r \pc v$ in $\aflowconstraint_2$.
	\hfill\qed
\end{example}

The main result is that path replacement is sound and complete for proving equivalence of transfer functions.

\begin{theorem}[Path Replacement Principle]\label{Theorem:PathReplacement}
	We have $\transformerof{\aflowconstraint_1}=\transformerof{\aflowconstraint_2}$ if and only if path replacement of $\aflowconstraint_1$ by $\aflowconstraint_2$ and of $\aflowconstraint_2$ by $\aflowconstraint_1$ hold.
\end{theorem}

The theorem is remarkable in several respects.
First, one would expect we have to replace the paths from all nodes in $\aflowconstraint_1$.
Instead, we can focus on the nodes where the outgoing edges changed.
Second, one would expect the replacing paths $\setpath$ start from arbitrary nodes in $\aflowconstraint_2$.
Such a set of paths would yield a transfer function of type $(\setvars\prall{\rightarrow}\amonoid)\prall{\rightarrow}\amonoid$.
Instead, we can work with a function of type ${\amonoid\prall{\rightarrow}\amonoid}$.
Even more, we can focus on paths starting in the same node as the path we intend to replace.
Finally, the paths we use for replacement come without any constraints, leaving room for heuristics.

The proof starts from a \emph{full path replacement condition} of $\aflowconstraint_1$ by $\aflowconstraint_2$, both over~$\setnodes$ and labeled by $\distfunof{\amonoid}$.
Full path replacement coincides with \Cref{Definition:PathReplacement} but draws $\anode$ from full $\setnodes$ rather than $\anode\in\outof{\aflowconstraint_1}{\aflowconstraint_2}$.
Full path replacement characterizes equivalence of the transfer functions in a monoid with idempotent addition in the case of continuous and distributive edge functions.

\begin{lemma}\label{Lemma:CharFullPathReplacement}
	We have $\transformerof{\aflowconstraint_1}=\transformerof{\aflowconstraint_2}$ if and only if full path replacement of $\aflowconstraint_1$ by $\aflowconstraint_2$ and of $\aflowconstraint_2$ by $\aflowconstraint_1$ hold.
\end{lemma}

The result is a consequence of \Cref{Theorem:ClosedForm}, which equates $\transformerof{\aflowconstraint_1}$ with the sum of the $\edges_{\apath}$ for all paths $\apath\in\setpathof{\aflowconstraint_1}{\anode}{\anodep, \anodepp}$ for all $\anode\in\setnodes$.
Full path replacement allows us to sum over $\edges_{\setpath}$ instead, for some $\setpath\subseteq\setpathof{\aflowconstraint_2}{\anode}{\anodep, \anodepp}$.
Over-approximating $\setpath$ with all paths $\setpathof{\aflowconstraint_2}{\anode}{\anodep, \anodepp}$, we obtain an upper bound for $\transformerof{\aflowconstraint_1}$.
It is easy to see that the resulting sum can be rewritten into the form of \Cref{Theorem:ClosedForm}, yielding $\transformerof{\aflowconstraint_1}\leq\transformerof{\aflowconstraint_2}$.
Analogously, we get $\transformerof{\aflowconstraint_1}\geq\transformerof{\aflowconstraint_2}$ and thus $\transformerof{\aflowconstraint_1}=\transformerof{\aflowconstraint_2}$ as required.
The reverse direction of the \namecref{Lemma:CharFullPathReplacement} is similar.

To conclude the proof of the path replacement principle in \Cref{Theorem:PathReplacement}, the following \namecref{Lemma:FullVsPathReplacement} states that full path replacement and (ordinary) path replacement of $\aflowconstraint_1$ by $\aflowconstraint_2$ coincide.
To see this, 
consider a path $\apath\in\setpathof{\aflowconstraint_1}{\anode}{\anodep,\anodepp}$ for any $\anode\in\setnodes$.
The goal is to show $\edges_\apath\leq\edges_\setpath$ for some $\setpath\in\setpathof{\aflowconstraint_2}{\anode}{\anodep,\anodepp}$.
To that end, decompose the path into $\apath=\apath_1 \pc \apath_2$ such that $\anode'\defeq\firstof{\apath_2}$ is the first node in $\apath$ from $\outof{\aflowconstraint_1}{\aflowconstraint_2}$.
Ordinary path replacement yields $\setpathp\in\setpathof{\aflowconstraint_2}{\anode'}{\anodep,\anodepp}$ with $\edges_{\apath_2}\leq\edges_\setpathp$.
Now, choose $\setpath\defeq\setcond{\apath_1 \pc \apathp}{\apathp\in\setpathp}$.
Because $\apath_1$ exists in $\aflowconstraint_1$ and $\aflowconstraint_2$ with the exact same edge labels, we immediately obtain the desired $\edges_{\apath}\leq\edges_\setpath$.

\begin{lemma}\label{Lemma:FullVsPathReplacement}
	Full path replacement of $\aflowconstraint_1$ by $\aflowconstraint_2$ holds if and only if path replacement of $\aflowconstraint_1$ by $\aflowconstraint_2$ holds.
\end{lemma}

\smartparagraph{Decreasingness.}
We assume that the edge functions $\atfun:\amonoid\rightarrow\amonoid$ are not only continuous and distributive, but also \emph{decreasing}: $\atfunof{\amonval}\leq \amonval$ for all ${\amonval\in\amonoid}$.
The assumption of decreasing edge functions is justified by the fact that a program that traverses the flow graph builds up information about the status of the structure, and smaller flow values mean more information (as in classical data flow analysis).
We use $\distdecfunof{\amonoid}$ to refer to the set of all continuous, distributive, and decreasing transfer functions over~$\amonoid$; \Cref{Lemma:FunctionDomains} carries over to this set.
Addition in the monoid is still assumed idempotent.


If all edge functions are decreasing, every cycle in the flow graph is decreasing as well.
The key observation is that, given an idempotent addition, cycles with decreasing edge functions can be avoided when forming sums over sets of paths.

\begin{lemma}\label{Lemma:Decreasing}
	Let $\aflowconstraint$ be labeled over $\distdecfunof{\amonoid}$ and $\apath_1\pc\apath\pc\apath_2\in \setpathof{\aflowconstraint}{\anode}{\anodep, \anodepp}$ with $\lastof{\apath}=\firstof{\apath}$.
	Then $\apath_1\pc\apath_2\in \setpathof{\aflowconstraint}{\anode}{\anodep, \anodepp}$ and $\edges_{\apath_1\pc\apath\pc\apath_2}\leq \edges_{\apath_1\pc\apath_2}$.
\end{lemma}

Call a path \emph{simple} if it does not repeat a node and let $\setsimplepathof{\aflowconstraint}{\anode}{\anodep, \anodepp}$ denote the set of all simple paths through $\aflowconstraint$ from $\anode$ to $\anodep$ and leaving the flow graph towards $\anodepp$.
Note that a finite graph only admits finitely many simple paths.

\begin{theorem}[Simple Paths]
	\label{Theorem:SimplePaths}
	Assuming continuous, distributive, and decreasing edge functions, and assuming idempotent addition,
	Theorem~\ref{Theorem:ClosedForm} and Theorem~\ref{Theorem:PathReplacement} hold with every occurrency of $\setpathof{\aflowconstraint}{\anode}{\anodep, \anodepp}$ replaced by $\setsimplepathof{\aflowconstraint}{\anode}{\anodep, \anodepp}$.
\end{theorem}

In practice, path-counting flows, keyset flows, reachability flows, shortest-path flows, and priority inheritance flows are relevant~\cite{oopsla,DBLP:journals/pacmpl/KrishnaSW18,DBLP:conf/esop/KrishnaSW20,DBLP:conf/pldi/KrishnaPSW20} and compatible with our theory.


\newcommand{\evalsubsection}[1]{\subsubsection*{#1}}
\newcommand{\mGeneral}{\textsc{naive}\xspace}
\newcommand{\mDist}{\textsc{dist}\xspace}
\newcommand{\mNew}{\textsc{new}\xspace}
\newcommand{\evalReps}{$1000$\xspace}

\section{Evaluation}
\label{sec:eval}

We substantiate the practicality of our new approach by evaluating it on a real-world collection of flow graphs extracted from the literature.
We explain how we obtained our benchmarks and how we implemented and evaluated our approach.

\smartparagraph{Benchmark Suite.}
As alluded to in \cref{sec:intro}, the flow framework has been used to verify complex concurrent data structures.
More specifically, it has been used for automated proof construction by the \plankton tool \cite{oopsla,plankton}.
\plankton performs an exhaustive proof search over a separation logic with support for flows---and further advanced features for establishing linearizability that do not matter for the present evaluation.
In order to handle heap updates, \plankton generates a footprint $\aflowconstraint$ for the flow graph ${\aflowconstraint_1=\aflowconstraint \mstar \aflowconstraint_\mathit{frame}}$ of the current proof state (represented as an assertion in separation logic).
It then frames the non-footprint part $\aflowconstraint_\mathit{frame}$ of the flow graph $\aflowconstraint_1$ to compute the post state $\aflowconstraint'$ of the heap update locally for the footprint $\aflowconstraint$.
The result is the new flow graph ${\aflowconstraint_2=\aflowconstraint' \mstar \aflowconstraint_\mathit{frame}}$.
We consider the pair $(\aflowconstraint_1,\aflowconstraint_2)$ a \textit{benchmark} for our evaluation.

We adapt \plankton to export the flow graph pairs for which a footprint is constructed.
This way, we obtain $1272$ benchmarks from the heap updates occurring during proof construction for a collection of 10 concurrent set data structures.
All flow graphs in this benchmark suite contain at most $4$ nodes.

Our benchmark suite is limited by the capabilities and restrictions of \plankton.
In particular, we inherit the confinement to concurrent search structures.
This is due to the fact that \plankton integrates support only for the keyset flow (cf. \Cref{ex-keyset-flow}).
Our evaluation will compute footprints with respect to this flow.

\smartparagraph{Implementation.}
We implement the fixed point computation to find footprints for two given flow graphs $\aflowconstraint_1,\aflowconstraint_2$ from \cref{Section:Footprints} in a tool called \krill \cite{artifact}.
It integrates three methods for computing the transfer failure $\tfailof{\aflowconstraint_1}{\aflowconstraint_2}{\setnodespp}$ of a footprint candidate~$\setnodespp$:
\begin{compactenum}
	\item \textbf{\mGeneral:}
		A naive method that computes the flow within the footprint $\setnodespp$.
		Following \cite{DBLP:conf/esop/KrishnaSW20}, we require acyclicity of flow graphs for this method to avoid solving a fixed point equation when computing the flow.

	\item \textbf{\mNew:}
		Our new approach leveraging the path replacement condition (cf. \Cref{Theorem:PathReplacement}) for simple paths (cf. \Cref{Theorem:SimplePaths}).
		This method requires distributive and decreasing edge functions as well as idempotent addition in the underlying monoid.

	\item \textbf{\mDist:}
		A variation of our new approach leveraging the closed-form representation (cf. \Cref{Theorem:ClosedForm}).
		This method requires distributive edge functions as well as acyclicity of the flow graphs to avoid an unbounded sum over all paths in the closed-form representation.
\end{compactenum}
Our benchmark suite satisfies the requirements for all three methods.
The \mGeneral and \mDist methods include a (sufficient) check to ensure acyclicity in the updated flow graph to guarantee soundness of the resulting footprint.

All three methods encode the necessary equivalence checks among transfer functions as SMT formulas which are then discharged using the off-the-shelf SMT solver \atoolname{Z3} \cite{DBLP:conf/tacas/MouraB08}.
Our encodings use the theory of integers with quantifiers.
The \mGeneral method additionally uses free functions to encode sets of integers.

\begin{figure}[t]
	\newcommand{\tGraph}[5]{#1}%
	\newcommand{\msec}{\mkern+2mu\mathrm{ms}}%
	\newcommand{\tTime}[2]{$#1\msec$}%
	\newcommand{\SpUp}[1]{$#1\%$}%
	\newcolumntype{Z}{>{\raggedleft\arraybackslash}m{1.1cm}}%
	\newcolumntype{Y}{>{\raggedleft\arraybackslash}m{1cm}}%
	\vspace{-8mm}
	\begin{minipage}{.65\textwidth}
		\begin{table}[H]
			\small
			\smaller
			\begin{tabularx}{\textwidth}{X r *{3}{Z} Y}
				\toprule
				Structure & ~\#FG & \mGeneral & \mDist & \mNew & Speedup \\ 
				\midrule
				Fine set \cite{DBLP:books/daglib/0020056}
					& \tGraph{  12}{  2}{  2}{  8}{  0}
					& \tTime{ 75 }{ 555.06 } & \tTime{ 48 }{ 352.15 } & \tTime{ 46 }{ 296.37 }
					& \SpUp{39}
					\\
				Lazy set \cite{DBLP:conf/opodis/HellerHLMSS05}
					& \tGraph{  14}{  4}{  4}{  6}{  0}
					& \tTime{ 73 }{ 289.16 } & \tTime{ 52 }{ 207.36 } & \tTime{ 51 }{ 208.70 }
					& \SpUp{30}
					\\
				ORVYY set \cite{DBLP:conf/podc/OHearnRVYY10}
					& \tGraph{  20}{  4}{  6}{ 10}{  0}
					& \tTime{ 106 }{ 252.01 } & \tTime{ 76 }{ 180.64 } & \tTime{ 74 }{ 154.92 }
					& \SpUp{30}
					\\
				VY DCAS set \cite{DBLP:conf/pldi/VechevY08}
					& \tGraph{  19}{  2}{  6}{ 11}{  0}
					& \tTime{ 109 }{ 225.73 } & \tTime{ 74 }{ 164.04 } & \tTime{ 73 }{ 142.30 }
					& \SpUp{33}
					\\
				VY CAS set \cite{DBLP:conf/pldi/VechevY08}
					& \tGraph{  28}{  4}{ 14}{ 10}{  0}
					& \tTime{ 139 }{ 678.17 } & \tTime{ 104 }{ 475.98 } & \tTime{ 102 }{ 402.79 }
					& \SpUp{27}
					\\
				Michael set \cite{DBLP:conf/spaa/Michael02}
					& \tGraph{ 225}{  4}{ 55}{166}{  0}
					& \tTime{ 1216 }{ 1177.44 } & \tTime{ 887 }{ 892.28 } & \tTime{ 874 }{ 949.32 }
					& \SpUp{28}
					\\
				Michael set (wait-free)
					& \tGraph{ 186}{  4}{ 47}{135}{  0}
					& \tTime{ 996 }{ 1300.99 } & \tTime{ 731 }{ 1009.06 } & \tTime{ 721 }{ 1027.49 }
					& \SpUp{27}
					\\
				Harris set \cite{DBLP:conf/wdag/Harris01}
					& \tGraph{ 352}{  4}{ 46}{144}{158}
					& \tTime{ 2242 }{ 3161.26 } & \tTime{ 1490 }{ 2434.80 } & \tTime{ 1443 }{ 2719.26 }
					& \SpUp{36}
					\\
				Harris set (wait-free)
					& \tGraph{ 296}{  4}{ 42}{122}{128}
					& \tTime{ 1859 }{ 2093.17 } & \tTime{ 1242 }{ 1666.86 } & \tTime{ 1205 }{ 2010.62 }
					& \SpUp{35}
					\\
				FEMRS tree \cite{DBLP:conf/wdag/FeldmanE0RS18}
					& \tGraph{ 120}{100}{ 20}{  0}{  0}
					& \tTime{ 519 }{ 581.00 } & \tTime{ 409 }{ 482.69 } & \tTime{ 407 }{ 494.47 }
					& \SpUp{22}
					\\
				\midrule
				Total
					& \tGraph{1272}{132}{242}{612}{286}
					& \tTime{ 7335 }{ 4582.30 } & \tTime{ 5114 }{ 3484.42 } & \tTime{ 4996 }{ 3901.20 }
					& \SpUp{32}
					\\
				\bottomrule
			\end{tabularx}
		\end{table}
	\end{minipage}
	\hfill
	\begin{minipage}{.323\textwidth}
		\begin{figure}[H]
			\begin{tikzpicture}
				\useasboundingbox (-.5,-.375) rectangle (3.445,4.09);
				\scriptsize
				\node[] at (-.2,3.15) {ms};
				\begin{axis}[
						legend style={at={(0.5,1.217)}, anchor=north, legend columns=-1},
						enlargelimits={abs=0.5},
						ybar=0pt,
						bar width=0.25,
						xtick={0.5,1.5,...,5.5},
						xticklabels={$\top$,1,2,3},
						x tick label as interval,
						x tick label style={yshift=1mm},
						ytick={0.5,1.0,...,7.0},
						width=5.025cm,
						height=4.9cm,
						legend cell align={left},
						area legend,
					]
					\addplot+[postaction={opacity=.375,pattern=crosshatch dots},error bars/.cd,y dir=both,y explicit] coordinates {
						(1, 5.892347) +- (0.0, 1.274778)
						(2, 3.92564) +- (0.0, 0.124725)
						(3, 6.181668) +- (0.0, 0.09928)
						(4, 6.628586) +- (0.0, 0.238186)
					};
					\addplot+[postaction={opacity=.375,pattern=north east lines},error bars/.cd,y dir=both,y explicit] coordinates {
						(1, 4.078146) +- (0.0, 0.272622)
						(2, 3.365953) +- (0.0, 0.075988)
						(3, 3.843186) +- (0.0, 0.028072)
						(4, 4.326741) +- (0.0, 0.123034)
					};
					\addplot+[postaction={opacity=.375,pattern=north west lines},error bars/.cd,y dir=both,y explicit] coordinates { 
						(1, 4.002394) +- (0.0, 0.167081)
						(2, 3.350851) +- (0.0, 0.075516)
						(3, 3.779711) +- (0.0, 0.03127)
						(4, 4.168013) +- (0.0, 0.112655)
					};
					\legend{
						\footnotesize{\mGeneral},
						\footnotesize{\mDist},
						\footnotesize{\mNew}
					}
				\end{axis}
			\end{tikzpicture}
		\end{figure}
	\end{minipage}
	\vspace{-1mm}
	\caption{
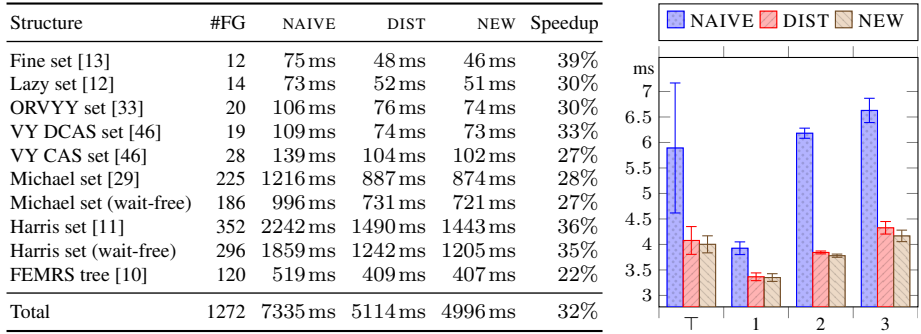
		Experimental results averaged over \evalReps repeated runs, conducted on an Apple M1 Pro.
		\textbf{(left)} Total runtime for computing footprints for flow graphs occurring during automated proof construction for highly concurrent set data structures.
		The speedup gives the relative performance improvement of \mNew over \mGeneral.
		\textbf{(right)} Average runtime for computing a single footprint, partitioned by footprint size ($\top$ indicates failure).%
		\label{fig:footprint-benchmark}%
		\label{fig:footprint-size}%
	}
	\vspace{-2mm}
\end{figure}

\smartparagraph{Experiments.}
We ran \krill on our benchmark suite and compared the runtime of the three different methods for computing the transfer failure.
Our results are summarized in \cref{fig:footprint-benchmark}(left).
For every search structure that we extracted benchmarks from, the figure lists:
\begin{inparaenum}[(i)]
	\item the number \#FG of flow graph pairs extracted,
	\item each method's total runtime for computing the footprints of all flow graph pairs, and
	\item the speedup of \mNew over \mGeneral in percent.
\end{inparaenum}
The experiments were conducted on an Apple M1 Pro.

\Cref{fig:footprint-benchmark}(left) shows that the runtime for all methods is roughly linear in the number of computed footprints.
Moreover, the absolute time for computing footprints is small, making the approaches practical.
The figure also shows that our \mNew and \mDist methods have a performance advantage over the \mGeneral method.
The \mNew method is between $22\%$ and $39\%$ faster than the \mGeneral method.
We believe that the difference is relatively small only because the acyclicity assumption avoids a potentially non-terminating fixed point computation.
Avoiding this fixed point in the presence of cycles is a major advantage that our \mNew method has over the \mGeneral and \mDist methods.
The performance difference for \mDist and \mNew are negligible because the acyclicity check is negligible.

We also factorized the runtimes of our benchmarks along the size of the resulting footprint.
\Cref{fig:footprint-size}(right) gives the average runtime and standard deviation for computing a single footprint, broken down by footprint size.
If no footprint could be found, its size is listed as $\top$.
These failed footprint constructions are consistent with \plankton's method and would not lead to verification failure.


\section{Related Work}
\label{sec:related}

Two alternative meta theories for the flow framework have been proposed in prior work~\cite{DBLP:journals/pacmpl/KrishnaSW18,DBLP:conf/esop/KrishnaSW20}.
Like in our setup, the original flow framework~\cite{DBLP:journals/pacmpl/KrishnaSW18} demands that the flow domain is an $\omega$-cpo to obtain a least fixed point semantics.
However, it proposes a different flow graph composition that leads to a notion of contextual equivalence relying on inflow equivalence classes.
This complicates proof automation.
In addition, the flow domain is assumed to be a semiring and edge functions are restricted to multiplication with a constant.
This limits expressivity.

As discussed in \cref{sec:intro}, the revised flow framework proposed in~\cite{DBLP:conf/esop/KrishnaSW20} requires that the flow monoid is cancellative but not an $\omega$-cpo.
This means that uniqueness of flows is not guaranteed per se.
Instead, uniqueness is obtained by imposing additional conditions on the edge functions.
However, these conditions are more restrictive than those imposed in our framework.
The \emph{capacity} of a flow graph introduced in \cite{DBLP:conf/esop/KrishnaSW20} closely relates to our notion of transfer function.
A closed-form representation based on sums over paths is used to check equivalence of capacities.
However, this reasoning is restricted to acyclic graphs.
Also, \cite{DBLP:conf/esop/KrishnaSW20} provides no algorithm for computing flow footprints. 

In a sense, our work strikes a balance between the two prior meta theories by guaranteeing unique flows without sacrificing expressivity and, at the same time, enabling better proof automation.
That said, we believe that the framework proposed in~\cite{DBLP:conf/esop/KrishnaSW20} remains of independent interest, in particular if the application does not require unique flows (i.e., does not impose lower bounds on flows that may trivially hold in the presence of vanishing flows).
Cancellativity allows one to aggregate inflows and outflows to unary functions, which can lead to smaller flow footprints (i.e., more local proofs).

The benchmark suite for our evaluation is obtained from \plankton \cite{plankton,oopsla}, a tool for verifying concurrent search structures using keyset flows.
When the program mutates the symbolic heap, \plankton creates a flow graph for the mutated nodes plus all nodes with a distance of $k$ or less from those nodes.
This flow graph is considered to be the footprint and contextual equivalence is checked.
The check is basically the same as for \mGeneral.
However, the paper does not present the meta theory for the underlying notion of flow graphs, nor does it provide any justification for the correctness of the implemented algorithms used to reason about flow graphs.

Flow graphs form a separation algebra.
Hence, the developed theory can be used in combination with any existing separation logic that is parametric in the underlying separation algebra such as~\cite{DBLP:conf/lics/CalcagnoOY07, DBLP:conf/popl/Dinsdale-YoungBGPY13, DBLP:conf/ecoop/PintoDG14, DBLP:conf/pldi/SergeyNB15, DBLP:journals/jfp/JungKJBBD18, oopsla}.
Identifying footprints of updates relates to the frame inference problem in separation logic, which has been studied extensively~\cite{DBLP:conf/aplas/BerdineCO05,DBLP:conf/popl/CalcagnoDOY09,DBLP:conf/cav/PiskacWZ13,DBLP:conf/cav/PiskacWZ14,DBLP:conf/cpp/RoweB17,DBLP:conf/tacas/Le0Q18,DBLP:conf/ecoop/HolikPRSVZ22}.
However, existing work focuses on frame inference for assertions that are expressed in terms of inductive predicates.
These techniques are not well-suited for reasoning about programs manipulating general graphs, including overlayed structures, which are often used in practice and easily expressed using flows.
A common approach to reason about general heap graphs in separation logic is to use iterated separating conjunction~\cite{Yang01ShorrWaite,DBLP:conf/popl/HoborV13,DBLP:conf/pldi/SergeyNB15, DBLP:conf/aplas/RaadHVG16} to abstract the heap by a \emph{pure} graph that does not depend on the program state.
Though, the verification of specifications that rely on inductive properties of the pure graph then resorts back to classical first-order reasoning and is difficult to automate.
An exception is~\cite{DBLP:journals/pacmpl/Ter-GabrielyanS19} which uses SMT solvers to frame binary reachability relations in graphs that are described by iterated separating conjunctions. 
However, the technique is restricted to such reachability properties only.

Unbounded footprints have been encountered early on when computing the post image for recursive predicates \cite{DBLP:conf/tacas/DistefanoOY06}.
This has spawned interest in separation logic fragments for which the reasoning can be efficiently automated \cite{DBLP:conf/fsttcs/BerdineCO04,DBLP:conf/fmco/BerdineCO05,DBLP:conf/cav/PiskacWZ13,DBLP:conf/atva/IosifRV14,DBLP:conf/nfm/EneaLSV17,DBLP:conf/vmcai/QiuW19,DBLP:conf/lpar/KatelaanZ20}.
A limitation that underlies all these works is an assumption of tree-regularity of the heap, in one way or another, which flows have been designed to overcome.
In cases where the program (or ghost code) traverses the unbounded footprint (before or after the update), recent works \cite{DBLP:conf/esop/KrishnaSW20,oopsla} have found a way to reduce the reasoning to bounded footprint chunks.

As pointed out in \cref{sec:intro}, the definition of a flow closely resembles the classical formulation of a forward data flow analysis.
In particular, the fact that the least fixed point of the flow equation for distributive edge functions can be characterized as a join over all paths in the flow graph mirrors dual results for greatest fixed points in data flow analysis~\cite{DBLP:conf/popl/Kildall73,DBLP:journals/acta/KamU77}.
In a similar vein, the notion of contextual equivalence of flow graphs relates to contextual program equivalence and fully abstract models in denotational semantics~\cite{DBLP:journals/tcs/Milner77,DBLP:journals/tcs/Plotkin77,DBLP:journals/iandc/HylandO00}.
In fact, Beki\'c's Lemma~\cite{DBLP:conf/ibm/Bekic84e}, which we use in the proofs of \Cref{Theorem:FramePreserving,Lemma:Inflow}, was originally motivated by the study of such models.
Flow graphs can serve as abstractions of programs (rather than just program states).
We therefore believe that our results could also be of interest for developing incremental and compositional data flow analysis frameworks.


\subsection*{Data Availability Statement}

The \krill artifact and dataset generated and/or analysed in the present paper
are available in the Zenodo repository \cite{artifact}, \url{https://zenodo.org/record/7566204}.

\subsection*{Acknowledgments}

This work is funded in part by NSF grant 1815633.
The first author was supported by the DFG project \emph{EDS@SYN: Effective Denotational Semantics for Synthesis}. 
The third author is supported by a Junior Fellowship from the Simons Foundation (855328, SW).

	\bibliographystyle{splncs04}
	\bibliography{bib}
	
	\clearpage
	\appendix
	{

\section{Evaluation Details}

\smartparagraph{Detailed Runtimes}
We repeat an upscaled version of \cref{fig:footprint-size}(right) in \cref{fig:details:footprints-overall}, showing the average runtime for computing a single footprint, partitioned by footprint size.
In \cref{fig:details:footprints-individual} we additionally partition the runtime for computing a single footprint by the data structure the footprint is taken from.

\begin{figure}[bth]
	\centering
	\begin{tikzpicture}
		\begin{axis}[xticklabels={\(\top\),1,2,3},legend pos=outer north east,legend cell align={left},enlargelimits={abs=0.5},ybar=0pt,xtick={0.5,1.5,...,100},x tick label as interval,ytick={0,1,...,8},ymin=0,ymax=8,axis x line shift=-.5,width=9cm,height=7cm,area legend]
			\addplot+[postaction={opacity=.375,pattern=crosshatch dots},error bars/.cd,y dir=both,y explicit] coordinates {
				(1, 5.892347) +- (0.0, 1.274778)
				(2, 3.92564) +- (0.0, 0.124725)
				(3, 6.181668) +- (0.0, 0.09928)
				(4, 6.628586) +- (0.0, 0.238186)
			};
			\addplot+[postaction={opacity=.375,pattern=north east lines},error bars/.cd,y dir=both,y explicit] coordinates {
				(1, 4.078146) +- (0.0, 0.272622)
				(2, 3.365953) +- (0.0, 0.075988)
				(3, 3.843186) +- (0.0, 0.028072)
				(4, 4.326741) +- (0.0, 0.123034)
			};
			\addplot+[postaction={opacity=.375,pattern=north west lines},error bars/.cd,y dir=both,y explicit] coordinates {
				(1, 4.002394) +- (0.0, 0.167081)
				(2, 3.350851) +- (0.0, 0.075516)
				(3, 3.779711) +- (0.0, 0.03127)
				(4, 4.168013) +- (0.0, 0.112655)
			};
			\legend{\mGeneral,\mDist,\mNew}
		\end{axis}
	\end{tikzpicture}
	\caption{%
		Average runtime for computing a single footprint, partitioned by footprint size ($\top$ indicates failure).
		Repeated from \cref{fig:footprint-size}(right).
		\label{fig:details:footprints-overall}
	}
\end{figure}
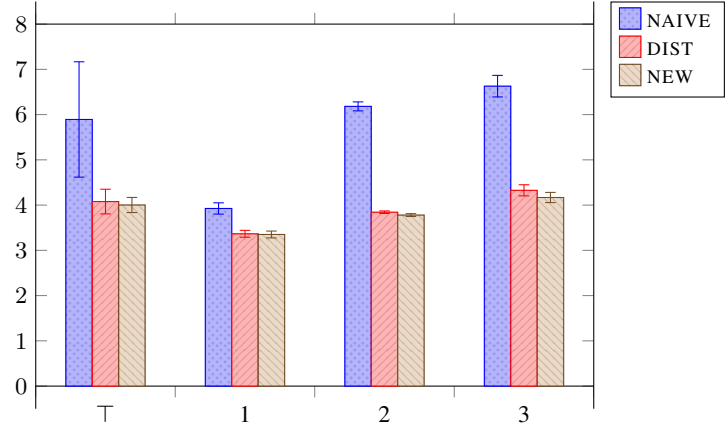

\begin{figure}[ptbh]
	\centering
	\begin{subfigure}[t]{.3\textwidth}
		\centering
		\begin{tikzpicture}
			\begin{axis}[xticklabels={\(\top\),1,2,3},width=5cm, height=5.25cm,legend pos=outer north east,legend cell align={left},enlargelimits={abs=0.5},bar width=.2,x tick label style={yshift=1mm},ytick={0,1,...,10},ybar=0pt,xtick={0.5,1.5,...,100},x tick label as interval,ymin=0,ymax=9,axis x line shift=-.5,area legend]
				\addplot+[postaction={opacity=.375,pattern=crosshatch dots},error bars/.cd,y dir=both,y explicit] coordinates {
					(1, 8.583695) +- (0.0, 0.088646)
					(2, 3.863978) +- (0.0, 0.161595)
					(3, 0.0) +- (0.0, 0.0)
					(4, 6.423856) +- (0.0, 0.076595)
				};
				\addplot+[postaction={opacity=.375,pattern=north east lines},error bars/.cd,y dir=both,y explicit] coordinates {
					(1, 4.522023) +- (0.0, 0.051342)
					(2, 3.343981) +- (0.0, 0.066975)
					(3, 0.0) +- (0.0, 0.0)
					(4, 4.186293) +- (0.0, 0.048885)
				};
				\addplot+[postaction={opacity=.375,pattern=north west lines},error bars/.cd,y dir=both,y explicit] coordinates {
					(1, 4.137858) +- (0.0, 0.049104)
					(2, 3.32257) +- (0.0, 0.060737)
					(3, 0.0) +- (0.0, 0.0)
					(4, 4.033509) +- (0.0, 0.045047)
				};
			\end{axis}
		\end{tikzpicture}
		\vspace{-6mm}
		\caption{Fine set \cite{DBLP:books/daglib/0020056}.}
		\vspace{2mm}
	\end{subfigure}
	~~~~
	\begin{subfigure}[t]{.3\textwidth}
		\centering
		\begin{tikzpicture}
			\begin{axis}[xticklabels={\(\top\),1,2,3},width=5cm, height=5.25cm,legend pos=outer north east,legend cell align={left},enlargelimits={abs=0.5},bar width=.2,x tick label style={yshift=1mm},ytick={0,1,...,10},ybar=0pt,xtick={0.5,1.5,...,100},x tick label as interval,ymin=0,ymax=9,axis x line shift=-.5,area legend]
				\addplot+[postaction={opacity=.375,pattern=crosshatch dots},error bars/.cd,y dir=both,y explicit] coordinates {
					(1, 8.542236) +- (0.0, 0.038354)
					(2, 3.806961) +- (0.0, 0.122405)
					(3, 0.0) +- (0.0, 0.0)
					(4, 6.439963) +- (0.0, 0.064944)
				};
				\addplot+[postaction={opacity=.375,pattern=north east lines},error bars/.cd,y dir=both,y explicit] coordinates {
					(1, 4.504589) +- (0.0, 0.024203)
					(2, 3.3358) +- (0.0, 0.068385)
					(3, 0.0) +- (0.0, 0.0)
					(4, 4.196519) +- (0.0, 0.040164)
				};
				\addplot+[postaction={opacity=.375,pattern=north west lines},error bars/.cd,y dir=both,y explicit] coordinates {
					(1, 4.125423) +- (0.0, 0.021101)
					(2, 3.32106) +- (0.0, 0.070379)
					(3, 0.0) +- (0.0, 0.0)
					(4, 4.046613) +- (0.0, 0.034265)
				};
			\end{axis}
		\end{tikzpicture}
		\vspace{-6mm}
		\caption{Lazy set \cite{DBLP:conf/opodis/HellerHLMSS05}.}
		\vspace{2mm}
	\end{subfigure}
	~~~~
	\begin{subfigure}[t]{.3\textwidth}
		\centering
		\begin{tikzpicture}
			\begin{axis}[xticklabels={\(\top\),1,2,3},width=5cm, height=5.25cm,legend pos=outer north east,legend cell align={left},enlargelimits={abs=0.5},bar width=.2,x tick label style={yshift=1mm},ytick={0,1,...,10},ybar=0pt,xtick={0.5,1.5,...,100},x tick label as interval,ymin=0,ymax=9,axis x line shift=-.5,area legend]
				\addplot+[postaction={opacity=.375,pattern=crosshatch dots},error bars/.cd,y dir=both,y explicit] coordinates {
					(1, 0.0) +- (0.0, 0.0)
					(2, 3.95107) +- (0.0, 0.127434)
					(3, 6.181668) +- (0.0, 0.09928)
					(4, 0.0) +- (0.0, 0.0)
				};
				\addplot+[postaction={opacity=.375,pattern=north east lines},error bars/.cd,y dir=both,y explicit] coordinates {
					(1, 0.0) +- (0.0, 0.0)
					(2, 3.325771) +- (0.0, 0.05764)
					(3, 3.843186) +- (0.0, 0.028072)
					(4, 0.0) +- (0.0, 0.0)
				};
				\addplot+[postaction={opacity=.375,pattern=north west lines},error bars/.cd,y dir=both,y explicit] coordinates {
					(1, 0.0) +- (0.0, 0.0)
					(2, 3.311252) +- (0.0, 0.056226)
					(3, 3.779711) +- (0.0, 0.03127)
					(4, 0.0) +- (0.0, 0.0)
				};
			\end{axis}
		\end{tikzpicture}
		\vspace{-6mm}
		\caption{FEMRS tree \cite{DBLP:conf/wdag/FeldmanE0RS18}.}
		\vspace{2mm}
	\end{subfigure}
	\\
	\begin{subfigure}[t]{.3\textwidth}
		\centering
		\begin{tikzpicture}
			\begin{axis}[xticklabels={\(\top\),1,2,3},width=5cm, height=5.25cm,legend pos=outer north east,legend cell align={left},enlargelimits={abs=0.5},bar width=.2,x tick label style={yshift=1mm},ytick={0,1,...,10},ybar=0pt,xtick={0.5,1.5,...,100},x tick label as interval,ymin=0,ymax=9,axis x line shift=-.5,area legend]
				\addplot+[postaction={opacity=.375,pattern=crosshatch dots},error bars/.cd,y dir=both,y explicit] coordinates {
					(1, 0.0) +- (0.0, 0.0)
					(2, 3.858988) +- (0.0, 0.126553)
					(3, 0.0) +- (0.0, 0.0)
					(4, 6.751128) +- (0.0, 0.427192)
				};
				\addplot+[postaction={opacity=.375,pattern=north east lines},error bars/.cd,y dir=both,y explicit] coordinates {
					(1, 0.0) +- (0.0, 0.0)
					(2, 3.319179) +- (0.0, 0.045031)
					(3, 0.0) +- (0.0, 0.0)
					(4, 4.231862) +- (0.0, 0.06334)
				};
				\addplot+[postaction={opacity=.375,pattern=north west lines},error bars/.cd,y dir=both,y explicit] coordinates {
					(1, 0.0) +- (0.0, 0.0)
					(2, 3.30448) +- (0.0, 0.043372)
					(3, 0.0) +- (0.0, 0.0)
					(4, 4.114569) +- (0.0, 0.066999)
				};
			\end{axis}
		\end{tikzpicture}
		\vspace{-6mm}
		\caption{ORVYY set \cite{DBLP:conf/podc/OHearnRVYY10}.}
		\vspace{2mm}
	\end{subfigure}
	~~~~
	\begin{subfigure}[t]{.3\textwidth}
		\centering
		\begin{tikzpicture}
			\begin{axis}[xticklabels={\(\top\),1,2,3},width=5cm, height=5.25cm,legend pos=outer north east,legend cell align={left},enlargelimits={abs=0.5},bar width=.2,x tick label style={yshift=1mm},ytick={0,1,...,10},ybar=0pt,xtick={0.5,1.5,...,100},x tick label as interval,ymin=0,ymax=9,axis x line shift=-.5,area legend]
				\addplot+[postaction={opacity=.375,pattern=crosshatch dots},error bars/.cd,y dir=both,y explicit] coordinates {
					(1, 7.84314) +- (0.0, 0.044636)
					(2, 3.865066) +- (0.0, 0.13175)
					(3, 0.0) +- (0.0, 0.0)
					(4, 6.546148) +- (0.0, 0.059598)
				};
				\addplot+[postaction={opacity=.375,pattern=north east lines},error bars/.cd,y dir=both,y explicit] coordinates {
					(1, 4.513352) +- (0.0, 0.023521)
					(2, 3.339117) +- (0.0, 0.042195)
					(3, 0.0) +- (0.0, 0.0)
					(4, 4.200501) +- (0.0, 0.025313)
				};
				\addplot+[postaction={opacity=.375,pattern=north west lines},error bars/.cd,y dir=both,y explicit] coordinates {
					(1, 4.391452) +- (0.0, 0.024305)
					(2, 3.324845) +- (0.0, 0.040393)
					(3, 0.0) +- (0.0, 0.0)
					(4, 4.077002) +- (0.0, 0.02092)
				};
			\end{axis}
		\end{tikzpicture}
		\vspace{-6mm}
		\caption{VY DCAS set \cite{DBLP:conf/pldi/VechevY08}}
		\vspace{2mm}
	\end{subfigure}
	~~~~
	\begin{subfigure}[t]{.3\textwidth}
		\centering
		\begin{tikzpicture}
			\begin{axis}[xticklabels={\(\top\),1,2,3},width=5cm, height=5.25cm,legend pos=outer north east,legend cell align={left},enlargelimits={abs=0.5},bar width=.2,x tick label style={yshift=1mm},ytick={0,1,...,10},ybar=0pt,xtick={0.5,1.5,...,100},x tick label as interval,ymin=0,ymax=9,axis x line shift=-.5,area legend]
				\addplot+[postaction={opacity=.375,pattern=crosshatch dots},error bars/.cd,y dir=both,y explicit] coordinates {
					(1, 5.188886) +- (0.0, 0.044836)
					(2, 3.849791) +- (0.0, 0.109965)
					(3, 0.0) +- (0.0, 0.0)
					(4, 6.413726) +- (0.0, 0.073684)
				};
				\addplot+[postaction={opacity=.375,pattern=north east lines},error bars/.cd,y dir=both,y explicit] coordinates {
					(1, 3.846818) +- (0.0, 0.034153)
					(2, 3.348115) +- (0.0, 0.061744)
					(3, 0.0) +- (0.0, 0.0)
					(4, 4.188216) +- (0.0, 0.043285)
				};
				\addplot+[postaction={opacity=.375,pattern=north west lines},error bars/.cd,y dir=both,y explicit] coordinates {
					(1, 3.821085) +- (0.0, 0.038864)
					(2, 3.33383) +- (0.0, 0.06149)
					(3, 0.0) +- (0.0, 0.0)
					(4, 4.04434) +- (0.0, 0.036908)
				};
			\end{axis}
		\end{tikzpicture}
		\vspace{-6mm}
		\caption{VY CAS set \cite{DBLP:conf/pldi/VechevY08}.}
		\vspace{2mm}
	\end{subfigure}
	\\
	\begin{subfigure}[t]{.37\textwidth}
		\centering
		\begin{tikzpicture}
			\begin{axis}[xticklabels={\(\top\),1,2,3},width=5cm, height=5.25cm,legend pos=outer north east,legend cell align={left},enlargelimits={abs=0.5},bar width=.2,x tick label style={yshift=1mm},ytick={0,1,...,10},ybar=0pt,xtick={0.5,1.5,...,100},x tick label as interval,ymin=0,ymax=9,axis x line shift=-.5,area legend]
				\addplot+[postaction={opacity=.375,pattern=crosshatch dots},error bars/.cd,y dir=both,y explicit] coordinates {
					(1, 5.271261) +- (0.0, 0.132186)
					(2, 3.855127) +- (0.0, 0.10697)
					(3, 0.0) +- (0.0, 0.0)
					(4, 6.393847) +- (0.0, 0.105489)
				};
				\addplot+[postaction={opacity=.375,pattern=north east lines},error bars/.cd,y dir=both,y explicit] coordinates {
					(1, 3.923933) +- (0.0, 0.069222)
					(2, 3.354053) +- (0.0, 0.06018)
					(3, 0.0) +- (0.0, 0.0)
					(4, 4.192077) +- (0.0, 0.030996)
				};
				\addplot+[postaction={opacity=.375,pattern=north west lines},error bars/.cd,y dir=both,y explicit] coordinates {
					(1, 3.899414) +- (0.0, 0.069679)
					(2, 3.336985) +- (0.0, 0.06003)
					(3, 0.0) +- (0.0, 0.0)
					(4, 4.014082) +- (0.0, 0.035119)
				};
			\end{axis}
		\end{tikzpicture}
		\vspace{-2mm}
		\caption{Michael set \cite{DBLP:conf/spaa/Michael02}.}
		\vspace{2mm}
	\end{subfigure}
	\begin{subfigure}[t]{.37\textwidth}
		\centering
		\begin{tikzpicture}
			\begin{axis}[xticklabels={\(\top\),1,2,3},width=5cm, height=5.25cm,legend pos=outer north east,legend cell align={left},enlargelimits={abs=0.5},bar width=.2,x tick label style={yshift=1mm},ytick={0,1,...,10},ybar=0pt,xtick={0.5,1.5,...,100},x tick label as interval,ymin=0,ymax=9,axis x line shift=-.5,area legend]
				\addplot+[postaction={opacity=.375,pattern=crosshatch dots},error bars/.cd,y dir=both,y explicit] coordinates {
					(1, 5.230356) +- (0.0, 0.098413)
					(2, 3.849244) +- (0.0, 0.108708)
					(3, 0.0) +- (0.0, 0.0)
					(4, 6.351297) +- (0.0, 0.072141)
				};
				\addplot+[postaction={opacity=.375,pattern=north east lines},error bars/.cd,y dir=both,y explicit] coordinates {
					(1, 3.919059) +- (0.0, 0.069411)
					(2, 3.347219) +- (0.0, 0.059892)
					(3, 0.0) +- (0.0, 0.0)
					(4, 4.189643) +- (0.0, 0.03092)
				};
				\addplot+[postaction={opacity=.375,pattern=north west lines},error bars/.cd,y dir=both,y explicit] coordinates {
					(1, 3.895874) +- (0.0, 0.068642)
					(2, 3.332001) +- (0.0, 0.060373)
					(3, 0.0) +- (0.0, 0.0)
					(4, 4.013991) +- (0.0, 0.031615)
				};
			\end{axis}
		\end{tikzpicture}
		\vspace{-2mm}
		\caption{Michael set (wait-free \texttt{search}).}
		\vspace{2mm}
	\end{subfigure}
	\\
	\begin{subfigure}[t]{.37\textwidth}
		\centering
		\begin{tikzpicture}
			\begin{axis}[xticklabels={\(\top\),1,2,3},width=5cm, height=5.25cm,legend pos=outer north east,legend cell align={left},enlargelimits={abs=0.5},bar width=.2,x tick label style={yshift=1mm},ytick={0,1,...,10},ybar=0pt,xtick={0.5,1.5,...,100},x tick label as interval,ymin=0,ymax=9,axis x line shift=-.5,area legend]
				\addplot+[postaction={opacity=.375,pattern=crosshatch dots},error bars/.cd,y dir=both,y explicit] coordinates {
					(1, 6.635127) +- (0.0, 1.574003)
					(2, 3.967854) +- (0.0, 0.089939)
					(3, 0.0) +- (0.0, 0.0)
					(4, 6.708154) +- (0.0, 0.209438)
				};
				\addplot+[postaction={opacity=.375,pattern=north east lines},error bars/.cd,y dir=both,y explicit] coordinates {
					(1, 4.270402) +- (0.0, 0.309034)
					(2, 3.439689) +- (0.0, 0.056974)
					(3, 0.0) +- (0.0, 0.0)
					(4, 4.382282) +- (0.0, 0.107722)
				};
				\addplot+[postaction={opacity=.375,pattern=north west lines},error bars/.cd,y dir=both,y explicit] coordinates {
					(1, 4.133218) +- (0.0, 0.161991)
					(2, 3.424632) +- (0.0, 0.057577)
					(3, 0.0) +- (0.0, 0.0)
					(4, 4.223564) +- (0.0, 0.085101)
				};
			\end{axis}
		\end{tikzpicture}
		\vspace{-2mm}
		\caption{Harris set \cite{DBLP:conf/wdag/Harris01}.}
		\vspace{2mm}
	\end{subfigure}
	\begin{subfigure}[t]{.37\textwidth}
		\centering
		\begin{tikzpicture}
			\begin{axis}[xticklabels={\(\top\),1,2,3},width=5cm, height=5.25cm,legend pos=outer north east,legend cell align={left},enlargelimits={abs=0.5},bar width=.2,x tick label style={yshift=1mm},ytick={0,1,...,10},ybar=0pt,xtick={0.5,1.5,...,100},x tick label as interval,ymin=0,ymax=9,axis x line shift=-.5,area legend]
				\addplot+[postaction={opacity=.375,pattern=crosshatch dots},error bars/.cd,y dir=both,y explicit] coordinates {
					(1, 6.562291) +- (0.0, 1.568479)
					(2, 3.960168) +- (0.0, 0.09142)
					(3, 0.0) +- (0.0, 0.0)
					(4, 6.695179) +- (0.0, 0.210555)
				};
				\addplot+[postaction={opacity=.375,pattern=north east lines},error bars/.cd,y dir=both,y explicit] coordinates {
					(1, 4.245625) +- (0.0, 0.300222)
					(2, 3.435692) +- (0.0, 0.057233)
					(3, 0.0) +- (0.0, 0.0)
					(4, 4.36363) +- (0.0, 0.106542)
				};
				\addplot+[postaction={opacity=.375,pattern=north west lines},error bars/.cd,y dir=both,y explicit] coordinates {
					(1, 4.117147) +- (0.0, 0.156339)
					(2, 3.419877) +- (0.0, 0.058568)
					(3, 0.0) +- (0.0, 0.0)
					(4, 4.20958) +- (0.0, 0.083487)
				};
			\end{axis}
		\end{tikzpicture}
		\vspace{-2mm}
		\caption{Harris set (wait-free \texttt{search}).}
		\vspace{2mm}
	\end{subfigure}
	\vspace{-4mm}
	\caption{%
		Average runtime for computing a single footprint, partitioned by data structure and footprint size ($\top$ indicates failure).
		\label{fig:details:footprints-individual}
	}
\end{figure}
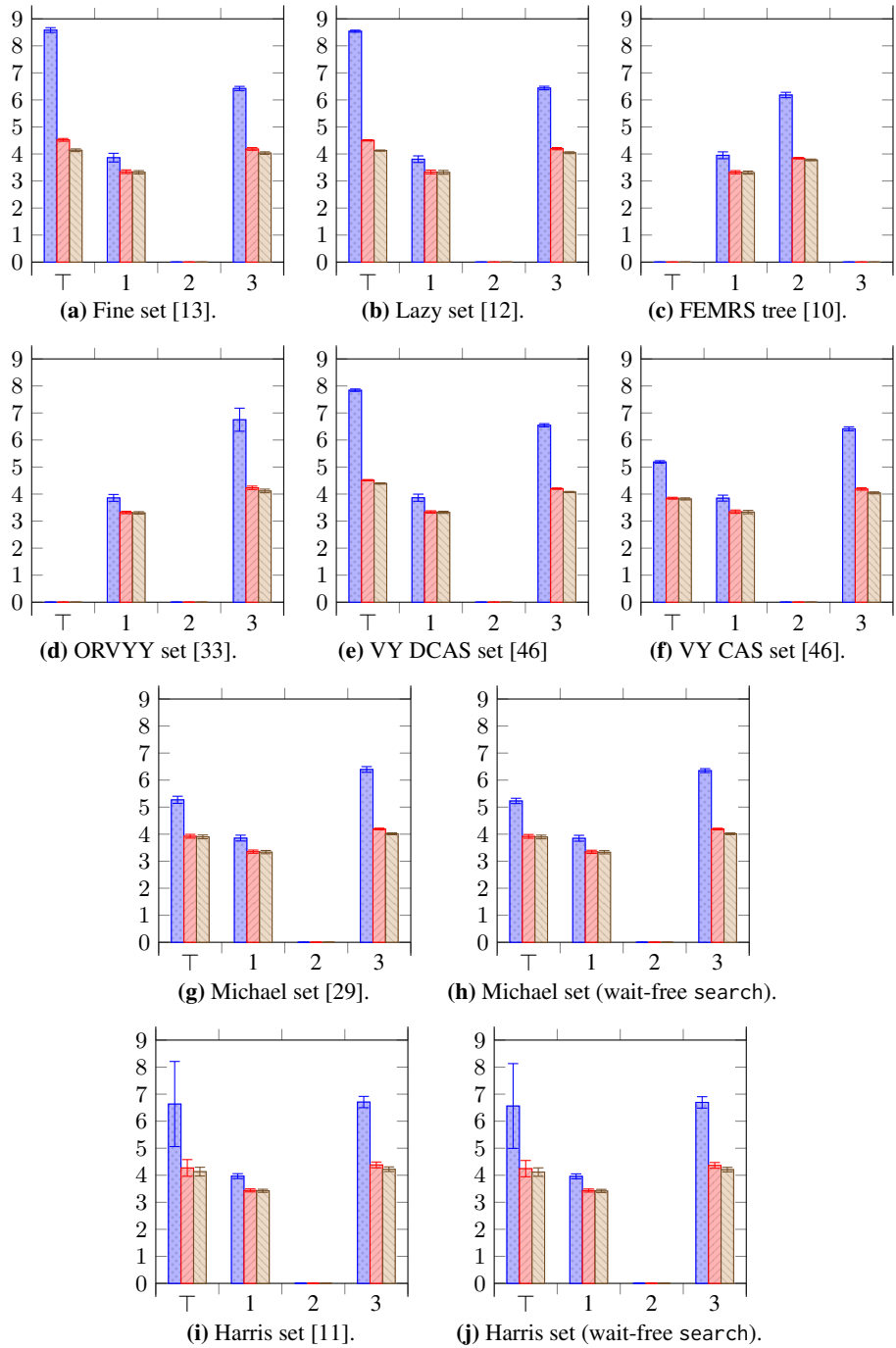

\newcommand{\objof}[1]{\mathsf{Obj}(#1)}

\smartparagraph{Symbolic Flow Graphs}
Recall from \cref{sec:eval} that for our evaluation we extracted flow graphs for which the \plankton tool computes footprints.
In \plankton, flow graphs are encoded symbolically by separation logic assertions $\aform$ of the form $\aform\defeq\afform\mstar\bigmstar_{\anode\in\somenodes}\objof{\anode}$ where $\afform$ can be thought of as the \emph{pure stack} (it does not contain resources, $\afform\mstar\afform=\afform$) and the iterated separating conjunction contains the predicate $\objof{\anode}$ reflecting the resource for every object $\anode\in\somenodes$ in the heap.
(This means that $\somenodes$ denotes a potentially unbounded set.)
The $\objof{\anode}$ predicate is, roughly, a points-to predicate containing the fields of object $\anode$.
For example, if $\anode$ is of type $\mathtt{T}$ with the following definition: \[
	\texttt{struct T \{ int key; bool mark; T* next; \}}
\]
then the resource $\objof{\anode}$ for $\anode$ is defined by $\objof{\anode}\defeq\anode\pointsto{}M,k,m,n$ where $M$, $k$, $m$, and $n$ are existentially quantified (at the level of the assertion) symbolic variables referring to $\anode$'s flow, $\mathtt{key}$ field, $\mathtt{mark}$ field, and $\mathtt{next}$ field, respectively.
Note that the flow $M$ is ghost state: it is not part of the node's physical representation.
The values of these symbolic variables may or may not be constrained by the stack $\afform$.

In order to turn an assertion $\aform$ into a flow graph $\aflowconstraint=(\setnodes, \edges, \inflow)$, we need to define how $\aform$ induces the nodes $\setnodes$, the edges $\edges$, and the inflow $\inflow$.
For $\setnodes$ we simply choose the objects in the heap, $\setnodes=\somenodes$.
The edges mimic the heap structure.
That is, if an object $\anode$'s pointer field references an object $\anodep$, then we add this pointer as an edge to $\edges$.
In the above example, we have $(\anode, \anodep, f)\in\edges$ if $\aform \models \anode{\pointsto{}}M,k,m,n ~\mstar~ n{=}\anodep ~\mstar~ \mathit{true}$ and the edge function $f$ is chose appropriately.
We discuss what it means for $f$ to be chosen appropriately.
In \plankton, edge functions are induced by the resources $\objof{-}$ and the stack $\afform$.
More precisely, the user specifies, for each pointer field $\mathtt{next}$ of each type $\mathtt{T}$, a generator function $\mathit{gen}_{\mathtt{T},\mathtt{next}}$ that produces the edge function $f$ for an edge between $\anode$ and $\anodep$.
To do so, the generator is given as input the valuation of all of $\anode$'s fields (the physical fields, not the ghost state).
Continuing with the example, we use a generator of type \[
	\mathit{gen}_{\mathtt{T},\mathtt{next}} ~:~
	\ZZ \times \BB \times \somenodes
	\rightarrow \contfunof{\amonoid\to\amonoid}
\]
which takes as input an object's $\mathtt{key}$, $\mathtt{mark}$, and $\mathtt{next}$ field (in that order).
Then, we can define the generator as follows in order to obtain, e.g, the edge labeling from \Cref{ex-keyset-flow}/\cref{fig-keyset-flow}: \[
	\mathit{gen}_{\mathtt{T},\mathtt{next}}(k,m,n)
	~\defeq~
	\begin{cases}
		\lambda_{-\infty} &\text{if } m = \mathit{true} \\
		\lambda_{k} &\text{otherwise} \ .
	\end{cases}
\]
It is worth pointing out that the stack may not constrain $m$.
In this case $\aform$ denotes multiple flow graphs, one where we have $\mathit{gen}_{\mathtt{T},\mathtt{next}}(k,m,n)=\lambda_{-\infty}$ and one where we have $\mathit{gen}_{\mathtt{T},\mathtt{next}}(k,m,n)=\lambda_{k}$.
For the inflow, we choose any function $\inflow$ that results in the flow being consistent with the stack $\afform$.
Again, this yields multiple, potentially unboundedly many choices for $\inflow$.

With the above understanding of how assertions denote flow graphs, the flow graph from \cref{fig-keyset-flow}(left) is described by the following assertion \[
	l\prall{\pointsto{}}M_l,6,\mathit{false},t
	~~\mstar~~
	t\prall{\pointsto{}}M_t,k_t,\mathit{true},r
	~~\mstar~~
	r\prall{\pointsto{}}M_r,8,\mathit{false},v
	~~\mstar~~
	M_l=(3,\infty)
\]
where we inlined some values into the points-to predicates for the assertion to be easier to read.
Note that we only specify the flow in $l$, $M_l=(3,\infty)$.
For the flow in $t$ and $r$ the assertion together with the above generator induces $M_t=(6,\infty)=M_r$.

The benchmark suite that we extracted from \plankton employs a semi-symbolic representation.
This means we extracted the nodes $\setnodes$ of the flow graphs $\aflowconstraint$ together with their (symbolic) field values.
Instead of storing the edges $\edges$ and the inflow $\inflow$ explicitly, however, we record the stack $\afform$ and the edge function generators.
Our tool \krill then applies the techniques discussed in \cref{Section:Footprints} to this semi-symbolic representation.
We use \atoolname{Z3} to discharge the SMT queries that this procedure involves.

		\renewenvironment{proof}[1][\proofname]{\subsubsection*{\bfseries Proof #1.}}{}

\section{Proofs of Section~\ref{Section:Flows}}
\label{Section:FlowsProofs}
\begin{proof}[of Lemma~\ref{Lemma:FlowAlgebra}]
	The laws of commutativity and units follow immediately from the definition of composition.
	For associativity, we show that $\aflowconstraint_2\statemultdef \aflowconstraint_3$ and $\aflowconstraint_1\statemultdef(\aflowconstraint_2\mstar\aflowconstraint_3)$ hold if and only if $\aflowconstraint_1\statemultdef\aflowconstraint_2$ and $(\aflowconstraint_1\mstar\aflowconstraint_2)\statemultdef\aflowconstraint_3$ hold.
	We prove the direction from left to right.
	The reverse direction holds by symmetry of definedness.

	Since $\aflowconstraint_2\statemultdef\aflowconstraint_3$, we have $(\aflowconstraint_2\discup\aflowconstraint_3).\fval\geq\aflowconstraint_2.\fval\discup\aflowconstraint_3.\fval$.
	Moreover, because of $\aflowconstraint_1\statemultdef(\aflowconstraint_2\mstar\aflowconstraint_3)$ we have $(\aflowconstraint_1\discup(\aflowconstraint_2\mstar\aflowconstraint_3)).\fval\geq\aflowconstraint_1.\fval\discup(\aflowconstraint_2\mstar\aflowconstraint_3).\fval$.
	Observe that we have $(\aflowconstraint_1\discup(\aflowconstraint_2\mstar\aflowconstraint_3)).\fval=(\aflowconstraint_1\discup\aflowconstraint_2\discup\aflowconstraint_3).\fval$.

	To show $\aflowconstraint_1\statemultdef\aflowconstraint_2$, we have to argue that $(\aflowconstraint_1\discup\aflowconstraint_2).\fval\geq \aflowconstraint_1.\fval\discup\aflowconstraint_2.\fval$.
	To see this, note that $(\aflowconstraint_1\discup\aflowconstraint_2).\fval\geq (\aflowconstraint_1\discup \aflowconstraint_2\discup\aflowconstraint_3).\fval|_{(\aflowconstraint_1\discup\aflowconstraint_2).\setnodes}\geq\aflowconstraint_1.\fval\discup\aflowconstraint_2.\fval$.
	The latter inequality is by the assumptions.
	For the former inequality, we note that the fixed point iteration for $(\aflowconstraint_1\discup\aflowconstraint_2).\fval$ starts with a contribution from $\aflowconstraint_3$ (given as inflow) that the iteration for $(\aflowconstraint_1\discup \aflowconstraint_2\discup\aflowconstraint_3).\fval$ only receives when reaching the fixed point.
	By monotonicity, every fixed point approximant to the left is then larger than the corresponding approximant to the right, and so is the fixed point.

	For  $(\aflowconstraint_1\mstar\aflowconstraint_2)\statemultdef\aflowconstraint_3$, we note that
	\begin{align*}
	 	((\aflowconstraint_1\mstar\aflowconstraint_2)\discup\aflowconstraint_3).\fval
	 	=
	 	(\aflowconstraint_1\discup\aflowconstraint_2\discup\aflowconstraint_3).\fval
	 	&\geq
	 	\aflowconstraint_1.\fval\discup\aflowconstraint_2.\fval\discup\aflowconstraint_3.\fval
	 	\\&\geq
	 	(\aflowconstraint_1\discup\aflowconstraint_2).\fval\discup\aflowconstraint_3.\fval
	 	\ .
	\end{align*}
	The first inequality is by the above assumptions.
	The second always holds, as remarked above.
	\qed
\end{proof}



\section{Proofs of Section~\ref{Section:Framing}}
\label{Section:FramingProofs}

\begin{proof}[of Theorem~\ref{Theorem:FramePreserving}(i)]
	For $\aflowconstraint_2.\setnodes\cap\aflowconstraint.\setnodes=\emptyset$, we use $\aflowconstraint_1.\setnodes\cap\aflowconstraint.\setnodes=\emptyset$ by $\aflowconstraint_1\statemultdef\aflowconstraint$ and $\aflowconstraint_1.\setnodes=\aflowconstraint_2.\setnodes$.
	To see that the outflow of $\aflowconstraint_2$ matches the inflow of $\aflowconstraint$, we consider $\anode\in\aflowconstraint_2.\setnodes$ and $\anodep\in\aflowconstraint.\setnodes$ and reason as follows:
	\begin{align*}
		\aflowconstraint_2.\outflowof{\anode, \anodep}=\aflowconstraint_1.\outflowof{\anode, \anodep}=\aflowconstraint.\inflowof{\anode, \anodep}. 
	\end{align*}
	The former equality is by $\aflowconstraint_2.\inflow=\aflowconstraint_1.\inflow$ and $\transformerof{\aflowconstraint_1}=_{\aflowconstraint_1.\inflow}\transformerof{\aflowconstraint_2}$.
	The second equality is by $\aflowconstraint_1\statemultdef\aflowconstraint$.
	The inflow is preserved by the assumption, hence we obtain the following equality: $\aflowconstraint.\outflowof{\anodep, \anode}=\aflowconstraint_1.\inflowof{\anodep, \anode}=\aflowconstraint_2.\inflowof{\anodep, \anode}$.

	It remains to show $(\aflowconstraint_2\discup\aflowconstraint).\fval=\aflowconstraint_2.\fval\discup\aflowconstraint.\fval$.
	We use Beki\'c's Lemma~\cite{DBLP:conf/ibm/Bekic84e}.
	Define the target pairing of two functions $f:A\rightarrow B$ and $g:A\rightarrow C$ over the same domain $A$ as the function $\pairingof{f}{g}:A\rightarrow B\times C$ with $\pairingof{f}{g}(a) \defeq (f(a), g(a))$.
	We compute the flow of $\aflowconstraint_2\discup \aflowconstraint$ as the least fixed point of a target pairing $\pairingof{f}{g}$ with
	\begin{align*}
		f:&\; ((\aflowconstraint_2.\setnodes\discup\aflowconstraint.\setnodes) \rightarrow \amonoid)\rightarrow \aflowconstraint_2.\setnodes\rightarrow \amonoid\\
		g:&\; ((\aflowconstraint_2.\setnodes\discup\aflowconstraint.\setnodes) \rightarrow \amonoid)\rightarrow \aflowconstraint.\setnodes\rightarrow \amonoid\; .  
	\end{align*}
	Function $f$ updates the flow of the nodes in $\aflowconstraint_2$ depending on the flow in/inflow from $\aflowconstraint$.
	Function $g$ is responsible for the flow of the nodes in $\aflowconstraint$.
	The inflow from the nodes outside $\aflowconstraint_2\discup\aflowconstraint$ is constant.
	The definition guarantees $(\aflowconstraint_2\discup\aflowconstraint).\fval=\lfpof{\pairingof{f}{g}}$.
	We curry the former function, 
	\begin{align*}
		f:\; (\aflowconstraint.\setnodes\rightarrow \amonoid)\rightarrow (\aflowconstraint_2.\setnodes\rightarrow \amonoid)\rightarrow \aflowconstraint_2.\setnodes\rightarrow \amonoid\; ,
	\end{align*}
	and obtain, for every $\cval:\aflowconstraint.\setnodes\rightarrow \amonoid$, the function
	\begin{align*}
		f(\cval):\; (\aflowconstraint_2.\setnodes\rightarrow \amonoid)\rightarrow \aflowconstraint_2.\setnodes\rightarrow \amonoid\; .
	\end{align*} 
	This function is still monotonic and therefore has a least fixed point.
	Hence, the function 
	\begin{align*}
		f^{\dagger}: (\aflowconstraint.\setnodes\rightarrow \amonoid)\rightarrow \aflowconstraint_2.\setnodes\rightarrow \amonoid\; 
	\end{align*}
	mapping valuation $\cval:\aflowconstraint.\setnodes\rightarrow \amonoid$ to the least fixed point $\lfpof{f(\cval)}$ is well-defined. 

	Beki\'c's Lemma tells us how to compute least fixed points of target pairings like $\pairingof{f}{g}$ above by successive elimination of the variables.
	We first determine $f^{\dagger}$, which is a function in $\aflowconstraint.\setnodes\rightarrow\amonoid$.
	We plug this function into $g$ to obtain a function solely in $\aflowconstraint.\setnodes\rightarrow\amonoid$.
	To be precise, since $g$ expects a function from $(\aflowconstraint_2.\setnodes\discup\aflowconstraint.\setnodes)\rightarrow\amonoid$, we pair $f^{\dagger}$ with $\myid=\myidof{\aflowconstraint.\setnodes\rightarrow\amonoid}$ and obtain
	\begin{align*}
		\pairingof{f^{\dagger}}{\myid}:(\aflowconstraint.\setnodes\rightarrow\amonoid)\rightarrow (\aflowconstraint_2.\setnodes\discup\aflowconstraint.\setnodes)\rightarrow\amonoid\; . 
	\end{align*}
	We compose this function with $g$ and get 
	\begin{align*}
	g\circ \pairingof{f^{\dagger}}{\myid}:(\aflowconstraint.\setnodes\rightarrow\amonoid)\rightarrow \aflowconstraint.\setnodes\rightarrow\amonoid\; .
	\end{align*}
	We compute the least fixed point of this composition to obtain the values of the least fixed point of interest on $\aflowconstraint.\setnodes$.
	For the values on $\aflowconstraint_2.\setnodes$, we reinsert the $\aflowconstraint.\setnodes$-values into $f^{\dagger}$.
	Beki\'c's Lemma guarantees the correctness of this successive elimination procedure: 
	\begin{align*}
	\lfpof{\pairingof{f}{g}}\; =\; (f^{\dagger}(\cval), \cval)\qquad\text{with}\qquad \cval\; =\; \lfpof{g\circ \pairingof{f^{\dagger}}{\myid}}\; .
	\end{align*}

	To conclude the proof, we recall that $\transformerof{\aflowconstraint_2}=_{\aflowconstraint_1.\inflow}\transformerof{\aflowconstraint_1}$.
	Further, for all $\anodep \in \aflowconstraint.\setnodes$ and $\anode \in \aflowconstraint_1.\setnodes$ we have $\aflowconstraint_1.\inflow(\anodep,\anode) = \aflowconstraint.\outflow(\anodep,\anode) = \aflowconstraint.\edges_{(\anodep, \anode)}(\aflowconstraint.\fvalof{\anode})$.
	Together with monotonicity of the edge functions, this implies $g\circ \pairingof{f^{\dagger}}{\myid}=_{\aflowconstraint.\fval} g\circ \pairingof{e^{\dagger}}{\myid}$.
	Here, $e: ((\aflowconstraint_2.\setnodes\discup\aflowconstraint.\setnodes) \rightarrow \amonoid)\rightarrow \aflowconstraint_2.\setnodes\rightarrow \amonoid$ is the transformer derived from $\aflowconstraint_1$ in the same way $f$ was derived from $\aflowconstraint_2$.
	We thus have for all $\anode\in\aflowconstraint.\setnodes$: 
	\begin{align*}
		&\quad(\aflowconstraint_2\discup\aflowconstraint).\fvalof{\anode}\\
		\text{\small(Definition $\fval$, $f$, $g$)}\quad =&\quad \lfpof{\pairingof{f}{g}}(\anode)\\
		\text{\small(Beki\'c's lemma, $\anode\in\aflowconstraint.\setnodes$)}\quad =&\quad \lfpof{g\circ \pairingof{f^{\dagger}}{\myid}}(\anode)\\
		\text{\small($\transformerof{\aflowconstraint_2}=_{\aflowconstraint_1.\inflow}\transformerof{\aflowconstraint_1}$, see above)}\quad=&\quad \lfpof{g\circ \pairingof{e^{\dagger}}{\myid}}(\anode)\\
		\text{\small(Beki\'c's lemma, $\anode\in\aflowconstraint.\setnodes$)}\quad =&\quad \lfpof{\pairingof{e}{g}}(\anode)\\
		\text{\small(Definition flow, $f$, $g$)}\quad =&\quad (\aflowconstraint_1\discup\aflowconstraint).\fvalof{\anode}\\
		\text{\small($\aflowconstraint_1\statemultdef\aflowconstraint$, $\anode\in\aflowconstraint.\setnodes$)}\quad=&\quad\aflowconstraint.\fvalof{\anode}.
	\end{align*}
	We argue that also for nodes $\anode\in\aflowconstraint_2.\setnodes$ we have $(\aflowconstraint_2\discup\aflowconstraint).\fvalof{\anode}=\aflowconstraint_2.\fvalof{\anode}$:
	\begin{align*}
		&\quad(\aflowconstraint_2\discup\aflowconstraint).\fvalof{\anode}\\
		\text{\small(Definition $\fval$, $f$, $g$)}\quad =&\quad \lfpof{\pairingof{f}{g}}(\anode)\\
		\text{\small(Beki\'c's lemma, $\anode\in\aflowconstraint_2.\setnodes$)}\quad =&\quad [f^{\dagger}(\cval)](\anode)\\
		\text{\small(See above)}\quad =&\quad [f^{\dagger}(\aflowconstraint.\fval)](\anode)\\
		\text{\small(See below)}\quad=&\quad \aflowconstraint_2.\fvalof{\anode}.
	\end{align*}
	To see  the last equality, note that $\aflowconstraint_2.\inflowof{\anodep, \anode}=\aflowconstraint.\outflowof{\anodep, \anode}=\aflowconstraint.\edges_{(\anodep, \anode)}(\aflowconstraint.\fvalof{\anode})$ holds for all $\anodep\in\aflowconstraint.\setnodes$.
	Hence, $\aflowconstraint_2.\fval$, which we compute from the inflow, is 
	the least fixed point of $f$ computed with $\aflowconstraint.\fval$ fixed.
	\qed
\end{proof}

\begin{proof}[of Theorem~\ref{Theorem:FramePreserving}(ii)]
	Follows with a similar but simpler application of Beki\'c's lemma than Theorem~\ref{Theorem:FramePreserving}(i).
	\qed
\end{proof}

\begin{proof}[of Lemma~\ref{Lemma:Restriction}]
	We use $\setnodes$ for the set of nodes $\aflowconstraint.\setnodes$.
	\begin{asparaenum}[(i)]
		\item[(i)]
		We immediately have $\sqsupseteq$. 
		For $\sqsubseteq$, we use the fact that the flow is formulated as a least fixed point as made explicit in Lemma~\ref{Lemma:Transfer}(ii):  
		$\restrictto{\aflowconstraint}{\setnodesp}.\fval = \lfpof{(\restrictto{\aflowconstraint}{\setnodesp}[\restrictto{\aflowconstraint}{\setnodesp}.\inflow])}$ and $\restrictto{\aflowconstraint.\fval}{\setnodesp} = \restrictto{(\lfpof{\aflowconstraint[\aflowconstraint.\inflow]})}{\setnodesp}$. 
		The fixed points are computed with a Kleene iteration: $ \lfpof{(\restrictto{\aflowconstraint}{\setnodesp}[\restrictto{\aflowconstraint}{\setnodesp}.\inflow])}$ is the join of the ascending chain 
		\begin{align*}
			(\restrictto{\aflowconstraint}{\setnodesp}[\restrictto{\aflowconstraint}{\setnodesp}.\inflow])^0(\bot)\sqsubseteq (\restrictto{\aflowconstraint}{\setnodesp}[\restrictto{\aflowconstraint}{\setnodesp}.\inflow])(\bot)\sqsubseteq(\restrictto{\aflowconstraint}{\setnodesp}[\restrictto{\aflowconstraint}{\setnodesp}.\inflow])^2(\bot)\sqsubseteq\ldots
		\end{align*} 
		where the value for no iteration is $(\restrictto{\aflowconstraint}{\setnodesp}[\restrictto{\aflowconstraint}{\setnodesp}.\inflow])^0(\bot)\defeq\bot$ and for iteration $i+1$ we have $(\restrictto{\aflowconstraint}{\setnodesp}[\restrictto{\aflowconstraint}{\setnodesp}.\inflow])^{i+1}(\bot)\defeq(\restrictto{\aflowconstraint}{\setnodesp}[\restrictto{\aflowconstraint}{\setnodesp}.\inflow])((\restrictto{\aflowconstraint}{\setnodesp}[\restrictto{\aflowconstraint}{\setnodesp}.\inflow])^{i}(\bot))$. 
		We will now show that each of these so-called Kleene approximants is dominated by $\restrictto{\aflowconstraint.\fval}{\setnodesp}$. 
		The restricted flow is thus an upper bound for all elements in the sequence, and hence larger than or equal to the join. 

		In the base case, we have $\bot\sqsubseteq \restrictto{\aflowconstraint.\fval}{\setnodesp}$. 
		Assume $(\restrictto{\aflowconstraint}{\setnodesp}[\restrictto{\aflowconstraint}{\setnodesp}.\inflow])^{i}(\bot)\sqsubseteq\restrictto{\aflowconstraint.\fval}{\setnodesp}$. 
		We have 
		\begin{align*}
			(\restrictto{\aflowconstraint}{\setnodesp}[\restrictto{\aflowconstraint}{\setnodesp}.\inflow])^{i+1}(\bot) &=(\restrictto{\aflowconstraint}{\setnodesp}[\restrictto{\aflowconstraint}{\setnodesp}.\inflow])((\restrictto{\aflowconstraint}{\setnodesp}[\restrictto{\aflowconstraint}{\setnodesp}.\inflow])^{i}(\bot))\\
			&\sqsubseteq (\restrictto{\aflowconstraint}{\setnodesp}[\restrictto{\aflowconstraint}{\setnodesp}.\inflow])(\restrictto{\aflowconstraint.\fval}{\setnodesp})\\
			&= \restrictto{\aflowconstraint.\fval}{\setnodesp}\ . 
		\end{align*} 
		The first equality is by the definition of Kleene approximants.
		The following inequality is monotonicity of the edge functions combined with the induction hypothesis. 
		For the last equality, we use that the flow is the least fixed point, $\aflowconstraint[\aflowconstraint.\inflow](\aflowconstraint.\fval)=\aflowconstraint.\fval$, and the definition of the inflow for the restricted flow graph.

		\item[(ii)]
		We first show definedness of the composition. 
		Disjointness of the sets of nodes is immediate. 
		Equality of the flows is~(i). 
		For the compatibility of inflow and outflow, 
		let $\anodep\in \setnodes\cap\setnodesp$ and $\anode\in\setnodes\setminus\setnodesp$. 
		We show $\restrictto{\aflowconstraint}{\setnodesp}.\outflow(\anodep, \anode) = \restrictto{\aflowconstraint}{\setnodes\setminus\setnodesp}.\inflow(\anodep, \anode)$. 
		We have 
		\begin{align*}
			\restrictto{\aflowconstraint}{\setnodesp}.\outflow(\anodep, \anode) &= \restrictto{\aflowconstraint}{\setnodesp}.\edges_{(\anodep, \anode)}(\restrictto{\aflowconstraint}{\setnodesp}.\fvalof{\anodep})\\
			&=\aflowconstraint.\edges_{(\anodep, \anode)}(\restrictto{\aflowconstraint}{\setnodesp}.\fvalof{\anodep})\\
			&=\aflowconstraint.\edges_{(\anodep, \anode)}(\aflowconstraint.\fvalof{\anodep})\ .
		\end{align*}
		The first equality is the definition of the outflow, the next is by the definition of restriction, the last is~(i). 
		Also by the definition of restriction, we have: 
		\begin{align*}
			\restrictto{\aflowconstraint}{\setnodes\setminus\setnodesp}.\inflow(\anodep, \anode) &=\aflowconstraint.\edges_{(\anodep, \anode)}(\aflowconstraint.\fvalof{\anodep})\ . 
		\end{align*}
		The two values coincide, as required. 
		The argumentation in the reverse direction is symmetric.
		It remains to show that the composition satisfies $\restrictto{\aflowconstraint}{\setnodesp}\statemult\restrictto{\aflowconstraint}{\setnodes\setminus\setnodesp}=\aflowconstraint$.
		The sets of nodes immediately coincide and so do the sets of edges. 
		The inflow from outside $\aflowconstraint.\setnodes$ is the same by the definition of restriction. 
		The inflow from the other flow graph present in $\restrictto{\aflowconstraint}{\setnodesp}$ resp. $\restrictto{\aflowconstraint}{\setnodes\setminus\setnodesp}$ is removed by the composition, as required.

		\item[(iii)]
		The set of nodes after the restriction is in both cases $\aflowconstraint.\setnodes\cap \setnodesp\cap \setnodespp$ to which we refer as $\setnodes'$. 
		Also the edges coincide as the restriction of the original edges to $\setnodes'\times\nat$. 
		The inflow from a node outside $\aflowconstraint.\setnodes$ to $\setnodes'$ immediately coincides for $\restrictto{(\restrictto{\aflowconstraint}{\setnodesp})}{\setnodespp}$ and $\restrictto{\aflowconstraint}{\setnodesp\cap\setnodespp}$.
		Consider the inflow from a node $\anodepp\in\aflowconstraint.\setnodes\setminus \setnodes'$ to $\anode\in\setnodes'$. 
		We have
		\begin{align*}
		\restrictto{\aflowconstraint}{\setnodesp\cap\setnodespp}.\inflow(\anodepp, \anode)=\aflowconstraint.\edges_{(\anodepp, \anode)}(\aflowconstraint.\fvalof{\anodepp})\ .
		\end{align*}
		In $\restrictto{(\restrictto{\aflowconstraint}{\setnodesp})}{\setnodespp}$, there are two cases. 
		If $\anodepp\in \aflowconstraint.\setnodes\cap(\setnodesp\setminus\setnodespp)$, we have
		\begin{align*}
		\restrictto{(\restrictto{\aflowconstraint}{\setnodesp})}{\setnodespp}.\inflow(\anodepp, \anode)=\restrictto{\aflowconstraint}{\setnodesp}.\edges_{(\anodepp, \anode)}(\restrictto{\aflowconstraint}{\setnodesp}.\fvalof{\anodepp})
		=\aflowconstraint.\edges_{(\anodepp, \anode)}(\aflowconstraint.\fvalof{\anodepp})\ .
		\end{align*}
		The first equation is the definition of restriction.
		In the next equation, equality of the edges is also by the definition of restriction. 
		The flow coincides by (i). 
		If $\anodepp\in \aflowconstraint.\setnodes\setminus \setnodesp$, we have
		\begin{align*}
		\restrictto{(\restrictto{\aflowconstraint}{\setnodesp})}{\setnodespp}.\inflow(\anodepp, \anode)&=\restrictto{\aflowconstraint}{\setnodesp}.\inflow(\anodepp, \anode)\\
		&=\aflowconstraint.\edges_{(\anodepp, \anode)}(\aflowconstraint.\fvalof{\anodepp})\ .
		\end{align*} 
		Both equations are by the definition of restriction. 
		The reasoning shows that the inflows coincide as well.
		\qed
	\end{asparaenum}
\end{proof}

\begin{proof}[of Lemma~\ref{Lemma:FootprintMonotonicity}]
	Let $\setnodespp\in\setflowfootprintof{\aflowconstraint_1}{\aflowconstraint_2}$ and $\setnodespp\subseteq\setnodesp\subseteq\setnodes\defeq\aflowconstraint_1.\setnodes$.
	We first show $\restrictto{\aflowconstraint_1}{\setnodesp}\ctxequiv \restrictto{\aflowconstraint_2}{\setnodesp}$. 
	For $i=1, 2$, the fact that $\setnodespp\subseteq\setnodesp$ and Lemma~\ref{Lemma:Restriction}(iii) yield: 
	\begin{align*}
	\restrictto{\aflowconstraint_i}{\setnodespp}=
	\restrictto{\aflowconstraint_i}{\setnodespp\cap\setnodesp} =
	\restrictto{(\restrictto{\aflowconstraint_i}{\setnodesp})}{\setnodespp}\ . 
	\end{align*}
	Combined with $\restrictto{\aflowconstraint_1}{\setnodespp}\ctxequiv\restrictto{\aflowconstraint_2}{\setnodespp}$ by the fact that $\setnodespp$ is a flow footprint, we get 
	\begin{align}
	\restrictto{(\restrictto{\aflowconstraint_1}{\setnodesp})}{\setnodespp}\ctxequiv\restrictto{(\restrictto{\aflowconstraint_2}{\setnodesp})}{\setnodespp}\ .\label{Equation:CTX}
	\end{align} 
	Also with the fact that $\setnodespp$ is a footprint, we have $\restrictto{\aflowconstraint_1}{\setnodes\setminus\setnodespp}=\restrictto{\aflowconstraint_2}{\setnodes\setminus\setnodespp}$. 
	This implies $\restrictto{(\restrictto{\aflowconstraint_1}{\setnodes\setminus\setnodespp})}{\setnodesp}=\restrictto{(\restrictto{\aflowconstraint_2}{\setnodes\setminus\setnodespp})}{\setnodesp}$. 
	We use Lemma~\ref{Lemma:Restriction}(iii) and $\setnodesp\subseteq\setnodes$ to derive for $i=1, 2$:
	\begin{align*}
	\restrictto{(\restrictto{\aflowconstraint_i}{\setnodes\setminus\setnodespp})}{\setnodesp}=\restrictto{\aflowconstraint_i}{(\setnodes\setminus\setnodespp)\cap\setnodesp}=\restrictto{\aflowconstraint_i}{\setnodesp\setminus\setnodespp}=\restrictto{(\restrictto{\aflowconstraint_i}{\setnodesp})}{\setnodesp\setminus\setnodespp}\ . 
	\end{align*}
	We thus have 
	\begin{align}
	\restrictto{(\restrictto{\aflowconstraint_1}{\setnodesp})}{\setnodesp\setminus\setnodespp}=\restrictto{(\restrictto{\aflowconstraint_2}{\setnodesp})}{\setnodesp\setminus\setnodespp}\ . \label{Equation:Equal}
	\end{align}
	Equations~\eqref{Equation:CTX} and~\eqref{Equation:Equal} combined with the definedness in Lemma~\ref{Lemma:Restriction}(ii) and Theorem~\ref{Theorem:FramePreserving} yield
	\begin{align} 
	\restrictto{(\restrictto{\aflowconstraint_1}{\setnodesp})}{\setnodespp}\statemult\restrictto{(\restrictto{\aflowconstraint_1}{\setnodesp})}{\setnodesp\setminus\setnodespp} \ctxequiv \restrictto{(\restrictto{\aflowconstraint_2}{\setnodesp})}{\setnodespp}\statemult\restrictto{(\restrictto{\aflowconstraint_2}{\setnodesp})}{\setnodesp\setminus\setnodespp}\ .\label{Equation:KeyCTX}
	\end{align}
	Lemma~\ref{Lemma:Restriction}(ii) shows for $i=1, 2$:
	\begin{align*}
	\restrictto{(\restrictto{\aflowconstraint_i}{\setnodesp})}{\setnodespp}\statemult\restrictto{(\restrictto{\aflowconstraint_i}{\setnodesp})}{\setnodesp\setminus\setnodespp} = \restrictto{\aflowconstraint_i}{\setnodesp}. 
	\end{align*}
	Combined with Equation~\eqref{Equation:KeyCTX}, this yields the desired
	\begin{align*} 
	\restrictto{\aflowconstraint_1}{\setnodesp} \ctxequiv \restrictto{\aflowconstraint_2}{\setnodesp}\ .
	\end{align*}

	It remains to prove $\restrictto{\aflowconstraint_1}{\setnodes\setminus\setnodesp}= \restrictto{\aflowconstraint_2}{\setnodes\setminus\setnodesp}$. 
	Since $\restrictto{\aflowconstraint_1}{\setnodes\setminus\setnodespp} = \restrictto{\aflowconstraint_2}{\setnodes\setminus\setnodespp}$, we have
	\begin{align}
	\restrictto{(\restrictto{\aflowconstraint_1}{\setnodes\setminus\setnodespp})}{\setnodes\setminus\setnodesp} = \restrictto{(\restrictto{\aflowconstraint_2}{\setnodes\setminus\setnodespp})}{\setnodes\setminus\setnodesp}\ . \label{Equality:KeyEqual}
	\end{align}
	By Lemma~\ref{Lemma:Restriction}(iii) and $\setnodespp\subseteq\setnodesp$, we get for $i=1, 2$:
	\begin{align*}
	\restrictto{(\restrictto{\aflowconstraint_i}{\setnodes\setminus\setnodespp})}{\setnodes\setminus\setnodesp} =
	\restrictto{\aflowconstraint_i}{(\setnodes\setminus\setnodespp)\cap(\setnodes\setminus\setnodesp)} = \restrictto{\aflowconstraint_i}{\setnodes\setminus\setnodesp}\ .
	\end{align*}
	Combined with Equation~\eqref{Equality:KeyEqual}, this yields
	\begin{align*}
	\restrictto{\aflowconstraint_1}{\setnodes\setminus\setnodesp} = \restrictto{\aflowconstraint_2}{\setnodes\setminus\setnodesp}\ . 
	\end{align*}
	This concludes the proof.
	\qed
\end{proof}



\section{Proofs of Section~\ref{Section:Footprints}}
\label{Section:FootprintProofs}

\begin{proof}[of Lemma~\ref{Lemma:Inflow}]
  The inflow from nodes outside $\setnodes$ is the same for both flow graphs, $\inflow\defeq\aflowconstraint_1.\inflow=\aflowconstraint_2.\inflow$. 
  To show that the inflow from $\setnodes\setminus\setnodespp$ coincides, we show that the flow in $\setnodes\setminus\setnodespp$ coincides. 

  We begin by computing the flow in $\aflowconstraint_i$ as an exchange of values between $\restrictto{\aflowconstraint_i}{\setnodespp}$ and $\restrictto{\aflowconstraint_i}{\setnodes\setminus\setnodespp}$. 
  By Lemma~\ref{Lemma:Restriction}, we have $\aflowconstraint_i = \restrictto{\aflowconstraint_i}{\setnodespp}\mstar\restrictto{\aflowconstraint_i}{\setnodes\setminus\setnodespp}$. 
  We can understand $\restrictto{\aflowconstraint_i}{\setnodespp}$ and $\restrictto{\aflowconstraint_i}{\setnodes\setminus\setnodespp}$ as functions
  \begin{align*}
  f_{i}:&\; (\setnodes \rightarrow \amonoid)\rightarrow \setnodespp\rightarrow \amonoid\\
  g_i:&\; (\setnodes \rightarrow \amonoid)\rightarrow (\setnodes\setminus\setnodespp)\rightarrow \amonoid\; .  
  \end{align*}
  Function $f_i$ computes the flow in the nodes from $\setnodespp$ depending on the current flow in $\setnodespp$ and $\setnodes\setminus\setnodespp$. 
  The inflow from outside $\setnodes$ is constantly $\inflow$. 
  Similarly, $g_i$ computes the flow in $\setnodes\setminus\setnodespp$ depending on the current flow in $\setnodes\setminus\setnodespp$ and $\setnodespp$, also with $\inflow$ constant.  
  By the fact that $\outof{\aflowconstraint_1}{\aflowconstraint_2}\subseteq\setnodespp$, the edges originating from nodes in $\setnodes\setminus\setnodespp$ coincide and we have $g_1=g_2$. 
  We refer to this function as $g$.
  We now consider the target pairing
  \begin{align*}
  \pairingof{f_i}{g}:(\setnodes\rightarrow\amonoid)\rightarrow [(\setnodespp\rightarrow\amonoid) \times ((\setnodes\setminus\setnodespp)\rightarrow\amonoid)]=_{\mathit{iso}}(\setnodes\rightarrow\amonoid)\ . 
  \end{align*}
  By Lemma~\ref{Lemma:Transfer}(ii) and the definition of $f_i$ and $g$, we have 
  \begin{align}
  \aflowconstraint_i.\fval\quad=\quad\mathit{lfp}.\aflowconstraint_i[\inflow]\quad=\quad\mathit{lfp}.\pairingof{f_i}{g}\ . \label{Equation:InflowFP}
  \end{align}
  Bekic's lemma~\cite{DBLP:conf/ibm/Bekic84e} explains how to compute the latter fixed point:
  \begin{align}
  \mathit{lfp}.\pairingof{f_i}{g}\; =\; (f_i^\dagger(\cval_i), \cval_i)\quad\text{with}\quad\cval_i\;=\;\mathit{lfp}.g\circ\pairingof{f_i^\dagger}{\myid_{(\setnodes\setminus\setnodespp)\rightarrow\amonoid}}\ .\label{Equation:InflowBekic}
  \end{align}

  To define the function $f_i^{\dagger}$, we curry the function $f_i$ and obtain
  \begin{align*}
  f_{i}:\; ((\setnodes\setminus\setnodespp)\rightarrow\amonoid)\rightarrow (\setnodespp \rightarrow \amonoid)\rightarrow \setnodespp\rightarrow \amonoid\ .
  \end{align*}
  Given a function $\cval:(\setnodes\setminus\setnodespp)\rightarrow\amonoid$, we have
  \begin{align*}
  f_{i}(\cval):\; (\setnodespp \rightarrow \amonoid)\rightarrow \setnodespp\rightarrow \amonoid\ .
  \end{align*}
  The function is again continuous and hence has a least fixed point.
  We define
  \begin{align*}
  f_{i}^\dagger:\; ((\setnodes\setminus\setnodespp)\rightarrow\amonoid)&\rightarrow \setnodespp\rightarrow \amonoid\ \\
  \cval&\mapsto\mathit{lfp}.f_i(\cval)\ . 
  \end{align*}
  Function $f_i^{\dagger}$ has an intuitive meaning. 
  It yields the flow in $\setnodespp$ when the flow in $\setnodes\setminus\setnodespp$ is $\cval$. 
  Formulated differently, it yields the flow in $\setnodespp$ when the inflow 
  from $\setnodes\setminus\setnodespp$ is $\restrictto{\outflow(\cval)}{(\setnodes\setminus\setnodespp)\times\setnodespp}$ with $\outflow$ as defined for transfer functions and the inflow from $\nat\setminus\setnodes$ is $\inflow$. 
  With
  \begin{align*}
  \inflow_{\cval}\;\defeq\;\restrictto{\inflow}{(\nat\setminus\setnodes)\times \setnodespp}\cup \restrictto{\outflow(\cval)}{(\setnodes\setminus\setnodespp)\times\setnodespp}\ 
  \end{align*}
  we get
  \begin{align*}
  f_{i}^\dagger(\cval)\quad =\quad \mathit{lfp}.f_i(\cval)\quad=\quad\mathit{lfp}.\restrictto{\aflowconstraint_i}{\setnodespp}[\inflow_{\cval}]\ .
  \end{align*} 
  We compose $f_i^\dagger$ with $\outflow$ and apply Lemma~\ref{Lemma:Transfer}:
  \begin{align*}
  \outflow\circ f_{i}^\dagger(\cval)\quad =\quad\outflow\circ \mathit{lfp}.\restrictto{\aflowconstraint_i}{\setnodespp}[\inflow_\cval]\quad = \quad\transformerof{\restrictto{\aflowconstraint_i}{\setnodespp}[\inflow_\cval]}\ . 
  \end{align*}
  The fact that $\tfailof{\aflowconstraint_1}{\aflowconstraint_2}{\setnodespp}= \emptyset$ implies $\transformerof{\restrictto{\aflowconstraint_1}{\setnodespp}}=_{\restrictto{\aflowconstraint_1}{\setnodespp}.\inflow} \transformerof{\restrictto{\aflowconstraint_2}{\setnodes\setminus\setnodespp}}$. 
  Moreover, since $\restrictto{\aflowconstraint_1}{\setnodespp}.\inflow(\anode, \anodepp)=\restrictto{\aflowconstraint_2}{\setnodespp}.\inflow(\anode, \anodepp)$ for all $\anode\in\nat\setminus\setnodes$, $\anodepp\in\setnodespp$, and since the edge functions are monotonic, we have 
  \begin{align}
  \outflow\circ f_{1}^\dagger(\cval)\quad =\quad \outflow\circ f_{2}^\dagger(\cval)\quad\text{for all $\cval\leq\restrictto{\aflowconstraint_1}{\setnodes\setminus\setnodespp}.\fval$}\ . \label{Equation:InflowKey}
  \end{align}

  We will now show that the flow in $\setnodes\setminus\setnodespp$ coincides for $\aflowconstraint_1$ and $\aflowconstraint_2$. 
  We combine Lemma~\ref{Lemma:Restriction}(i) with Equations~\eqref{Equation:InflowFP} and~\eqref{Equation:InflowBekic} and get
  \begin{align}
  \restrictto{\aflowconstraint_i}{\setnodes\setminus\setnodespp}.\fval
  \; =\;\restrictto{(\aflowconstraint_i.\fval)}{\setnodes\setminus\setnodespp}
  \; =\;\restrictto{(\mathit{lfp}.\pairingof{f_i}{g})}{\setnodes\setminus\setnodespp}
  \; =\; \mathit{lfp}.g\circ\pairingof{f_i^\dagger}{\myid}\ .\label{Equation:InflowFlow}
  \end{align}
  We thus intend to show that $\mathit{lfp}.g\circ\pairingof{f_1^\dagger}{\myid}= \mathit{lfp}.g\circ\pairingof{f_2^\dagger}{\myid}$. 
  It will help to unfold the definition of these functions.
  Given a flow $\cval$ in $\setnodes\setminus\setnodespp$, they yield the following value for $\anode\in\setnodes\setminus\setnodespp$:
  \begin{align}
  (g\circ\pairingof{f_i^\dagger}{&\myid})(\cval)(\anode)\notag\\
  &=\; \sum_{\anodepp\in\setnodespp}\outflow(f_i^\dagger(\cval))(\anodepp, \anodep)+ \sum_{\anodep\in\setnodes\setminus\setnodespp} \aflowconstraint_i.\edges_{(\anodep, \anode)}(\cval(\anodep))\ + \inflow_{\anode}\ .\label{Equation:InflowDefinition}
  \end{align}
  Here, $\inflow_{\anode}= \sum_{\anodep\in\nat\setminus\setnodes}\inflow(\anodep, \anode)$ as in the definition of transfer functions. 
  Note that the only difference between $(g\circ\pairingof{f_1^\dagger}{\myid})(\cval)$ and $(g\circ\pairingof{f_2^\dagger}{\myid})(\cval)$ is the application of $\outflow\circ f_1^{\dagger}$ and $\outflow\circ f_2^{\dagger}$ to $\cval$, respectively. 
  These functions, however, coincide for $\cval\leq \restrictto{\aflowconstraint_1}{\setnodes\setminus\setnodespp}.\fval$ by Equation~\eqref{Equation:InflowKey}. 

  We now lift the suggested equality of $(g\circ\pairingof{f_1^\dagger}{\myid})(\cval)$ and $(g\circ\pairingof{f_2^\dagger}{\myid})(\cval)$ to the fixed points. 
  By Kleene's theorem, $\mathit{lfp}.g\circ\pairingof{f_i^\dagger}{\myid}$ is the join of the ascending chain
  \begin{align*}
  (g\circ\pairingof{f_i^\dagger}{\myid})^0(\bot)\sqsubseteq(g\circ\pairingof{f_i^\dagger}{\myid})(\bot)\sqsubseteq (g\circ\pairingof{f_i^\dagger}{\myid})^2(\bot)\sqsubseteq \ldots 
  \end{align*}
  with $(g\circ\pairingof{f_i^\dagger}{\myid})^0(\bot)\defeq \bot$ and $(g\circ\pairingof{f_i^\dagger}{\myid})^{j+1}(\bot)\defeq  (g\circ\pairingof{f_i^\dagger}{\myid})[(g\circ\pairingof{f_i^\dagger}{\myid})^j(\bot)]$. 
  We will show that for all $j$, we have $(g\circ\pairingof{f_i^\dagger}{\myid})^j(\bot)\leq \restrictto{\aflowconstraint_1}{\setnodes\setminus\setnodespp}.\fval$. 
  The above reasoning then shows that the sequence of Kleene approximants coincides. 
  This carries over to the join.

  We proceed by induction on $j$. 
  In the base case $j=0$ there is nothing to do as $(g\circ\pairingof{f_i^\dagger}{\myid})^0(\bot)=\bot$.
  To show $(g\circ\pairingof{f_i^\dagger}{\myid})^{j+1}(\bot)\leq \restrictto{\aflowconstraint_1}{\setnodes\setminus\setnodespp}.\fval$, we use the following implication:
  \begin{align*}
  \cval\leq \restrictto{\aflowconstraint_1}{\setnodes\setminus\setnodespp}.\fval\quad\text{implies}\quad(g\circ\pairingof{f_i^\dagger}{\myid})(\cval)\leq \restrictto{\aflowconstraint_1}{\setnodes\setminus\setnodespp}.\fval\ .
  \end{align*} 

  For $g\circ\pairingof{f_1^\dagger}{\myid}$, we use monotonicity of $g\circ\pairingof{f_1^\dagger}{\myid}$ plus $\cval\leq \restrictto{\aflowconstraint_1}{\setnodes\setminus\setnodespp}.\fval$ and the fact that  by Equation~\eqref{Equation:InflowFlow} $\restrictto{\aflowconstraint_1}{\setnodes\setminus\setnodespp}.\fval$ is a fixed point of $g\circ\pairingof{f_1^\dagger}{\myid}$:
  \begin{align*}
  (g\circ\pairingof{f_1^\dagger}{\myid})(\cval)\leq (g\circ\pairingof{f_1^\dagger}{\myid})(\restrictto{\aflowconstraint_1}{\setnodes\setminus\setnodespp}.\fval)=\restrictto{\aflowconstraint_1}{\setnodes\setminus\setnodespp}.\fval\ .
  \end{align*}
  For $g\circ\pairingof{f_2^\dagger}{\myid}$, we proceed similarly but add the observation made above that by Equation~\eqref{Equation:InflowDefinition} combined with Equation~\eqref{Equation:InflowKey} we have $(g\circ\pairingof{f_2^\dagger}{\myid})(\restrictto{\aflowconstraint_1}{\setnodes\setminus\setnodespp}.\fval)=(g\circ\pairingof{f_1^\dagger}{\myid})(\restrictto{\aflowconstraint_1}{\setnodes\setminus\setnodespp}.\fval)$ :  
  \begin{align*}
  (g\circ\pairingof{f_2^\dagger}{\myid})(\cval)&\leq (g\circ\pairingof{f_2^\dagger}{\myid})(\restrictto{\aflowconstraint_1}{\setnodes\setminus\setnodespp}.\fval)\\
  &=
  (g\circ\pairingof{f_1^\dagger}{\myid})(\restrictto{\aflowconstraint_1}{\setnodes\setminus\setnodespp}.\fval)=\restrictto{\aflowconstraint_1}{\setnodes\setminus\setnodespp}.\fval\ .
  \end{align*}
  \qed
\end{proof}

\begin{proof}[of Theorem~\ref{Theorem:Soundness}]
  Let $\afootprint\defeq\mathit{lfp}.\extend{\aflowconstraint_1}{\aflowconstraint_2}\neq\top$, meaning $\afootprint\subseteq\setnodes$.  
  We first show $\restrictto{\aflowconstraint_1}{\afootprint}\ctxequiv\restrictto{\aflowconstraint_2}{\afootprint}$. 
  The sets of nodes coincide by the definition of restriction.
  The inflows coincide by the fact that $\outof{\aflowconstraint_1}{\aflowconstraint_2}\subseteq\afootprint\subseteq\setnodes$, $\tfailof{\aflowconstraint_1}{\aflowconstraint_2}{\afootprint}=\emptyset$, and Lemma~\ref{Lemma:Inflow}. 
  We refer to this inflow as $\inflow_{\afootprint}\defeq\restrictto{\aflowconstraint_1}{\afootprint}.\inflow=\restrictto{\aflowconstraint_2}{\afootprint}.\inflow$. 
  The transfer functions coincide by $\tfailof{\aflowconstraint_1}{\aflowconstraint_2}{\afootprint}=\emptyset$. 
  This means that for all $\anode\in\nat\setminus\afootprint$, for every function $\inflow\leq \inflow_{\afootprint}$, and for all $\anodepp\in\afootprint$ we have $[\transformerof{\restrictto{\aflowconstraint_1}{\afootprint}}(\aninflow)](\anodepp, \anode) = [\transformerof{\restrictto{\aflowconstraint_2}{\afootprint}}(\aninflow)](\anodepp,\anode)$. 
  This, however, is the definition of $\transformerof{\restrictto{\aflowconstraint_1}{\afootprint}}=_{\inflow_{\afootprint}}\transformerof{\restrictto{\aflowconstraint_2}{\afootprint}}$. 

  It remains to show $\restrictto{\aflowconstraint_1}{\setnodes\setminus\afootprint}=\restrictto{\aflowconstraint_2}{\setnodes\setminus\afootprint}$. 
  The sets of nodes again coincide by the definition of restriction. 
  The sets of edges coincide by the fact that $\outof{\aflowconstraint_1}{\aflowconstraint_2}\subseteq\afootprint$ and so $\setnodes\setminus\afootprint\subseteq \setnodes\setminus\outof{\aflowconstraint_1}{\aflowconstraint_2}$. 
  The latter, in turn, is the largest set of nodes on which the edges coincide. 
  To be precise, for every $\anode\in\setnodes\setminus\outof{\aflowconstraint_1}{\aflowconstraint_2}$ and every $\anodepp\in\nat$ we have $\aflowconstraint_1.\edges(\anode, \anodepp)=\aflowconstraint_2.\edges(\anode, \anodepp)$. 
  This, however, is equality of the edge functions, $\aflowconstraint_1.\edges=\aflowconstraint_2.\edges$. 
  The equality is preserved if we restrict the edge functions to $\afootprint\times\nat$. 
  For inflows that stem from nodes outside $\setnodes$, we have equality by the fact that $\aflowconstraint_1.\inflow=\aflowconstraint_2.\inflow$. 
  For $i=1, 2$, consider $\restrictto{\aflowconstraint_i}{\setnodes\setminus\afootprint}.\inflow(\anodepp, \anode)$ with $\anodepp\in\afootprint$. 
  By the definition of restriction, we have 
  \begin{align}
  \restrictto{\aflowconstraint_i}{\setnodes\setminus\afootprint}.\inflow(\anodepp, \anode)=\aflowconstraint_i.\edges_{(\anodepp, \anode)}(\aflowconstraint_i.\fvalof{\anodepp})\ .\label{Equation:SoundnessRestriction}
  \end{align} 
  We have $\aflowconstraint_i.\fvalof{\anodepp} = \restrictto{\aflowconstraint_i}{\afootprint}.\fvalof{\anodepp}$ by Lemma~\ref{Lemma:Restriction}(i). 
  By Lemma~\ref{Lemma:Transfer}(ii), we have 
  \begin{align}
  \restrictto{\aflowconstraint_i}{\afootprint}.\fvalof{\anodepp}=\mathit{lfp}.\restrictto{\aflowconstraint_i}{\afootprint}[\restrictto{\aflowconstraint_i}{\afootprint}.\inflow]\ .\label{Equation:SoundnessFlowLFP}
  \end{align} 
  With Lemma~\ref{Lemma:Transfer}(i), we get 
  \begin{align}
  \aflowconstraint_i.\edges_{(\anodepp, \anode)}(\mathit{lfp}.\restrictto{\aflowconstraint_i}{\afootprint}[\restrictto{\aflowconstraint_i}{\afootprint}.\inflow]) = [\transformerof{\restrictto{\aflowconstraint_i}{\afootprint}}(\restrictto{\aflowconstraint_i}{\afootprint}.\inflow)](\anodepp, \anode)\ .\label{Equation:SoundnessTF}
  \end{align}
  To sum up, the equations yield 
  \begin{align}
  \restrictto{\aflowconstraint_i}{\setnodes\setminus\afootprint}.\inflow(\anodepp, \anode)&\overset{\eqref{Equation:SoundnessRestriction}}{=}\aflowconstraint_i.\edges_{(\anodepp, \anode)}(\aflowconstraint_i.\fvalof{\anodepp})\notag\\
  &\overset{\eqref{Equation:SoundnessFlowLFP}}{=}\aflowconstraint_i.\edges_{(\anodepp, \anode)}(\mathit{lfp}.\restrictto{\aflowconstraint_i}{\afootprint}[\restrictto{\aflowconstraint_i}{\afootprint}.\inflow])\notag\\
  &\overset{\eqref{Equation:SoundnessTF}}{=} [\transformerof{\restrictto{\aflowconstraint_i}{\afootprint}}(\restrictto{\aflowconstraint_i}{\afootprint}.\inflow)](\anodepp, \anode)\ . \label{Equation:SoundnessOverall}
  \end{align} 
  We showed above that $\restrictto{\aflowconstraint_1}{\afootprint}\ctxequiv\restrictto{\aflowconstraint_2}{\afootprint}$ holds, and so we obtain $\restrictto{\aflowconstraint_1}{\afootprint}.\inflow=\restrictto{\aflowconstraint_2}{\afootprint}.\inflow$ as well as 
  $\transformerof{\restrictto{\aflowconstraint_1}{\afootprint}}=_{\restrictto{\aflowconstraint_1}{\afootprint}.\inflow}\transformerof{\restrictto{\aflowconstraint_2}{\afootprint}}$. 
  Combined with Equation~\eqref{Equation:SoundnessOverall}, this yields the desired 
  $\restrictto{\aflowconstraint_1}{\setnodes\setminus\afootprint}.\inflow(\anodepp, \anode)=\restrictto{\aflowconstraint_2}{\setnodes\setminus\afootprint}.\inflow(\anodepp, \anode)$. \qed
\end{proof}

\begin{proof}[of Lemma~\ref{Lemma:CharFullPathReplacement}]
  We start with the implication from left to right.
  Consider nodes $\anode, \anodep\in\setnodes$ and $\anodepp\in\nat\setminus\setnodes$ and a path $\apath\in\setpathof{\aflowconstraint_1}{\anode}{\anodep, \anodepp}$.
  If there is a set of paths suitable for replacement, then the full set
  $\setpath=\setpathof{\aflowconstraint_2}{\anode}{\anodep, \anodepp}$  will do.
  To prove $\edges_{\apath}\leq \edges_{\setpath}$, 
  consider a value~$\amonval\in\amonoid$.
  Let $\inflow^{\amonval}$ be the inflow that maps $\ahvar$ to~$\amonval$ and the remaining nodes to~$0$.
  For the left-hand side of the desired inequality, we have
  \begin{align*}
  \edges_{\apath}(\amonval)\; \leq\;  \sum_{\anode\in\setnodes} \sum_{\apath\in\setpathof{\aflowconstraint_1}{\anode}{\anodep, \anodepp}} \edges_{\apath}(\inflow^{\amonval}(\anode))\;  = \; [\transformerof{\aflowconstraint_1}(\inflow^{\amonval})](\anodep, \anodepp)\ .
  \end{align*}
  The latter equality uses Theorem~\ref{Theorem:ClosedForm}.
  For the right-hand side of the desired inequality, we have
  \begin{align*}
  \edges_{\setpath}(\amonval)\;  &=\;
  \sum_{\apath\in \setpathof{\aflowconstraint_2}{\anode}{\anodep, \anodepp}}
  \edges_{\apath}(\inflow^{\amonval}(\anode))+
  \sum_{\anode'\in\setnodes\setminus\set{\anode}} \sum_{\apath\in\setpathof{\aflowconstraint_2}{\anode'}{\anodep, \anodepp}} \edges_{\apath}(0)\\
  &= \; \sum_{\anode\in\setnodes} \sum_{\apath\in\setpathof{\aflowconstraint_2}{\anode}{\anodep, \anodepp}} \edges_{\apath}(\inflow^{\amonval}(\anode))
  = \; [\transformerof{\aflowconstraint_2}(\inflow^{\amonval})](\anodep, \anodepp)\; .
  \end{align*}
  The first equality is the definition of $\setpath$ and $\atfunof{0}=0$ for all edge functions.
  The second equality rearranges the sums and uses the definition of $\inflow^{\amonval}$.
  The last equality is again Theorem~\ref{Theorem:ClosedForm}.
  Since $[\transformerof{\aflowconstraint_1}(\inflow^{\amonval})](\anodep, \anodepp)=[\transformerof{\aflowconstraint_2}(\inflow^{\amonval})](\anodep, \anodepp)$ by assumption, we can conclude $\edges_{\apath}(\amonval)\leq \edges_{\setpath}(\amonval)$ as required.
  The argumentation for path replacement of $\aflowconstraint_2$ by $\aflowconstraint_1$ is symmetric.

  Assume full path replacement of $\aflowconstraint_1$ by $\aflowconstraint_2$ holds.
  We show $\transformerof{\aflowconstraint_1}\leq \transformerof{\aflowconstraint_2}$.
  The result follows.
  Consider an inflow $\aninflow$, a node $\anodep\in\setnodes$, and a node $\anodepp\in\nat\setminus\setnodes$.
  By Theorem~\ref{Theorem:ClosedForm}, we have
  \begin{align*}
  [\transformerof{\aflowconstraint_1}(\aninflow)](\anodep, \anodepp)\; =\; \sum_{\anode\in\setnodes} \sum_{\apath\in\setpathof{\aflowconstraint_1}{\anode}{\anodep, \anodepp}} \edges_{\apath}(\aninflow_{\anode})\; .
  \end{align*}
  Consider a node $\anode$ and a path $\apath\in\setpathof{\aflowconstraint_1}{\anode}{\anodep, \anodepp}$.
  We show that the following property holds: $\edges_{\apath}(\aninflow_{\anode})\leq [\transformerof{\aflowconstraint_2}(\aninflow)](\anodep, \anodepp)$.
  Then the inequality $[\transformerof{\aflowconstraint_1}(\aninflow)](\anodep, \anodepp)\leq [\transformerof{\aflowconstraint_1}(\aninflow)](\anodep, \anodepp)$ follows by idempotence, as desired.

  By path replacement, there is a set of paths $\setpath\subseteq\setpathof{\aflowconstraint_2}{\anode}{\anodep, \anodepp}$
  that allows us to derive the following:
  \begin{align*}
  \edges_{\apath}(\aninflow_{\anode})\; &\leq \;
  \edges_{\setpath}(\aninflow_{\anode})
   =\;
   \sum_{\apathp\in \setpath}
  \edges_{\apathp}(\aninflow_{\anode})\\
  &\leq \;
  \sum_{\apathp\in \setpathof{\aflowconstraint_2}{\anode}{\anodep, \anodepp}}
  \edges_{\apathp}(\aninflow_{\ahvar})+
  \sum_{\anode'\in\setnodes\setminus\set{\anode}}
  \sum_{\apathp\in\setpathof{\aflowconstraint_2}{\anode'}{\anodep, \anodepp}} \edges_{\apathp}(\inflow_{\anode'})\\
  &= \; [\transformerof{\aflowconstraint_2}(\inflow)](\anodep, \anodepp)\; .
  \end{align*}
  The first inequality is path replacement.
  The following equality is the definition of $\edges_{\setpath}$ for sets of paths.
  The next inequality adds further paths.
  The last equality is Theorem~\ref{Theorem:ClosedForm}.\qed
\end{proof}

\begin{proof}[of Lemma~\ref{Lemma:FullVsPathReplacement}]
    Full path replacement implies path replacement.
    For the reverse direction, we begin by considering nodes $\anode\in\outpof{\aflowconstraint_1}{\aflowconstraint_2}$, $\anodep\in\setnodes$, $\anodepp\in\nat\setminus\setnodes$, and a path $\apath\in\setpathof{\aflowconstraint_1}{\anode}{\anodep, \anodepp}$.
  By the assumption of path replacement, there is a set of paths $\setpath\subseteq\setpathof{\aflowconstraint_2}{\anode}{\anodep, \anodepp}$ with $\edges_{\apath}\leq \edges_{\setpath}$.

  For a node $\anode\in\setnodes\setminus \outpof{\aflowconstraint_1}{\aflowconstraint_2}$ and a path $\apath\in\setpathof{\aflowconstraint_1}{\anode}{\anodep, \anodepp}$, there are two cases.
  Either the path does not visit a node from $ \outpof{\aflowconstraint_1}{\aflowconstraint_2}$, or it does.
  In the former case, the path will still exist in~$\aflowconstraint_2$ and we can replace it by itself, choosing $\setpath=\set{\apath}$.
  In the latter case, we decompose the path into $\apath=\apath_1.\apath_2$ so that $\lastof{\apath_1}=\firstof{\apath_2}$ is the first time the path visits $\outpof{\aflowconstraint_1}{\aflowconstraint_2}$. 
  For~$\apath_2$, path replacement gives us a set of paths $\setpathp\subseteq\setpathof{\aflowconstraint_2}{\firstof{\apath_2}}{\anodep, \anodepp}$ with $\edges_{\apath_2}\leq \edges_{\setpathp}$.
  Now $\apath_1.\setpathp = \setcond{\apath_1.\apathp}{\apathp\in\setpathp}$ is a set of paths in $\aflowconstraint_2$.
    
    We show $\edges_{\apath_1.\setpathp}\geq \edges_{\apath}$ as follows: 
  \begin{align*}
  \edges_{\apath_1.\setpathp}\; &=\; \sum_{\apathp\in \setpathp}\edges_{\apath_1.\apathp}
   =\; \sum_{\apathp\in \setpathp}(\edges_{\apathp}\circ \edges_{\apath_1})
  =\; (\sum_{\apathp\in \setpathp}\edges_{\apathp})\circ \edges_{\apath_1}\\
  &=\; \edges_{\setpathp}\circ \edges_{\apath_1}
  \geq\; \edges_{\apath_2}\circ \edges_{\apath_1}
  =\; \edges_{\apath}\ .
  \end{align*}
  The first equality is the the generalization of edge functions to sets of paths combined with the definition of $\apath_1.\setpathp$. 
  The second equality is the definition of edge functions for paths. 
  The third equality is the fact that sums of functions are evaluated as $(\sum_{\apathp\in \setpathp}\edges_{\apathp})(x)=\sum_{\apathp\in \setpathp}\edges_{\apathp}(x)$.
  The fourth equality is again the definition of edge functions for sets of paths.
  The inequality is due to $\edges_{\apath_2}\leq \edges_{\setpathp}$.
  The last equality is again the definition of edge functions for paths and the decomposition of path~$\apath$.\qed
\end{proof}


	}
\end{document}